\documentstyle[prl,aps,psfig]{revtex}
\def\Bf#1{\mbox{\boldmath{$#1$}}}
\def\bF#1{\mbox{\scriptsize\boldmath{$#1$}}}

\def\Ieq#1#2{{\mathstrut_{#1}^{#2}}}

\def\wt#1{\widetilde{#1}}
\def\wh#1{\widehat{#1}}

\def\tJ{$t$-$J$ }
\begin{document} 
%.\draft
%.\twocolumn[\hsize\textwidth\columnwidth\hsize\csname
%.@twocolumnfalse\endcsname

\title{
On the (anisotropic) uniform metallic ground states of fermions 
interacting through arbitrary two-body potentials in $d$ 
dimensions }

\author{\sc Behnam Farid}

\address{Spinoza Institute, Department of Physics and Astronomy,
University of Utrecht,\\
Leuvenlaan 4, 3584 CE Utrecht, The Netherlands
\footnote{Electronic address: B.Farid@phys.uu.nl.} }

%\date{16 April 2003}
\date{Received 16 April 2003 and accepted 28 July 2003}

\maketitle

\begin{abstract}
\leftskip 54.8pt
\rightskip 54.8pt
We demonstrate that the skeleton of the Fermi surface 
${\cal S}_{{\sc f};\sigma}$ pertaining to a uniform metallic ground 
state (corresponding to fermions with spin index $\sigma$) is 
determined by the Hartree-Fock contribution 
$\Sigma_{\sigma}^{\sc hf}({\Bf k})$ to the dynamic self-energy 
$\Sigma_{\sigma}({\Bf k};\varepsilon)$. That is to say, in order 
for ${\Bf k} \in {\cal S}_{{\sc f};\sigma}$, it is {\sl necessary} 
(but for anisotropic ground states in general {\sl not} sufficient) 
that the following equation be satisfied:
\begin{eqnarray}
\varepsilon_{\bF k} + \hbar \Sigma_{\sigma}^{\sc hf}({\Bf k}) = 
\varepsilon_{\sc f},
\nonumber
\end{eqnarray}
where $\varepsilon_{\bF k}$ stands for the underlying non-interacting 
energy dispersion and $\varepsilon_{\sc f}$ for the exact interacting
Fermi energy. The Fermi surface ${\cal S}_{{\sc f};\sigma}$ consists 
of the set of ${\Bf k}$ points which in addition to satisfying the 
above equation fulfil
\begin{eqnarray}
S_{\sigma}({\Bf k}) = 0,
\nonumber
\end{eqnarray}
where
\begin{eqnarray}
S_{\sigma}({\Bf k}) {:=} \frac{1}{\pi}
\int_0^{\infty} \frac{ {\rm d}\varepsilon}{\varepsilon}\; 
{\rm Im}[\Sigma_{\sigma}({\Bf k};\varepsilon_{\sc f}+\varepsilon) 
+ \Sigma_{\sigma}({\Bf k};\varepsilon_{\sc f}-\varepsilon)]. 
\nonumber
\end{eqnarray} 
The set of ${\Bf k}$ points which satisfy the first of the above 
two equations but fail to satisfy the second constitute the 
{\sl pseudogap} region of the putative Fermi surface of the 
interacting system. We consider the behaviour of the ground-state 
momentum-distribution function ${\sf n}_{\sigma}({\Bf k})$ for 
${\Bf k}$ in the vicinity of ${\cal S}_{{\sc f};\sigma}$ and 
show that, whereas for the uniform metallic ground states of the 
conventional single-band Hubbard Hamiltonian, described in
terms of an on-site interaction, 
${\sf n}_{\sigma}({\Bf k}_{{\sc f};\sigma}^-) \ge \frac{1}{2}$ and
${\sf n}_{\sigma}({\Bf k}_{{\sc f};\sigma}^+) \le \frac{1}{2}$ (here
${\Bf k}_{{\sc f};\sigma}^-$ and ${\Bf k}_{{\sc f};\sigma}^+$ denote 
vectors infinitesimally close to ${\cal S}_{{\sc f};\sigma}$, located
respectively {\sl inside} and {\sl outside} the underlying Fermi sea),
for interactions of non-zero range these inequalities can be violated
(without thereby contravening the stability condition ${\sf n}_{\sigma}
({\Bf k}_{{\sc f};\sigma}^-) \ge {\sf n}_{\sigma}({\Bf k}_{{\sc f};
\sigma}^+)$). This aspect is borne out by the ${\sf n}_{\sigma}
({\Bf k})$ pertaining to the normal states of for instance liquid 
${}^3$He (corresponding to a range of applied pressure) as determined 
by means of quantum Monte Carlo calculations. We further demonstrate 
that for Fermi-liquid metallic states of fermions interacting through 
interaction potentials of non-zero range (e.g. the Coulomb potential), 
the zero-temperature {\sl limit} of ${\sf n}_{\sigma}({\Bf k})$ does 
{\sl not} need to be equal to $\frac{1}{2}$ for ${\Bf k} \in 
{\cal S}_{{\sc f};\sigma}$; this in strict contrast with the uniform 
{\sl Fermi-liquid} metallic states of the single-band Hubbard Hamiltonian 
(if such states at all exist). This aspect should be taken into account 
while analyzing the ${\sf n}_{\sigma}({\Bf k})$ deduced from the 
angle-resolved photoemission data concerning real materials. We
discuss, in the light of the findings of the present work, the 
growing experimental evidence with regard to the `frustration' of the 
kinetic energy of the charge carriers in the normal states of the 
copper-oxide-based high-temperature superconducting compounds. 
\end{abstract}

\vspace{1.5cm}
\noindent
\underline{\sf Preprint number: SPIN-2002/37 }
\vspace{1.0cm}

\noindent\underline{\sf arXiv:cond-mat/0304350}\\
\underline{\sc Philosophical Magazine, 2004, Vol.~84, No.~2, 
109-156}
%
%\pacs{} ]
\narrowtext
\twocolumn
%\vfill
%\pagebreak
%\widetext

% 0.
{\footnotesize
\contentsline {section}
 {\numberline {1.} Introduction}
{\pageref{s0}}
\contentsline {section}
 {\numberline {2.} Preliminaries}
{\pageref{s1}}
\contentsline {section}
 {\numberline {3.} Equations determining the Fermi surfaces of
uniform metallic ground states}
{\pageref{s2}}
\contentsline {section}
 {\numberline {4.} Exposing a further close relationship between
$\Sigma_{\sigma}({\Bf k};\varepsilon_{\sc f})$ and
${\cal S}_{{\sc f};\sigma}$ }
{\pageref{s3}}
\contentsline {section}
 {\numberline {5.} Ground-state exchange and correlation
potentials: a test case}
{\pageref{s4}}
\contentsline {section}
 {\numberline {6.} Behaviour of ${\sf n}_{\sigma}({\Bf k})$ for
${\Bf k}$ infinitesimally close to ${\cal S}_{{\sc f};\sigma}$ }
{\pageref{s5}}
\contentsline {subsection}
 {\numberline {6.1.} The case when $0 <\gamma < 1$, $\; 0 <\tau < 1$}
{\pageref{s5a}}
\contentsline {subsubsection}
 {\numberline {6.1.1.}{\hskip 3mm} $0 <\gamma <\tau < 1$}
{\pageref{s5aa}}
\contentsline {subsubsection}
 {\numberline {6.1.2.}{\hskip 3mm} $0 <\tau <\gamma < 1$}
{\pageref{s5ab}}
\contentsline {subsubsection}
 {\numberline {6.1.3.}{\hskip 3mm} $0 <\tau =\gamma \le 1\;\;$
(includes Fermi liquids)}
{\pageref{s5ac}}
\contentsline {subsection}
 {\numberline {6.2.} The case when $\gamma = 1$, $\; \tau = 1$:
general}
{\pageref{s5b}}
\contentsline {subsubsection}
 {\numberline {6.2.1.}{\hskip 3mm} Considering 
${\sf n}_{\sigma}({\Bf k}_{{\sc f};\sigma}^-) \equiv 
\Lambda_{\sigma}^-/(1 + \Lambda_{\sigma}^-)$}
{\pageref{s5ba}}
\contentsline {subsubsection}
 {\numberline {6.2.2.}{\hskip 3mm} Considering 
${\sf n}_{\sigma}({\Bf k}_{{\sc f};\sigma}^+) \equiv
\Lambda_{\sigma}^+/(1 + \Lambda_{\sigma}^+)$}
{\pageref{s5bb}}
\contentsline {subsubsection}
 {\numberline {6.2.2.1.}{\hskip 6mm} The case when $\;\; d_{\sigma}^+ < 0$}
{\pageref{s5bba}}
\contentsline {subsubsection}
 {\numberline {6.2.2.2.}{\hskip 6mm} The case when $\;\; d_{\sigma}^+ > 0$}
{\pageref{s5bbb}}
\contentsline {subsection}
 {\numberline {6.3.} The case when $\gamma = 1$, $\; \tau = 1$:
Fermi liquids}
{\pageref{s5c}}
\contentsline {section}
 {\numberline {7.} Summary and concluding remarks}
{\pageref{s6}}
\contentsline {section}
 {\numberline {{}} Acknowledgements}
{\pageref{s7}}
\contentsline {section}
 {\numberline {{}} {Appendix~A:}
Estimation of error in $S_{\sigma}(\lowercase{\Bf k})$
within the framework of a finite-order many-body perturbation
theory for $\Sigma_{\sigma}(\lowercase{\Bf k};\varepsilon)$ }
{\pageref{s8}}
\contentsline {section}
 {\numberline {{}} References}
{\pageref{s9}}
}
%.
%\pagebreak

\section*{\S~1.~Introduction}
\label{s0}

Theoretical investigation of anisotropic interacting metallic 
systems is specifically demanding through the possibility, even 
for isotropic interaction potentials, of the deformation of the 
Fermi surface in these systems with respect to the underlying 
non-interacting Fermi surface \cite{KL60,PN64,NO88}. For recent 
approaches to this problem we refer the reader to \cite{DD02,LK02} 
and the references herein. 

Building on recent work \cite{BF02a} concerning the uniform metallic 
ground states (GSs) of the conventional single-band Hubbard Hamiltonian 
\cite{PWA59,R62,MG63,JHI63} for arbitrary spatial dimensions $d$, in 
this work we consider similar states for {\sl arbitrary} two-body
interaction potentials (provided only that these possess Fourier 
transform) corresponding to systems of fermions whose non-interacting 
energy dispersion $\varepsilon_{\bF k}$ can be both bounded (as is the 
case for the single-band Hubbard Hamiltonian) and unbounded from above
(as in the case of the free-fermion model where $\varepsilon_{\bF k} = 
\hbar^2 \|{\Bf k}\|^2/[2 {\sf m}]$, in which ${\sf m}$ stands for the 
fermion bare mass). Our results, of which a number we have presented 
in the abstract, make explicit some of the fundamental differences 
between systems in which the interaction potential $v({\Bf r}-{\Bf r}')$ 
is of contact type, of strictly bounded and of unbounded range. 
Following the general considerations leading to the equations defining 
${\cal S}_{{\sc f};\sigma}$, presented above, in this paper we deduce 
the most general expression for ${\sf n}_{\sigma}({\Bf k})$ in the close 
vicinity of ${\cal S}_{{\sc f};\sigma}$ and for specific cases deal 
with this expression in considerable detail. Amongst others, we show
that, in contrast with the case of the conventional single-band Hubbard 
Hamiltonian, ${\sf n}_{\sigma}({\Bf k})$ can become smaller (greater) 
than $\frac{1}{2}$ for ${\Bf k}$ inside (outside) the Fermi sea and 
infinitesimally close to ${\cal S}_{{\sc f};\sigma}$, without
infringing on the condition ${\sf n}_{\sigma}({\Bf k}_{{\sc f};\sigma}^-)
-{\sf n}_{\sigma}({\Bf k}_{{\sc f};\sigma}^+) \ge 0$ implied by the 
assumption of the stability of the GS of the system under consideration. 
This aspect is significant in that in cases where ${\sf n}_{\sigma}
({\Bf k}_{{\sc f};\sigma}^-) < \frac{1}{2}$, by continuity ${\sf n}_{\sigma}
({\Bf k})$ is expected to be more suppressed for ${\Bf k}$ {\sl inside} 
the Fermi sea (at least in the vicinity of ${\cal S}_{{\sc f};\sigma}$) 
in comparison with cases where ${\sf n}_{\sigma}({\Bf k}_{{\sc f};
\sigma}^-) \ge \frac{1}{2}$, implying thus an increase in the 
contribution of the correlated kinetic energy $\sum_{{\bF k},\sigma} 
\varepsilon_{\bF k} {\sf n}_{\sigma}({\Bf k})$ to the GS total energy 
of the system under consideration (see later). In this connection, note 
that $\sum_{{\bF k},\sigma} {\sf n}_{\sigma}({\Bf k})$ is equal to the 
total number $N$ of particles in the GS and therefore independent of 
the magnitude of the coupling constant of interaction.

For a number of reasons of contemporary interest, a reliable
description of ${\sf n}_{\sigma}({\Bf k})$ seems timely and 
desirable. Foremost amongst these are investigations regarding 
the determination of the mechanism of superconductivity in the 
copper-oxide-based high-temperature superconducting materials
(hereafter referred to as {\sl cuprates} or {\sl cuprate compounds}); 
for recent reviews see \cite{OM00,CEKO02,NP03}. Both the earlier 
angle-resolved photoemission results \cite{SWF98} (see in particular 
Fig.~3(F) herein) and the recent in-plane optical-conductivity data 
concerning Bi$_2$Sr$_2$CaCu$_2$O$_{8+\delta}$ \cite{SS01,MPMKL02} (see 
also \cite{MMPS03}) reveal {\sl decrease} in the in-plane kinetic 
energy of the system upon entering into the superconducting state, 
by an amount commensurate with the expected superconducting 
condensation energy. 
\footnote{\label{f1}
This aspect has been explicitly quantified in 
\protect\cite{SS01,MPMKL02,MMPS03}. }
A similar trend has been observed in the measured out-of-plane (i.e. 
the $c$-axis) optical conductivities of Tl$_2$Ba$_2$CuO$_{6+x}$, 
La$_{2-x}$Sr$_x$CuO$_4$, and YBa$_2$Cu$_3$O$_{6.6}$ \cite{BAS99}. 
\footnote{\label{f2}
The `interlayer tunnelling' theory by Anderson and co-workers 
\protect\cite{WHA88,PWA97} (see also \protect\cite{SC98}) ascribes 
the superconducting condensation energy wholly to the reduction 
in the {\sl kinetic} energy of the cuprate compounds in consequence of 
the possibility of coherent inter-layer tunnelling of the {\sl paired} 
electrons in the superconducting states, a process deemed infeasible 
in the normal states of these materials (owing to the incoherent 
nature of the underlying single-particle excitations). Although, 
among other things (see \protect\cite{PWA98} and 
\protect\cite{PWA97}), the measured $c$-axis optical conductivity 
in Tl$_2$Ba$_2$CuO$_{6+x}$, La$_{2-x}$Sr$_x$CuO$_4$, and 
YBa$_2$Cu$_3$O$_{6.6}$ \cite{BAS99} may be interpreted within the 
framework of the inter-layer tunnelling theory \protect\cite{BAS99} 
(see however \protect\cite{IM99,IM00}), some experiments 
\protect\cite{MKH98,TMMK98} appear \protect\cite{BF98} (see also 
\protect\cite{MK99}) not to support this theory; see however 
\protect\cite{CKA99}, \protect\cite{PCX00} (and \protect\cite{MLLK02}).
The significance to pairing of the {\sl in-plane} kinetic-energy 
frustration in the normal states of the cuprate compounds has been 
put forward and established by the model considerations of Hirsch 
and Marsiglio \protect\cite{JEHM89,JEHM00} and Hirsch 
\protect\cite{JEH92a,JEH92b} (see also \protect\cite{JEH02}). }
Assuming the sufficiency of the BCS variational wavefunction to 
describe the superconducting states of the cuprate compounds (an
assumption that has been widely anticipated), 
\footnote{\label{f3}
According to this assumption, the zero-temperature BCS occupation 
fraction $v_{\bF k}^2 \equiv 1-u_{\bF k}^2$ (see \cite{MT96}, 
Chapter 3 and in particular Fig.~3.1) plays the same role as 
${\sf n}_{\sigma}({\Bf k})$ does in relation to the underlying 
normal GS. For completeness, within the framework of the conventional 
BCS theory, and at zero temperature, the ${\sf n}_{\sigma}({\Bf k})$ 
pertaining to the normal GS coincides with the unit-step function, 
equal to unity for ${\Bf k}$ inside the Fermi sea and to zero for 
${\Bf k}$ outside. Consequently, at zero temperature the BCS state 
accommodates a {\sl higher} kinetic energy than the underlying 
uncorrelated normal GS (for a quantitative consideration see \S~3.4 
in \protect\cite{MT96}). Note in passing that Chester's 
\protect\cite{GVC56} findings do {\sl not} apply to a pairing, or 
reduced, Hamiltonian such as the one employed in the conventional 
BCS theory. }
one readily observes that ${\sf n}_{\sigma}({\Bf k}_{{\sc f};\sigma}^-) 
< \frac{1}{2}$, with ${\Bf k}_{{\sc f};\sigma} \in {\cal S}_{{\sc f};
\sigma}$, in combination with the assumption of continuity of 
${\sf n}_{\sigma}({\Bf k})$ for ${\Bf k}$ inside the Fermi sea
and in the vicinity of ${\cal S}_{{\sc f};\sigma}$, is tantamount to
an excess kinetic energy in the normal GS of the underlying system in 
comparison with the kinetic energy in the paired state (see the previous 
paragraph). 

The above-indicated experimental results on the one hand and the 
feasibility of ${\sf n}_{\sigma}({\Bf k}_{{\sc f};\sigma}^-) < 
\frac{1}{2}$ for the metallic states of fermions interacting through 
potentials of at least finite range on the other clearly signify the 
likelihood of the existence of a crucial relationship between the latter 
potentials and the unconventional properties of the metallic states 
of the cuprate compounds (see also \cite{AJL99,MTL03}). In this 
connection it is relevant to mention that it has been rigorously shown 
\cite{SS98} that for $d=2$ the conventional single-band Hubbard 
Hamiltonian does {\sl not} support ${\rm d}_{x^2-y^2}$ long-range 
pairing order at non-zero temperatures, 
\footnote{\label{f4}
The superconducting order parameter in the cuprate compounds is 
generally considered to have the ${\rm d}_{x^2-y^2}$ symmetry
\protect\cite{OM00,CEKO02,NP03}. Note in passing that in practice the 
inter-planar coupling of the electronic degrees of freedom is capable 
of stabilizing a ${\rm d}$-wave superconducting state at finite 
temperatures, such as obtained within the framework of the so-called 
`cluster dynamical mean-field' approximation \protect\cite{LK00,MJPK00} 
(see also \protect\cite{GS01,MJP01}). }
an aspect strongly anticipated by earlier Monte Carlo results (which 
incidentally also do {\sl not} support extended ${\rm s}$-wave pairing) 
on finite systems \cite{HL88,MIH89,MI91,AM92,ZCG97}. Further, the 
view is increasingly gaining prominence that the long range of the 
Coulomb potential, such as taken account of by means of a variety of 
`extended' Hubbard (and associated \tJ$\!\!$) Hamiltonians 
\cite{MRRT88,JEH90,HC90,JEH92b,XZY93,DR94,JEH97,CG97,YY98,SCCG98,%
MM00,JEHM00,HWL01,VV01,AHH02,CNP02,KK03}, plays a crucial role in 
stabilizing one from amongst a number of competing orders in the 
cuprate compounds over some relevant range of parameters. Of these, 
one concerns the periodic (but incommensurate with respect to the 
underlying lattice) ordering of spin and charge (also referred to 
as spin-charge {\sl stripes}; for general reviews see 
\cite{JZ99,NCMS03} and the references herein), which appears to 
compete against macroscopic phase separation \cite{EKL90}. This 
statement requires some qualification. Macroscopic phase separation 
is believed not to concern the Hubbard Hamiltonian \cite{ED94,WS00}, 
but solely the \tJ Hamiltonian. However, even for the \tJ Hamiltonian 
the actual situation is not firmly established; on the one hand, 
Green-function Monte Carlo results (concerning finite systems) have 
been interpreted as implying phase separation at hole dopings and 
interaction strengths appropriate to the cuprate superconductors 
\cite{HM97}, while, on the other hand, subsequent numerical 
calculations \cite{WS00} based on `density-matrix renormalization 
group' (DMRG) method performed on larger systems have upheld the 
viewpoint that the `stripes' state would have a lower energy than 
the phase-separated state. In spite of these opposing observations,
it is generally believed that the sensitivity of the results 
corresponding to the \tJ Hamiltonian on such details as, for instance, 
the shape and size of the clusters on which computations are carried 
out does not warrant an {\it a priori} neglect of the long range of 
the Coulomb interaction \cite{HWL01,AHH02,CNP02} (see also \cite{LL97}). 
Later in this paper we shall discuss and document the significance 
of a non-contact-type interaction in accurately describing the normal 
liquid state of ${}^3$He.

Using the experimental angle-resolved photoemission data it is 
possible to determine ${\sf n}_{\sigma}({\Bf k})$ for ${\Bf k}$ in 
the vicinity of ${\cal S}_{{\sc f};\sigma}$, as has been reported 
in \cite{CGD99} for Bi$_2$Sr$_2$CaCu$_2$O$_{8+\delta}$ (Bi2212) and 
Bi$_2$Sr$_2$Cu$_1$O$_{6+\delta}$ (Bi2201), in both the overdoped and 
the underdoped regime. In \cite{CGD99} the Fermi surface has been 
asserted as being the locus of the ${\Bf k}$ points at which, firstly,
the (experimental) single-particle spectral function $A_{\sigma}({\Bf k};
\varepsilon)$ is peaked for $\varepsilon=\varepsilon_{\sc f}$ and, 
secondly, ${\sf n}_{\sigma}({\Bf k})$ takes the value $\frac{1}{2}$
(see \cite{REA95}); 
\footnote{\label{f5}
Explicitly, in \protect\cite{CGD99} the authors introduced three 
criteria to be obeyed ``at a true FS [Fermi surface] crossing'',
of which the second reads: ``(2) $n(k)$ [${\sf n}_{\sigma}({\Bf k})$]
should be near 50\% of its maximal value, or equivalently the maximal 
gradient point''. Elsewhere these workers stated: ``a FS [Fermi
surface] crossing should occur when $n(k)$ loses about half of its 
maximum value (excluding the background)''. As we indicate in the 
main text, these workers fixed the scale of the experimentally 
determined ${\sf n}_{\sigma}({\Bf k})$ by identifying with unity 
the maximum value of the latter function over all the ${\Bf k}$ 
points probed experimentally. }
in \cite{CGD99} the absolute value of ${\sf n}_{\sigma}({\Bf k})$ 
is determined by equating with unity the maximum value achieved 
by the experimental ${\sf n}_{\sigma}({\Bf k})$, $\forall {\Bf k}$, 
a procedure which evidently adds uncertainty to the experimentally 
determined ${\sf n}_{\sigma}({\Bf k})$. As demonstrated in \cite{BF02a}, 
and in view of the findings in the present work, the requirement 
${\sf n}_{\sigma}({\Bf k}) = \frac{1}{2}$ for ${\Bf k} \in 
{\cal S}_{{\sc f};\sigma}$ is {\sl not} justified in general; 
\footnote{\label{f6}
For cases in which ${\sf n}_{\sigma}({\Bf k})$ is discontinuous at 
${\Bf k}\in {\cal S}_{{\sc f};\sigma}$, in the present context the 
value assigned to ${\sf n}_{\sigma}({\Bf k})$ for ${\Bf k}\in 
{\cal S}_{{\sc f};\sigma}$ is the arithmetic mean of ${\sf n}_{\sigma}
({\Bf k}')$ for ${\Bf k}'$ approaching ${\Bf k}$ from inside and 
outside the Fermi sea. This mean value coincides with the 
zero-temperature {\sl limit} of ${\sf n}_{\sigma}({\Bf k})$ for 
${\Bf k}\in {\cal S}_{{\sc f};\sigma}$ \protect\cite{BF02a}. }
firstly, within the framework of the conventional single-band Hubbard 
model, the latter is solely a {\sl necessary} (and {\sl not} sufficient) 
condition for metallic states to be Fermi-liquid (FL) and, secondly, 
for metallic states of fermions interacting through potentials of 
non-zero range the condition ${\sf n}_{\sigma}({\Bf k}) = \frac{1}{2}$ 
for ${\Bf k} \in {\cal S}_{{\sc f};\sigma}$ is {\sl not} demanded even 
for FL metallic states. In the light of these and neglecting for the 
moment the above-mentioned uncertainty concerning the absolute values 
of the experimentally determined ${\sf n}_{\sigma}({\Bf k})$, the results 
depicted in Fig.~2 of \cite{CGD99} have far-reaching implications. For 
instance, these results clearly show that a contact-type interaction 
{\sl cannot} be capable of reliably describing Bi2212 and Bi2201 in 
the over-doped and the underdoped regimes (at {\sl no} peak position
of $A_{\sigma}({\Bf k};\varepsilon_{\sc f})$ does ${\sf n}_{\sigma}
({\Bf k})$ acquire a value reasonably close to $\frac{1}{2}$); a number 
of these results further show that the pertinent single-particle spectral 
functions $A_{\sigma}({\Bf k};\varepsilon)$, corresponding to the 
systems considered, are most probably free from quasi-particle 
$\delta$-function contributions (establishing the breakdown of the 
FL state) at regions of ${\cal S}_{{\sc f};\sigma}$ neighbouring the 
$\overline{M}$ points of the underlying Brillouin zones (compare for 
instance Figs.~2(g) and 2(i) in \cite{CGD99} with the results in 
Eq.~(\ref{e109}) below). 

The organization of this paper is as follows. In \S~2 we introduce the 
Hamiltonian on which the considerations in this paper are based; this
Hamiltonian, which is expressed in terms of a single spin-degenerate 
energy dispersion $\varepsilon_{\bF k}$, is defined on the continuum; 
however, through some minor adjustments, which we explicitly specify, 
this reduces to the conventional single-band Hubbard Hamiltonian, defined 
on a lattice. In \S~3 we present the details underlying the equations 
from which ${\cal S}_{{\sc f};\sigma}$ pertaining to the uniform 
GSs of the Hamiltonian introduced in \S~2 is determined. In \S~4 
we expose an intimate relationship between ${\cal S}_{{\sc f};
\sigma}$ and $\Sigma_{\sigma}^{\sc hf}({\Bf k})$ mediated through the 
single-particle density matrix $\varrho_{\sigma}({\Bf r},{\Bf r}')$ 
which for uniform GSs, where the latter function depends on ${\Bf r}
-{\Bf r}'$, coincides with the Fourier transform of the GS momentum 
distribution function ${\sf n}_{\sigma}({\Bf k})$; the information 
with regard to ${\cal S}_{{\sc f};\sigma}$, through the singular 
nature of ${\sf n}_{\sigma}({\Bf k})$ for ${\Bf k} \in 
{\cal S}_{{\sc f};\sigma}$, is shown to be already fully reflected in 
the leading term in the asymptotic series of $\varrho_{\sigma}
({\Bf r},{\Bf r}')$ for $\|{\Bf r}-{\Bf r}'\| \to\infty$. In \S~5 we 
examine the equation for ${\cal S}_{{\sc f};\sigma}$ in terms of 
$\Sigma_{\sigma}^{\sc hf}({\Bf k})$ by applying this equation to an 
isotropic metallic GS for which ${\cal S}_{{\sc f};\sigma}$ is 
rotationally invariant and is fully determined by the concentration 
of the fermions in the system. By doing so we deduce expressions for 
the exchange potential $\mu_{\rm x;\sigma}$ and correlation potential 
$\mu_{\rm c;\sigma}$, as encountered within the framework 
of the GS density-functional theory \cite{HK64,KS65,DG90}. Our 
expression for $\mu_{\rm x;\sigma}$ identically reproduces the exact 
$\mu_{\rm x;\sigma}$ corresponding to the Coulomb-interacting 
uniform-electron-gas system. Making use of ${\sf n}_{\sigma}^{\sc rpa}
({\Bf k})$, the random-phase approximation (RPA) to ${\sf n}_{\sigma}
({\Bf k})$, which for systems of Coulomb-interacting fermions 
asymptotically approaches the exact ${\sf n}_{\sigma}({\Bf k})$
in the high-density regime, we calculate $\mu_{\rm c;\sigma}$ which 
in the latter regime up to an unimportant additive constant reproduces 
the $\mu_{\rm c;\sigma}$ deduced from quantum Monte-Carlo calculations. 
In \S~6 we consider the behaviour of ${\sf n}_{\sigma}({\Bf k})$ for 
${\Bf k}$ in the vicinity of ${\cal S}_{{\sc f};\sigma}$, both inside 
and outside the Fermi sea. Rather than being exhaustive, in this 
Section we attempt to contrast the important consequences that the 
range of interaction can have on the behaviour of ${\sf n}_{\sigma}
({\Bf k})$ in the mentioned region of the ${\Bf k}$ space. In \S~7 we 
present a summary of the main aspects of this paper. Here we also 
briefly relate our theoretical findings with some available results 
concerning correlated fermion systems, specifically liquid ${}^3$He 
in the normal state. Finally, in appendix A which follows the main 
body of this paper we present an analysis which exposes the nature 
of inaccuracy in $S_{\sigma}({\Bf k})$ as calculated on the basis of 
a finite-order perturbation series for the self-energy operator. For 
reasons that will be clarified in appendix A, the defining 
expression for ${\cal S}_{{\sc f};\sigma}$ as introduced in this 
paper (i.e. Eq.~(\ref{e53}) below) more reliably reproduces
${\cal S}_{{\sc f};\sigma}$ than the conventional defining 
expression (i.e. Eq.~(\ref{e45})).

\section*{\S~2.~Preliminaries}
\label{s1}

In a recent work \cite{BF02a} we have in some detail considered the 
uniform metallic GSs of the single-band Hubbard Hamiltonian. Here 
we extend our investigations by considering uniform metallic GSs of 
the following Hamiltonian
\begin{equation}
\label{e1}
\wh{H} = \wh{H}_0 + {\sf g}\,\wh{\sf H}_1,
\end{equation}
where
\begin{equation}
\label{e2}
\wh{H}_0 = 
\sum_{{\bF k},\sigma} 
\varepsilon_{{\bF k}}\,
{\hat c}_{{\bF k};\sigma}^{\dag}
{\hat c}_{{\bF k};\sigma},
\end{equation}
\begin{equation}
\label{e3}
\wh{\sf H}_1 =
\frac{1}{2 \Omega} \sum_{\sigma,\sigma'}\,
\sum_{{\bF k},{\bF p},{\bF q}} 
{\bar w}(\| {\Bf q}\|)\, 
{\hat c}_{{\bF k}+{\bF q};\sigma}^{\dag}
{\hat c}_{{\bF p}-{\bF q};\sigma'}^{\dag}
{\hat c}_{{\bF p};\sigma'}
{\hat c}_{{\bF k};\sigma}.
\end{equation}
In Eq.~(\ref{e1}), ${\sf g}$ (which has the dimension of energy) stands 
for the coupling-constant of interaction, $\varepsilon_{\bF k}$ for a 
spin-degenerate single-particle energy dispersion (which may be equated 
with $\hbar^2 \| {\Bf k}\|^2/[2 {\sf m}]$, the energy dispersion of 
non-interacting free fermions with mass ${\sf m}$), ${\hat c}_{{\bF k};
\sigma}^{\dag}$ and ${\hat c}_{{\bF k};\sigma}$ for the canonical 
creation and annihilation operators respectively for fermions with 
spin index $\sigma$, ${\sf g}\, {\bar w}(\|{\Bf q}\|) \equiv 
{\bar v}(\|{\Bf q}\|)$ for the Fourier transform of the two-body 
interaction potential $v({\Bf r}-{\Bf r}')$ (assumed to be isotropic), 
and $\Omega = L^d$ for the volume of the (macroscopic) $d$-dimensional 
hyper-cubic box occupied by the system. The wave-vector sums in 
Eqs.~(\ref{e2}) and (\ref{e3}) are over a regular lattice (the 
underlying lattice constant being equal to $2\pi/L$) covering in 
principle the entire reciprocal space. The Hamiltonian in Eq.~(\ref{e1}) 
can be fruitfully employed to investigate the consequences of interaction 
and interplay between this and anisotropy (as implied by 
$\varepsilon_{\bF k}$) on the observable quantities in many-particle 
systems.

On effecting the replacements
\begin{equation}
\label{e4}
{\sf g} \rightharpoonup U,\;\;\;\;
{\bar w}(\| {\Bf q}\|) \rightharpoonup 
\frac{\Omega}{N_{\sc l}},
\end{equation}
and restricting the wave-vector sums to $N_{\sc l}$ wavevectors 
uniformly distributed over the first Brillouin zone (1BZ) associated 
with the Bravais lattice spanned by $\{ {\Bf R}_{j} \| j=1, \dots,
N_{\sc l} \}$, the Hamiltonian in Eq.~(\ref{e1}) goes over into the 
conventional single-band Hubbard Hamiltonian \cite{PWA59,R62,MG63,JHI63}
$\wh{\cal H}$ corresponding to the on-site interaction energy $U$; 
in cases where ${\Bf k}+{\Bf q}$ and ${\Bf p}-{\Bf q}$ on the
right-hand side (RHS) of Eq.~(\ref{e3}) lie outside the 1BZ, these 
vectors are to be identified with the vectors inside the 1BZ obtained 
from ${\Bf k}+{\Bf q}$ and ${\Bf p}-{\Bf q}$ through Umklapp processes.
On relaxing the replacements in Eq.~(\ref{e4}), while maintaining
the above-mentioned restrictions concerning the wave vectors,
the Hamiltonian in Eq.~(\ref{e1}) coincides with an `extended' 
Hubbard Hamiltonian.

Along the lines considered in \cite{BF02a}, it can be shown that
for the Fermi surface ${\cal S}_{{\sc f};\sigma}$, pertaining to the
uniform metallic GS of $\wh{H}$, one has
\begin{equation}
\label{e5}
{\cal S}_{{\sc f};\sigma} =
\{ {\Bf k}\,\|\, 
(\varepsilon_{\sc f} - \varepsilon_{{\bF k};\sigma}^{<})
(\varepsilon_{\sc f}^+ - \varepsilon_{{\bF k};\sigma}^{>}) = 0\},
\end{equation}
where $\varepsilon_{\sc f}^+ =\varepsilon_{\sc f} + 0^+$
and
\begin{equation}
\label{e6}
\varepsilon_{{\bF k};\sigma}^{\Ieq<>}
= \varepsilon_{\bF k} + {\sf g}\,
\frac{ \beta_{{\bF k};\sigma}^{\Ieq<>} }
{ \nu_{\sigma}^{\Ieq<>}({\Bf k}) }.
\end{equation}
On the basis of similar arguments as presented in \cite{BF02a}, it can 
be shown that $(\varepsilon_{\sc f} - \varepsilon_{{\bF k};\sigma}^{<})
=0$ implies that $(\varepsilon_{\sc f}^+ - \varepsilon_{{\bF k};
\sigma}^{>})=0$, and vice versa so that, for ${\Bf k} \in 
{\cal S}_{{\sc f}; \sigma}$, $\varepsilon_{{\bF k};\sigma}^{<}$ 
and $\varepsilon_{{\bF k};\sigma}^{>}$ are up to an infinitesimal
difference equal. In Eq.~(\ref{e6}),
\begin{eqnarray}
\label{e7}
\nu_{\sigma}^{\Ieq><}({\Bf k})
{:=} \left\{ \begin{array}{l}
{\sf n}_{\sigma}({\Bf k}), \\
1-{\sf n}_{\sigma}({\Bf k}),
\end{array} \right.
\end{eqnarray}
\begin{eqnarray}
\label{e8}
&&\beta_{{\bF k};\sigma}^{<}
= \frac{-1}{\Omega} \sum_{\sigma'}
\sum_{{\bF p}',{\bF q}'} {\bar w}(\|{\Bf q}'\|)
\nonumber\\
&&\;\;\;\;\;\;
\times\langle\Psi_{N;0}\vert
{\hat c}_{{\bF k};\sigma}^{\dag}
{\hat c}_{{\bF p}'+{\bF q}';\sigma'}^{\dag}
{\hat c}_{{\bF k}+{\bF q}';\sigma}
{\hat c}_{{\bF p}';\sigma'}
\vert\Psi_{N;0}\rangle,
\end{eqnarray}
\begin{equation}
\label{e9}
\beta_{{\bF k};\sigma}^{>} = \vartheta_{{\bF k};\sigma}
- \beta_{{\bF k};\sigma}^{<},
\end{equation}
where $\vert\Psi_{N;0}\rangle$ stands for the $N$-particle GS
of $\wh{H}$, 
\begin{equation}
\label{e10}
{\sf n}_{\sigma}({\Bf k}) {:=}
\langle\Psi_{N;0}\vert 
{\hat c}_{{\bF k};\sigma}^{\dag}
{\hat c}_{{\bF k};\sigma} \vert\Psi_{N;0}\rangle
\end{equation}
for the GS momentum distribution function, and
\begin{equation}
\label{e11}
\vartheta_{{\bF k};\sigma} {:=}
-\langle\Psi_{N;0}\vert
\big[ {\hat c}_{{\bF k};\sigma}^{\dag},
[ \wh{\sf H}_1, {\hat c}_{{\bF k};\sigma} ]_{-} \big]_{+}
\vert\Psi_{N;0}\rangle,
\end{equation}
in which $[{\;},{\;}]_{-}$ and $[{\;},{\;}]_{+}$ denote commutation 
and anti-commutation respectively. Making use of the canonical 
anti-commutation relations for ${\hat c}_{{\bF k};\sigma}^{\dag}$ 
and ${\hat c}_{{\bF k};\sigma}$, one readily obtains
\begin{eqnarray}
\label{e12}
\vartheta_{{\bF k};\sigma} &=&
\frac{N}{\Omega}\, {\bar w}(0) - 
\frac{1}{\Omega} \sum_{{\bF p}}
{\bar w}(\|{\Bf p}\|) {\sf n}_{\sigma}({\Bf k}+{\Bf p})
\nonumber\\ 
& &\;\;\;\;\;\;\;\;\;\;\;\;\;\;\;\;\;\;\;\;\;\;\;\;\;
\mbox{\rm (for bounded ${\bar w}(0)$)}.
\end{eqnarray}
Note that $N/\Omega {=:} n_0$ is the total concentration of the 
fermions in the $N$-particle GS of $\wh{H}$. For the Hubbard 
Hamiltonian, making use of the prescription in Eq.~(\ref{e4}) and 
the fact that $\sum_{\bF p} {\sf n}_{\sigma}({\Bf k}+{\Bf p})=
N_{\sigma}$ (recall that in the case of the Hubbard or an
extended Hubbard Hamiltonian the latter sum is restricted to 
points inside the 1BZ), one obtains \cite{BF02a}
\begin{equation}
\label{e13}
\vartheta_{{\bF k};\sigma} = n - n_{\sigma} \equiv n_{\bar\sigma}
\;\;\;\mbox{\rm (Hubbard Hamiltonian)},
\end{equation}
where 
\begin{equation}
\label{e14}
n_{\sigma} {:=} \frac{N_{\sigma}}{N_{\sc l}}
\end{equation}
stands for the number of fermions with spin index $\sigma$ per
site in the uniform $N$-particle GS of the Hubbard Hamiltonian;
thus $n {:=} N/N_{\sc l}$. In Eq.~(\ref{e13}), $\bar\sigma$ denotes 
the spin index complementary to $\sigma$, that is for $\sigma=
\uparrow$, $\bar\sigma=\downarrow$ and vice versa.

In cases where ${\bar w}(\|{\Bf q}\|)$ is unbounded for $\|{\Bf q}\| 
=0$ (such as is the case for the two-body Coulomb potential), the 
expression in Eq.~(\ref{e12}) has to be regularized; this is achieved 
by formally taking into account the interaction of the fermions with 
a positively charged uniform background through supplementing the 
Hamiltonian in Eq.~(\ref{e1}) with ${\sf g}\,\wh{\sf H}'$, where
\begin{equation}
\label{e15}
\wh{\sf H}' {:=} -n_0\, {\bar w}(\kappa) \wh{N},\;\;\;
\kappa\downarrow 0,
\end{equation}
in which $\wh{N}$ stands for the total number operator. Following
this, one obtains
\begin{eqnarray}
\label{e16}
\vartheta_{{\bF k};\sigma} &=&
- \frac{1}{\Omega} \sum_{\bF p}'
{\bar w}(\|{\Bf p}\|) {\sf n}_{\sigma}({\Bf k}+{\Bf p}),
\nonumber\\ 
& &\;\;\;\;\;\;\;\;\;\;\;\;\;\;\;\;\;\;\;\;\;\;\;\;\;
\mbox{\rm (for unbounded ${\bar w}(0)$)},
\end{eqnarray} 
where $\sum_{\bF p}'$ leaves out ${\Bf p}={\Bf 0}$. For systems in the 
thermodynamic limit, sums of the form $\Omega^{-1}\sum_{\bF q}$ 
(or $\Omega^{-1}\sum_{\bF q}'$) can be replaced by $(2\pi)^{-d} \int 
{\rm d}^dq$. In any of the above cases, whether one deals 
with the Hubbard Hamiltonian or with a more general Hamiltonian,
from the expressions in Eqs.~(\ref{e12}) and (\ref{e16}) it is 
easily verified that
\begin{equation}
\label{e17}
\vartheta_{{\bF k};\sigma} \equiv \frac{\hbar}{\sf g}\, 
\Sigma_{\sigma}^{\sc hf}({\Bf k}),
\end{equation}
where $\Sigma_{\sigma}^{\sc hf}({\Bf k})$ stands for the
Hartree-Fock self-energy. From Eqs.~(\ref{e6}), (\ref{e7}),
(\ref{e9}) and (\ref{e17}), one readily deduces that ({\it cf}. 
Eq.~(56) in \cite{BF02a})
\begin{equation}
\label{e18}
{\sf n}_{\sigma}({\Bf k}) \varepsilon_{{\bF k};\sigma}^{<}
+ \big(1 - {\sf n}_{\sigma}({\Bf k})\big) 
\varepsilon_{{\bF k};\sigma}^{>} =
\varepsilon_{\bF k} + \hbar\Sigma_{\sigma}^{\sc hf}({\Bf k}).
\end{equation}
This result will prove useful in our later considerations in
this paper.

By introducing the decomposition
\begin{equation}
\label{e19}
\beta_{{\bF k};\sigma}^{<} \equiv {\sf n}_{\sigma}({\Bf k})\,
\xi_{{\bF k};\sigma},
\end{equation}
which implies no restriction so long as ${\sf n}_{\sigma}({\Bf k})
\not=0$, one can express $\varepsilon_{{\bF k};\sigma}^{<}$, 
introduced in Eq.~(\ref{e6}) above, as follows:
\begin{equation}
\label{e20}
\varepsilon_{{\bF k};\sigma}^{<} = \varepsilon_{\bF k}
+ {\sf g}\, \xi_{{\bF k};\sigma}.
\end{equation}
From this, Eq.~(\ref{e6})  and the results in Eqs.~(\ref{e18}) 
and (\ref{e17}), one similarly has
\begin{equation}
\label{e21}
\varepsilon_{{\bF k};\sigma}^{>} = \varepsilon_{\bF k}
+ {\sf g}\,
\frac{ \vartheta_{{\bF k};\sigma} - {\sf n}_{\sigma}({\Bf k})\,
\xi_{{\bF k};\sigma} }
{1 - {\sf n}_{\sigma}({\Bf k}) }.
\end{equation}

From the expression in Eq.~(\ref{e5}) above and the ensuing details 
(see the text subsequent to Eq.~(\ref{e6})) one infers the following 
relationships from the expressions in Eqs.~(\ref{e20}) and (\ref{e21}) 
(for ${\Bf k}_{{\sc f};\sigma} \in {\cal S}_{{\sc f};\sigma}$) 
\begin{eqnarray}
\label{e22}
&&\xi_{{\bF k};\sigma} \sim \frac{1}{\sf g}
(\varepsilon_{\sc f} - \varepsilon_{{\bF k}_{{\sc f};\sigma}}),\;\;
{\Bf k} \to {\Bf k}_{{\sc f};\sigma},\\
\label{e23}
&&\vartheta_{{\bF k};\sigma} \sim \frac{1}{\sf g}
(\varepsilon_{\sc f} - \varepsilon_{{\bF k}_{{\sc f};\sigma}}),\;\;
{\Bf k} \to {\Bf k}_{{\sc f};\sigma}.
\end{eqnarray}
In other words, to the leading order in $({\Bf k}
-{\Bf k}_{{\sc f};\sigma})$, one has
\begin{equation}
\label{e24}
\xi_{{\bF k};\sigma} \sim \vartheta_{{\bF k};\sigma}\;\;\;
\mbox{\rm for}\;\;\; {\Bf k} \to {\cal S}_{{\sc f};\sigma}.
\end{equation}
Note that here, as well as in Eqs.~(\ref{e22}) and (\ref{e23}),
${\cal S}_{{\sc f};\sigma}$ is considered as given, so that 
the expressions in these equations reflect some properties of 
${\cal S}_{{\sc f};\sigma}$ rather than necessarily fully 
defining it. 

In analogy with the expression in Eq.~(\ref{e19}) above, we introduce 
the auxiliary function $\eta_{{\bF k};\sigma}$, satisfying
\begin{equation}
\label{e25}
{\sf n}_{\sigma}({\Bf k})\,\eta_{{\bF k};\sigma} {:=} 
\vartheta_{{\bF k};\sigma} - \frac{1}{\sf g}\,
(\varepsilon_{\sc f} - \varepsilon_{{\bF k}_{{\sc f};\sigma}}).
\end{equation} 
This expression is general for all ${\Bf k}$ for which 
${\sf n}_{\sigma}({\Bf k})\not=0$. In the light of Eq.~(\ref{e23}), 
and since, for interacting GSs, ${\sf n}_{\sigma}({\Bf k})\not=0$ 
in a neighbourhood of ${\cal S}_{{\sc f};\sigma}$ (see later), 
we must have 
\begin{equation}
\label{e26}
\eta_{{\bF k};\sigma} \sim 0 \;\;\;
\mbox{\rm for}\;\;\; {\Bf k} \to {\cal S}_{{\sc f};\sigma}.
\end{equation}
Defining $\zeta_{{\bF k};\sigma}$ as follows
\begin{equation}
\label{e27}
\zeta_{{\bF k};\sigma} {:=}
\xi_{{\bF k};\sigma} - \frac{1}{\sf g}\,
(\varepsilon_{\sc f} - \varepsilon_{{\bF k}_{{\sc f};\sigma}}),
\end{equation}
on account of Eq.~(\ref{e22}) we similarly have
\begin{equation}
\label{e28}
\zeta_{{\bF k};\sigma} \sim 0 \;\;\;
\mbox{\rm for}\;\;\; {\Bf k} \to {\cal S}_{{\sc f};\sigma}.
\end{equation}
On the basis of the above expressions, from Eqs.~(\ref{e20}) and
(\ref{e21}) for ${\Bf k} \to {\Bf k}_{{\sc f};\sigma} 
\in {\cal S}_{{\sc f};\sigma}$ we deduce
\begin{equation}
\label{e29}
\varepsilon_{{\bF k};\sigma}^{<} \sim \mu
+ {\Bf a}({\Bf k}_{{\sc f};\sigma})
\cdot ({\Bf k}-{\Bf k}_{{\sc f};\sigma}) 
+ {\sf g}\,\zeta_{{\bF k};\sigma},
\end{equation}
\begin{equation}
\label{e30}
\varepsilon_{{\bF k};\sigma}^{>} \sim \mu
+ {\Bf a}({\Bf k}_{{\sc f};\sigma})
\cdot ({\Bf k}-{\Bf k}_{{\sc f};\sigma}) 
- {\sf g}\, \Lambda_{\sigma}({\Bf k})\,
(\zeta_{{\bF k};\sigma} -\eta_{{\bF k};\sigma}),
\end{equation}
where
\begin{equation}
\label{e31}
{\Bf a}({\Bf k}_{{\sc f};\sigma}) {:=} \left.
{\Bf\nabla}_{\bF k}\varepsilon_{\bF k}\right|_{{\bF k}
={\bF k}_{{\sc f};\sigma}},
\end{equation}
\begin{equation}
\label{e32}
\Lambda_{\sigma}({\Bf k}) {:=} 
\frac{ {\sf n}_{\sigma}({\Bf k})}
{1 - {\sf n}_{\sigma}({\Bf k}) }.
\end{equation}
In Eqs.~(\ref{e29}) and (\ref{e30}), $\mu$ stands for the `chemical 
potential' which for metallic states is infinitesimally greater 
than the Fermi energy $\varepsilon_{\sc f}$ (for details see
\cite{BF02a,BF02b}).

Making use of the equation of motion for the operator 
${\hat c}_{{\bF k};\sigma}$ in the Heisenberg picture, that is
\begin{equation}
\label{e33}
i\hbar \frac{\partial}{\partial t} 
{\hat c}_{{\bF k};\sigma}(t) = 
\big[ {\hat c}_{{\bF k};\sigma}(t), \wh{H}\big]_{-},
\end{equation}
from the defining expression in Eq.~(\ref{e8}), one obtains 
\begin{equation}
\label{e34}
\beta_{{\bF k};\sigma}^{<} =  \frac{1}{\sf g}
\Big\{\left.\!\hbar\frac{\partial}{\partial t}
G_{\sigma}({\Bf k};t-t')\right|_{t'\downarrow t}
-\varepsilon_{\bF k} {\sf n}_{\sigma}({\Bf k}) \Big\},
\end{equation}
where the single-particle Green function in the time domain is 
defined as follows:
\begin{equation}
\label{e35}
G_{\sigma}({\Bf k};t-t') {:=}
-i\,\langle\Psi_{N;0}\vert
{\cal T}\big\{ 
{\hat c}_{{\bF k};\sigma}(t)
{\hat c}_{{\bF k};\sigma}^{\dag}(t') \big\}
\vert\Psi_{N;0}\rangle,
\end{equation}
in which ${\cal T}$ stands for the fermion time-ordering operator.
We have (below $\eta\downarrow 0$)
\begin{eqnarray}
\label{e36}
\left. \hbar \frac{\partial}{\partial t}
G_{\sigma}({\Bf k};t-t')\right|_{t'\downarrow t}
&=&\frac{1}{\hbar} 
\int_{-\infty}^{\infty} \frac{{\rm d}\varepsilon}{2\pi i}\;
{\rm e}^{i\varepsilon\eta/\hbar}\,
\varepsilon\, G_{\sigma}({\Bf k};\varepsilon) \nonumber\\
&\equiv&
\frac{1}{\hbar} \int_{-\infty}^{\mu}
{\rm d}\varepsilon\; \varepsilon\, A_{\sigma}({\Bf k};\varepsilon),
\end{eqnarray}
where $A_{\sigma}({\Bf k};\varepsilon)$ stands for the 
single-particle spectral function, defined in terms
of $\wt{G}_{\sigma}({\Bf k};z)$, the analytic continuation
of $G_{\sigma}({\Bf k};\varepsilon) =
\lim_{\eta\downarrow 0} \wt{G}_{\sigma}({\Bf k};\varepsilon
\pm i\eta)$, $\varepsilon\,\Ieq<>\, \mu$, as follows:
\begin{equation}
\label{e37}
A_{\sigma}({\Bf k};\varepsilon) {:=}
\frac{1}{2\pi i} \big\{
\wt{G}_{\sigma}({\Bf k};\varepsilon-i\eta) -
\wt{G}_{\sigma}({\Bf k};\varepsilon+i\eta) \big\},\;\;\;
\eta\downarrow 0.
\end{equation}

With (see Eq.~(\ref{e10}) above)
\begin{equation}
\label{e38}
{\sf n}_{\sigma}({\Bf k}) =
\frac{1}{\hbar} \int_{-\infty}^{\mu}
{\rm d}\varepsilon\; A_{\sigma}({\Bf k};\varepsilon),
\end{equation}
from Eqs.~(\ref{e19}), (\ref{e34}), (\ref{e36}) and (\ref{e38}) 
we deduce that
\begin{equation}
\label{e39}
\xi_{{\bF k};\sigma} \equiv
\frac{\beta_{{\bF k};\sigma}^{<}}{\nu_{\sigma}^{<}({\Bf k})} =
\frac{1}{\sf g} \Big\{
\frac{\int_{-\infty}^{\mu} {\rm d}\varepsilon\;
\varepsilon\, A_{\sigma}({\Bf k};\varepsilon)}
{\int_{-\infty}^{\mu} {\rm d}\varepsilon\;
A_{\sigma}({\Bf k};\varepsilon)} 
-\varepsilon_{\bF k} \Big\},
\end{equation}
which in combination with Eq.~(\ref{e22}) leads to
\begin{equation}
\label{e40}
\frac{\int_{-\infty}^{\mu} {\rm d}\varepsilon\;
\varepsilon\, A_{\sigma}({\Bf k};\varepsilon)}
{\int_{-\infty}^{\mu} {\rm d}\varepsilon\;
A_{\sigma}({\Bf k};\varepsilon)} \sim
\varepsilon_{\sc f}\;\;\;\mbox{\rm for}\;\;\;
{\Bf k}\to {\Bf k}_{{\sc f};\sigma}\in 
{\cal S}_{{\sc f};\sigma}. 
\end{equation}
With
\begin{equation}
\label{e41}
Z_{{\bF k}_{{\sc f};\sigma}} {:=}
{\sf n}_{\sigma}({\Bf k}_{{\sc f};\sigma}^-) -
{\sf n}_{\sigma}({\Bf k}_{{\sc f};\sigma}^+),
\end{equation}
from Eqs.~(\ref{e38}) and (\ref{e40}) we obtain
\begin{equation}
\label{e42}
Z_{{\bF k}_{{\sc f};\sigma}}
= \frac{1}{\hbar\varepsilon_{\sc f}}
\int_{-\infty}^{\mu} {\rm d}\varepsilon\;
\varepsilon\,\big[ 
A_{\sigma}({\Bf k}_{{\sc f};\sigma}^-;\varepsilon)-
A_{\sigma}({\Bf k}_{{\sc f};\sigma}^+;\varepsilon)\big],
\end{equation}
which is alternative to the standard expression
\begin{equation}
\label{e43}
Z_{{\bF k}_{{\sc f};\sigma}} = \frac{1}{\hbar}
\int_{-\infty}^{\mu} {\rm d}\varepsilon\;
\big[ A_{\sigma}({\Bf k}_{{\sc f};\sigma}^-;\varepsilon) -
A_{\sigma}({\Bf k}_{{\sc f};\sigma}^+;\varepsilon) \big],
\end{equation}
directly obtained from the spectral representation of 
${\sf n}_{\sigma}({\Bf k})$ in Eq.~(\ref{e38}) above. The 
expression in Eq.~(\ref{e42}) has been explicitly derived in 
\cite{BF02a} for the uniform metallic GSs of the single-band 
Hubbard Hamiltonian. Here as in \cite{BF02a} the single-particle 
energies $\varepsilon_{{\bF k};\sigma}^{\Ieq<>}$ (associated with 
some fictitious particles \cite{BF02a}) correspond to the 
single-particle spectral function ${\cal A}_{\sigma}({\Bf k};
\varepsilon)$, for which we have \cite{BF02a}
\begin{equation}
\label{e44}
{\cal A}_{\sigma}({\Bf k};\varepsilon) = \hbar \big\{
\nu_{\sigma}^{<}({\Bf k}) \delta(\varepsilon
-\varepsilon_{{\bF k};\sigma}^{<}) + 
\nu_{\sigma}^{>}({\Bf k}) \delta(\varepsilon
-\varepsilon_{{\bF k};\sigma}^{>}) \big\}.
\end{equation}
It is readily verified that the exact $Z_{{\bF k}_{{\sc f};\sigma}}$ 
is obtained by replacing $A_{\sigma}({\Bf k};\varepsilon)$ in 
Eqs.~(\ref{e42}) and (\ref{e43}) by ${\cal A}_{\sigma}({\Bf k};
\varepsilon)$. In this connection note that, for stable (uniform) 
GSs, $\varepsilon_{{\bF k};\sigma}^{\Ieq<>}\,\Ieq<>\, \mu$,
$\forall {\Bf k}$ \cite{BF02a}. The GS total energy of the interacting 
system is similarly exactly reproduced by replacing $A_{\sigma}
({\Bf k};\varepsilon)$ in the pertinent expression (see Eq.~(51) in 
\cite{BF02a}) by ${\cal A}_{\sigma}({\Bf k};\varepsilon)$.

\section*{\S~3.~Equations determining the Fermi surfaces of uniform 
metallic ground states}
\label{s2}

The Fermi surface ${\cal S}_{{\sc f};\sigma}$ is conventionally 
defined as follows \cite{GM58,L60,BF02a}
\begin{equation}
\label{e45}
{\cal S}_{{\sc f};\sigma}
{:=} \{ {\Bf k}\,\|\,
\varepsilon_{\bF k} + \hbar\Sigma_{\sigma}({\Bf k};\varepsilon_{\sc f})
= \varepsilon_{\sc f} \},
\end{equation}
where, on general grounds, ${\rm Im}[\Sigma_{\sigma}({\Bf k};
\varepsilon_{\sc f})]\equiv 0$, $\forall {\Bf k}$ 
\cite{L61,BF02a,BF02b}. From Eqs.~(\ref{e17}) and (\ref{e23}) 
we deduce
\begin{equation}
\label{e46}
\varepsilon_{\bF k} + 
\hbar\Sigma_{\sigma}^{\sc hf}({\Bf k})
= \varepsilon_{\sc f}\;\;\;
\mbox{\rm for}\;\;\; {\Bf k}\in {\cal S}_{{\sc f};\sigma}.
\end{equation}
Comparison of this expression with the defining expression for 
${\cal S}_{{\sc f};\sigma}$ in Eq.~(\ref{e45}) reveals that to the 
leading order in $({\Bf k}-{\Bf k}_{{\sc f};\sigma})$ we must have
\begin{equation}
\label{e47}
\Sigma_{\sigma}({\Bf k};\varepsilon_{\sc f}) \sim 
\Sigma_{\sigma}^{\sc hf}({\Bf k})\;\;\;\mbox{\rm as}\;\;\;
{\Bf k}\to {\Bf k}_{{\sc f};\sigma} \in 
{\cal S}_{{\sc f};\sigma}.
\end{equation}
The significance of this expression becomes more apparent by 
considering the Kramers-Kr\"onig relation for 
${\rm Re}[\Sigma_{\sigma}({\Bf k};\varepsilon)]$ in terms of 
${\rm Im}[\Sigma_{\sigma}({\Bf k};\varepsilon)]$; since 
\cite{BF99,BF02b}
\begin{equation}
\label{e48}
\Sigma_{\sigma}({\Bf k};\varepsilon) 
-\Sigma_{\sigma}^{\sc hf}({\Bf k})
= o(1) \;\;\; \mbox{\rm for}\;\;\;
\vert\varepsilon\vert\to\infty,
\end{equation} 
the above-mentioned Kramers-Kr\"onig relation reads
\begin{eqnarray}
\label{e49}
&&{\rm Re}[\Sigma_{\sigma}({\Bf k};\varepsilon)]
= \Sigma_{\sigma}^{\sc hf}({\Bf k}) \nonumber\\
&&\;\;\;\;\;\;\;\;\;\;\;
-\wp\!\int_{-\infty}^{\infty} 
\frac{{\rm d}\varepsilon'}{\pi}\,
\frac{ {\rm sgn}(\mu -\varepsilon')\,
{\rm Im}[\Sigma_{\sigma}({\Bf k};\varepsilon')]}
{\varepsilon' - \varepsilon}.
\end{eqnarray}
From the leading term in the asymptotic series of the integral on 
the RHS of Eq.~(\ref{e49}) for $\varepsilon\to\varepsilon_{\sc f}$, 
making use of the fact that ${\rm Im}[\Sigma_{\sigma}({\Bf k};
\varepsilon)] \sim 0$ as $\varepsilon\to \varepsilon_{\sc f}$, we
obtain \cite{BF99}
\begin{equation}
\label{e50}
\Sigma_{\sigma}({\Bf k};\varepsilon_{\sc f})
= \Sigma_{\sigma}^{\sc hf}({\Bf k}) + S_{\sigma}({\Bf k}),
\end{equation}
where
\begin{equation}
\label{e51}
S_{\sigma}({\Bf k}) {:=}
\frac{1}{\pi} \int_{0}^{\infty}
\frac{ {\rm d}\varepsilon}{\varepsilon}\;
{\rm Im}[\Sigma_{\sigma}({\Bf k};\varepsilon_{\sc f}+\varepsilon) + 
\Sigma_{\sigma}({\Bf k};\varepsilon_{\sc f}-\varepsilon)].
\end{equation}
Comparison of Eq.~(\ref{e47}) with Eq.~(\ref{e50}) reveals the 
necessity of the condition
\begin{equation}
\label{e52}
S_{\sigma}({\Bf k}) = 0
\end{equation}
in order for ${\Bf k}\in {\cal S}_{{\sc f};\sigma}$. In the light of 
the results in Eqs.~(\ref{e47}) and (\ref{e50}), we thus arrive at 
the following alternative defining expression for ${\cal S}_{{\sc f};
\sigma}$ ({\it cf}. Eq.~(\ref{e45}) above):
\begin{equation}
\label{e53}
{\cal S}_{{\sc f};\sigma} =
\{ {\Bf k}\,\|\, 
\varepsilon_{\bF k} + 
\hbar\Sigma_{\sigma}^{\sc hf}({\Bf k})
= \varepsilon_{\sc f}\;\wedge\;
S_{\sigma}({\Bf k}) = 0\}.
\end{equation}
The necessity of the additional condition $S_{\sigma}({\Bf k}) = 0$ 
in the defining expression for ${\cal S}_{{\sc f};\sigma}$ stems
from the fact that the equation in Eq.~(\ref{e46}) applies for all 
${\Bf k}$ on a pre-supposed ${\cal S}_{{\sc f};\sigma}$ so that, 
in general Eq.~(\ref{e46}) {\sl cannot} be sufficient fully to define
${\cal S}_{{\sc f};\sigma}$. The significance of this aspect can 
be easiest appreciated by specializing Eq.~(\ref{e46}) to the uniform 
metallic GSs of the single-band Hubbard Hamiltonian, in which case 
$\Sigma_{\sigma}^{\sc hf}({\Bf k})$ is {\sl independent} of ${\Bf k}$ 
and equal to $\frac{1}{\hbar} U n_{\bar\sigma}$ (see Eqs.~(\ref{e12}), 
(\ref{e17}) and (\ref{e4}) above). The sufficiency of 
$\varepsilon_{\bF k}+\hbar\Sigma_{\sigma}^{\sc hf}({\Bf k}) = 
\varepsilon_{\sc f}$ to define ${\cal S}_{{\sc f};\sigma}$ would in 
the present case incorrectly imply that {\sl all} solutions of the 
latter equation, which constitute the entire ${\cal S}_{{\sc f};
\sigma}^{(0)}$ (the Fermi surface corresponding to $\varepsilon_{\bF k}$ 
and the partial number $N_{\sigma}$ of fermions with spin index 
$\sigma$ in the exact interacting GS), were in general points 
located on ${\cal S}_{{\sc f};\sigma}$; the unconditional equality 
${\cal S}_{{\sc f};\sigma} = {\cal S}_{{\sc f};\sigma}^{(0)}$, rather 
than the relationship ${\cal S}_{{\sc f};\sigma} \subseteq 
{\cal S}_{{\sc f};\sigma}^{(0)}$ deduced in \cite{BF02a}, would rule 
out the feasibility that the pseudogap phenomenon is described in 
the space spanned by the uniform metallic GSs of the single-band 
Hubbard Hamiltonian \cite{BF02a}, contradicting the results in
\cite{KS90,SW92,VW93,BSW94,VT96}. Note that for isotropic GSs 
the condition $S_{\sigma}({\Bf k}) = 0$ is automatically satisfied 
at {\sl any} ${\Bf k}$ for which Eq.~(\ref{e46}) is satisfied, on 
account of the fact that the metallic nature of a GS implies that
$S_{\sigma}(\pm{\Bf k}) =0$ for at least one ${\Bf k}$ (and all 
points related to this by symmetry) on ${\cal S}_{{\sc f};\sigma}$, 
which by isotropy must remain valid for {\sl all} ${\Bf k}$ on 
${\cal S}_{{\sc f};\sigma}$. 

We note that along the lines of reasoning as in \cite{BF02a} it 
can be shown that ${\sf n}_{\sigma}({\Bf k})$ is continuous (see 
Eq.~(67) in \cite{BF02a}) at {\sl all} ${\Bf k}$ which satisfy 
Eq.~(\ref{e46}) but fail to satisfy Eq.~(\ref{e52}). We further 
note that, since for ${\Bf k}\in {\cal S}_{{\sc f};\sigma}$ up to 
infinitesimal corrections \cite{BF02a} we have $\varepsilon_{{\bF k};
\sigma}^{<} = \varepsilon_{{\bF k};\sigma}^{>} =\varepsilon_{\sc f}$, 
Eq.~(\ref{e18}) is seen to conform with the result in Eq.~(\ref{e46}) 
above.

It is interesting to rewrite the condition $S_{\sigma}({\Bf k}) =0$ 
for ${\Bf k}\in {\cal S}_{{\sc f};\sigma}$ in the following appealing 
form:
\begin{equation}
\label{e54}
\int_{-\infty}^{\varepsilon_{\sc f}} \!\!
{\rm d}\varepsilon\; 
\frac{ {\rm Im}[\Sigma_{\sigma}({\Bf k};\varepsilon)]}
{\varepsilon_{\sc f} - \varepsilon} =
\int_{\varepsilon_{\sc f}}^{\infty} \!
{\rm d}\varepsilon\; 
\frac{ {\rm Im}[\Sigma_{\sigma}({\Bf k};\varepsilon)]}
{\varepsilon_{\sc f} - \varepsilon}, \;\;\;
{\Bf k}\in {\cal S}_{{\sc f};\sigma}.
\end{equation}
The integrands on both sides of this expression are non-negative 
over the pertinent ranges of integration. In the light of this and 
of the inherent symmetry of these integrands for both $\vert\varepsilon
-\varepsilon_{\sc f}\vert\to 0$ and $\vert\varepsilon
-\varepsilon_{\sc f}\vert\to\infty$ (to the leading orders in
$(\varepsilon-\varepsilon_{\sc f})$ and
$1/(\varepsilon-\varepsilon_{\sc f})$ respectively \cite{BF02b}), 
Eq.~(\ref{e54}) can be viewed as implying some degree of similarity 
between $\vert {\rm Im}[\Sigma_{\sigma}({\Bf k};\varepsilon_{\sc f} - 
\varepsilon)]\vert$ and $\vert {\rm Im}[\Sigma_{\sigma}({\Bf k};
\varepsilon_{\sc f} +\varepsilon)]\vert$ at intermediate values of 
$\varepsilon$ (as measured with respect to $\varepsilon_{\sc f}$) 
when ${\Bf k}\in {\cal S}_{{\sc f};\sigma}$; for ${\Bf k}\not\in 
{\cal S}_{{\sc f};\sigma}$, this similarity should be, if not absent, 
less pronounced.
\footnote{\label{f7}
We believe that the apparent violation of the result in
Eq.~(\protect\ref{e54}) by the first-order self-energy in terms of 
the screened interaction function (as evident from the middle 
figures in Figs.~16 and 17 in \protect\cite{HL69}, corresponding 
to $k=1.0 k_{\sc f}$) signifies the shortcomings of this approximate 
self-energy, as explicitly discussed in \protect\cite{BF02b}
(for instance, compare the first-order result in Eq.~(278) with
the exact result in Eq.~(209) and in the light of the contrast
between the two results, consider the expression concerning 
${\rm Im}[\Sigma_{\sigma}({\Bf k};\varepsilon)]$ in Eq.~(234a); 
in this connection, recall that in the paramagnetic state, dealt 
with here, $n_{0;\bar\sigma}-n_{0;\sigma}$ as encountered in the 
latter equation is equal to zero). }
We point out that since $S_{\sigma}({\Bf k})=0$
applies for {\sl all} ${\Bf k}\in {\cal S}_{{\sc f};\sigma}$, 
in principle one has (see later)
\begin{equation}
\label{e55}
\left. {\Bf\nabla}_{{\bF k'}} 
S_{\sigma}({\Bf k}')\right|_{{\bF k}'={\bF k}^{\mp}}
= \lambda_{\sigma}({\Bf k}^{\mp})\, {\hat {\Bf n}}({\Bf k})
\;\;\; \mbox{\rm for}\;\;\; 
{\Bf k} \in {\cal S}_{{\sc f};\sigma},
\end{equation}
where ${\hat {\Bf n}}({\Bf k})$ stands for the (outward) unit 
vector normal to ${\cal S}_{{\sc f};\sigma}$ at ${\Bf k} \in 
{\cal S}_{{\sc f};\sigma}$; here ${\Bf k}^{-}$ and ${\Bf k}^{+}$ denote 
vectors infinitesimally close to ${\Bf k}$, with the former inside 
and the latter outside the Fermi sea, and $\lambda_{\sigma}({\Bf k})$ 
stands for a scalar function which {\sl cannot} be identically vanishing. 
For conciseness, in this work we shall denote the Fermi sea 
corresponding to fermions with spin index $\sigma$ by FS$_{\sigma}$ 
and its complementary part with respect to the entire available 
reciprocal space, by $\overline{\rm FS}_{\sigma}$ \cite{BF02a}. 

In writing the expression in Eq.~(\ref{e55}) we have taken account 
of the fact that $S_{\sigma}({\Bf k})$ may not be continuously 
differentiable in a neighbourhood of some or all points 
${\Bf k} \in {\cal S}_{{\sc f};\sigma}$ (whence ${\Bf k}^{\pm}$ in 
$\lambda_{\sigma}({\Bf k}^{\pm})$); however, in Eq.~(\ref{e55}) we have 
{\sl not} taken account of the possibility that $S_{\sigma}({\Bf k})$ 
may not be differentiable in a neighbourhood of ${\Bf k}\in 
{\cal S}_{{\sc f};\sigma}$. The latter possibility in fact arises 
when, for instance, fermions in $d$ spatial dimensions interact 
through the long-range Coulomb potential, in which case 
${\Bf\nabla}_{\bF k} \Sigma_{\sigma}^{\sc hf}({\Bf k})$ is 
logarithmically divergent for ${\Bf k}$ approaching any point 
${\Bf k}_{{\sc f};\sigma}$ on ${\cal S}_{{\sc f};\sigma}$ (both 
from inside and from outside the Fermi sea). It follows that, for 
metallic states of systems of Coulomb-interacting fermions for 
which quasi-particles are well-defined in the neighbourhood of 
${\cal S}_{{\sc f};\sigma}$,
\footnote{\label{f9} 
This implying the differentiability of $\Sigma_{\sigma}({\Bf k};
\varepsilon_{\sc f})$ {\sl and} continuity of ${\Bf\nabla}_{\bF k}
\Sigma_{\sigma}({\Bf k};\varepsilon_{\sc f})$ in the infinitesimal 
neighbourhood of ${\cal S}_{{\sc f};\sigma}$ \protect\cite{BF02a,BF99}. }
the above-mentioned singularity of 
${\Bf\nabla}_{\bF k}\Sigma_{\sigma}^{\sc hf}({\Bf k})$ must be 
fully cancelled by a singular contribution arising from 
${\Bf\nabla}_{\bF k} S_{\sigma}({\Bf k})$ (see Eq.~(\ref{e50}) above),
so that, for these metallic states, Eq.~(\ref{e55}) {\sl cannot} 
be meaningful.

\section*{\S~4.~Exposing a further close relationship between
$\Sigma_{\sigma}(\lowercase{\Bf k};\varepsilon_{\sc f})$ 
and ${\cal S}_{{\sc f};\sigma}$ }
\label{s3}

The fact that, for ${\Bf k}\to {\Bf k}_{{\sc f};\sigma} \in
{\cal S}_{{\sc f};\sigma}$, $\Sigma_{\sigma}({\Bf k};
\varepsilon_{\sc f}) \sim \Sigma_{\sigma}^{\sc hf}({\Bf k})$
to leading order in $({\Bf k}-{\Bf k}_{{\sc f};\sigma})$
(see Eq.~(\ref{e47}) above), deserves some closer inspection. 
From Eqs.~(\ref{e12}), (\ref{e16}) and (\ref{e17}) it is evident 
that $\Sigma_{\sigma}^{\sc hf}({\Bf k})$ is fully determined by 
${\sf n}_{\sigma}({\Bf k})$. For uniform GSs this function is 
the Fourier transform of the single-particle density matrix 
$\varrho_{\sigma}({\Bf r}-{\Bf r}')$, or
\begin{equation}
\label{e56}
\varrho_{\sigma}({\Bf r}) =
\int \frac{{\rm d}^d k}{(2\pi)^d}\;
{\rm e}^{i {\bF k}\cdot {\bF r}}\,
{\sf n}_{\sigma}({\Bf k}).
\end{equation}
The singular nature of ${\sf n}_{\sigma}({\Bf k})$ for ${\Bf k} 
\in {\cal S}_{{\sc f};\sigma}$ has far-reaching consequences for 
the behaviour of $\varrho_{\sigma}({\Bf r})$ for $\|{\Bf r}\|
\to\infty$ \cite{BF00a,BF02a} and consequently for that of
$\Sigma_{\sigma}^{\sc hf}({\Bf k})$ for ${\Bf k}$ in a neighbourhood 
of ${\cal S}_{{\sc f};\sigma}$. To illustrate this aspect, it is 
convenient to introduce 
\begin{equation}
\label{e57}
\hat{\bf e}_r {:=} \frac{\Bf r}{r}\;\;\;
\mbox{\rm where}\;\;\; r {:=} \|{\Bf r}\|.
\end{equation}
Without loss of generality, we explicitly consider the case when $d=2$ 
and assume that the angle $\varphi$ in the cylindrical coordinates 
$(k,\varphi)$ of ${\Bf k}$ is measured with respect to the direction 
specified by $\hat{\bf e}_r$. We thus re-write Eq.~(\ref{e56}) as 
follows:
\begin{equation}
\label{e58}
\varrho_{\sigma}({\Bf r})
= \frac{1}{(2\pi)^2}
\int_0^{\infty} {\rm d}k\; k\,
{\cal I}(k,r), 
\end{equation}
where
\begin{equation}
\label{e59}
{\cal I}_{\sigma}(k,r) {:=} \int_0^{2\pi} {\rm d}\varphi\;
{\rm e}^{i k r \cos(\varphi)}\, 
\tilde{\sf n}_{\sigma}(k,\varphi),
\end{equation}
in which $\tilde{\sf n}_{\sigma}(k,\varphi) \equiv
{\sf n}_{\sigma}({\Bf k})$.
Making use the stationary-phase method \cite{JDM84}, we readily
obtain
\begin{equation}
\label{e60}
{\cal I}_{\sigma}(k,r) \sim
2 \sqrt{\frac{2\pi}{k r}}\, {\sf n}_{\sigma}(k \hat{\bf e}_r)
\cos(k r-\pi/4),\;\;\;
r \to \infty,
\end{equation}
where we have made use of the fact that by the time-reversal 
symmetry of the GS, 
\begin{equation}
\label{e61}
\tilde{\sf n}_{\sigma}(k,\pi) =\tilde{\sf n}_{\sigma}(k,0) 
\equiv {\sf n}_{\sigma}(k \hat{\bf e}_r).
\end{equation}
Thus, from Eqs.~(\ref{e58}) and (\ref{e60}) we have
\begin{eqnarray}
\label{e62}
\varrho_{\sigma}(r \hat{\bf e}_r) &\sim&
\frac{2}{(2\pi)^{3/2} r^{1/2}}
\int_0^{\infty} {\rm d}k\; k^{1/2}\,
{\sf n}_{\sigma}(k \hat{\bf e}_r) \nonumber\\
& &\;\;\;\;\;\;\;\;\;\;\;\;\;\;\;\;\;\;\;\;\;\;
\times \cos(k r-\pi/4),\;\;
r\to\infty.
\end{eqnarray}
Let ${\Bf k}_{{\sc f};\sigma}$ point in the direction of ${\Bf r}$, 
i.e. ${\Bf k}_{{\sc f};\sigma} = k_{{\sc f};\sigma} \hat{\bf e}_r$.
The singular nature of ${\sf n}_{\sigma}(k \hat{\bf e}_r)$ at $k
=k_{{\sc f};\sigma}$ implies that for the further simplification of 
the expression on the RHS of Eq.~(\ref{e62}) it is necessary \cite{BF02b}
to expresses the $k$ integral over $[0,\infty)$ in terms of at least 
two $k$ integrals over the {\sl disjoint} intervals $[0,k_{{\sc f};
\sigma})$ and $(k_{{\sc f};\sigma},\infty)$; this subdivision should 
be carried out for all $k$ points at which ${\sf n}_{\sigma}
(k \hat{\bf e}_r)$ is singular; however, to avoid any complication 
unnecessary to our present considerations, here we deal with the case 
where the most dominant singularity of ${\sf n}_{\sigma}
(k \hat{\bf e}_r)$ is located at $k=k_{{\sc f};\sigma}$. Assuming 
${\sf n}_{\sigma}(k \hat{\bf e}_r)$ to be discontinuous at 
$k=k_{{\sc f};\sigma}$ and continuous elsewhere, through integration 
by parts, from Eq.~(\ref{e62}) we obtain \cite{BF00a,BF02b}
\begin{eqnarray}
\label{e63}
\varrho_{\sigma}(r \hat{\bf e}_r) &\sim&
\frac{ Z_{k_{{\sc f};\sigma} \hat{\bf e}_r} k_{{\sc f};\sigma}^{1/2}}
{2^{1/2} \pi^{3/2}}\,
\frac{\sin(k_{{\sc f};\sigma} r - \pi/4)}{r^{3/2}}\nonumber\\
&+& \frac{ {\sf n}_{\sigma}(k_{\rm z} \hat{\bf e}_{r}) 
k_{\rm z}^{1/2}}{2^{1/2} \pi^{3/2}}\,
\frac{\sin(k_{\rm z} r - \pi/4)}{r^{3/2}}, \;\;\;
r\to \infty,
\end{eqnarray}
where for systems defined on lattice $k_{\rm z} \hat{\bf e}_r$ denotes 
the vector from the origin to the Brillouin zone boundary (note that 
for these systems our assumption with regard to the uniformity of 
the underlying GSs implies that we should only consider ${\Bf r} \in 
\{ {\Bf R}_j\, \|\, j=1,\dots, N_{\sc l} \}$) and for those defined 
on the continuum $k_{\rm z}=\infty$ so that ${\sf n}_{\sigma}
(k_{\rm z} \hat{\bf e}_{r})=0$; for the former systems and for a given 
$\hat{\bf e}_r$, ${\sf n}_{\sigma}(k_{\rm z} \hat{\bf e}_{r})$ may or 
may not be vanishing. In what follows, for simplicity we set
${\sf n}_{\sigma}(k_{\rm z} \hat{\bf e}_{r})$ equal to zero.
One observes that the full information concerning the shape and 
dimensions of ${\cal S}_{{\sc f};\sigma}$ is already contained in 
the leading-order term in the large-$\|{\Bf r}\|$ asymptotic series 
of $\varrho_{\sigma}({\Bf r})$. 

For the Fock part $\Sigma_{\sigma}^{\sc f}$ of the self-energy
in the coordinate representation and for arbitrary GSs and two-body 
interaction functions $v({\Bf r}-{\Bf r}')$ we have \cite{BF02b}
\begin{equation}
\label{e64}
\Sigma_{\sigma}^{\sc f}({\Bf r},{\Bf r}')
= -\frac{1}{\hbar} v({\Bf r}-{\Bf r}')
\varrho_{\sigma}({\Bf r}',{\Bf r}),
\end{equation}
where in the case of uniform GSs $\varrho_{\sigma}({\Bf r}',{\Bf r})$ 
is to be replaced by $\varrho_{\sigma}({\Bf r}'-{\Bf r})$. 
It follows that the information concerning ${\cal S}_{{\sc f};
\sigma}$ contained in $\varrho_{\sigma}({\Bf r},{\Bf r}')$ is 
similarly contained in 
\footnote{\label{f10}
This observation corroborates the validity of the alternative 
definition for ${\cal S}_{{\sc f};\sigma}$ in Eq.~(\protect\ref{e53}) 
above. }
$\Sigma_{\sigma}^{\sc f}({\Bf r},{\Bf r}')$. In particular, for any 
two-body potential other than those for which $v({\Bf r}-{\Bf r}') 
\equiv 0$ when $\|{\Bf r}-{\Bf r}'\| \ge R > 0$, the behaviour of 
$\varrho_{\sigma}({\Bf r},{\Bf r}')$ for $\|{\Bf r} -{\Bf r}'\|
\to\infty$ directly determines that of $\Sigma_{\sigma}^{\sc f}
({\Bf r},{\Bf r}')$ for $\|{\Bf r}-{\Bf r}'\| \to\infty$. For instance, 
for the uniform GSs of systems of particles in two-dimensional space 
(i.e. for $d=2$) interacting through the Coulomb potential and for 
which ${\sf n}_{\sigma}(k \hat{\bf e}_r)$ is discontinuous at 
$k=k_{{\sc f};\sigma}$, from Eqs.~(\ref{e63}) and (\ref{e64}) we have 
(following our above convention, for uniform GSs we use the notation 
$\Sigma_{\sigma}^{\sc f}({\Bf r}-{\Bf r}')$ for what in the general 
case is denoted by $\Sigma_{\sigma}^{\sc f}({\Bf r},{\Bf r}')$)
\begin{equation}
\label{e65}
\Sigma_{\sigma}^{\sc f}(r \hat{\bf e}_r) \sim
-\frac{e^2}{h}
\frac{ Z_{k_{{\sc f};\sigma} \hat{\bf e}_r} k_{{\sc f};\sigma}^{1/2}}
{ (2\pi)^{3/2} \epsilon_0}\,
\frac{\sin(k_{{\sc f};\sigma} r - \pi/4)}{r^{5/2}},\;\;\;
r\to \infty,
\end{equation} 
where $e^2$ stands for the particle-charge squared and $\epsilon_0$ 
for the vacuum permittivity (note that for $-e$ the electron charge,
$e^2/h = 3.874\dots\times 10^{-5}$~S is the quantized Hall conductance). 
\footnote{\label{f11}
We note that the direct evaluation of the expression in 
Eq.~(\protect\ref{e65}) from the Fourier integral representation
of $\Sigma_{\sigma}^{\sc f}({\Bf r}-{\Bf r}')$ is not straightforward. 
This is partly because $\Sigma_{\sigma}^{\sc f}(k \hat{\bf e}_r)$ is 
non-analytically singular at $k=k_{{\sc f};\sigma}$ (recall that in 
determining the expression in Eq.~(\protect\ref{e63}) we have 
explicitly assumed that the leading singularity of ${\sf n}_{\sigma}
(k \hat{\bf e}_r)$ consists of a {\sl discontinuity} at 
$k=k_{{\sc f};\sigma}$) so that, for the determination of the 
large-$\|{\Bf r}-{\Bf r}'\|$ asymptotic expansion of 
$\Sigma_{\sigma}^{\sc f}({\Bf r}-{\Bf r}')$, one has to employ the 
Laplace method \protect\cite{JDM84} which requires analytic 
continuation of $\Sigma_{\sigma}^{\sc f}(k \hat{\bf e}_r)$ into the 
complex $k$ plane. In this connection we point out that the 
appearance of $\sin(k_{{\sc f};\sigma} r -\pi/4)$ on the RHS of 
Eq.~(\protect\ref{e65}), rather than $\cos(k_{{\sc f};\sigma} r -\pi/4)$, 
should be a reminder of the non-triviality of the result in 
Eq.~(\protect\ref{e65}). }
The fact that the oscillatory behaviour of the leading term in the
asymptotic series of $\Sigma_{\sigma}^{\sc f}(r\hat{\bf e}_r)$ for 
$r\to\infty$ is entirely determined by $k_{{\sc f};\sigma}$ indicates 
that, similar to ${\sf n}_{\sigma}({\Bf k})$, $\Sigma_{\sigma}^{\sc hf}
({\Bf k})$ must be singular for all ${\Bf k}$ on 
${\cal S}_{{\sc f};\sigma}$;
\footnote{\label{f12}
This statement can also be explicitly verified from the expression 
for $\Sigma_{\sigma}^{\sc hf}({\Bf k}) \equiv ({\sf g}/\hbar) 
\vartheta_{{\bF k};\sigma}$ (see Eq.~(\protect\ref{e17})), making use 
of Eq.~(\protect\ref{e16}). }
in the present case, where $\Sigma_{\sigma}^{\sc f}(r \hat{\bf e}_r)$
decays with one power of $1/r$ faster than $\varrho_{\sigma}
(r \hat{\bf e}_r)$ for $r\to\infty$, the assumed discontinuity of 
${\sf n}_{\sigma}({\Bf k})$ at ${\Bf k} = k_{{\sc f};\sigma} 
\hat{\bf e}_r$ implies singularity (explicitly, a logarithmic
divergence
\footnote{\label{f13}
For some pertinent details we refer the reader to an extensive 
analysis of the large-$r$ asymptotic behaviour of 
$\varrho_{\sigma}(r)$ pertaining to uniform and {\sl isotropic} 
GSs in appendix J of \protect\cite{BF02b}. From this analysis and 
the bounded result in Eq.~(\protect\ref{e65}) above one can readily 
infer the nature of the singularity (i.e. logarithmic divergence) of  
${\Bf\nabla}_{\bF k}\Sigma_{\sigma}^{\sc f}({\Bf k})$ at ${\Bf k}\in 
{\cal S}_{{\sc f};\sigma}$. })
in the behaviour of 
${\Bf\nabla}_{\bF k}\Sigma_{\sigma}^{\sc f}({\Bf k})$ at ${\Bf k} 
= k_{{\sc f};\sigma} \hat{\bf e}_r$.
These facts expose a fundamental distinction between systems of 
fermions interacting through potentials of finite range and potentials 
of infinite range. Notably, whereas for the uniform metallic GSs of 
the single-band Hubbard Hamiltonian one has ${\sf n}_{\sigma}({\Bf k}) 
\ge \frac{1}{2}$ for ${\Bf k}$ inside the Fermi sea and infinitesimally 
close to ${\cal S}_{{\sc f};\sigma}$ \cite{BF02a}, as we show later in 
this paper (\S~6), this is {\sl not} necessarily the case when the 
interaction potential is of longer range than the intra-atomic 
potential encountered in the Hubbard Hamiltonian. In this connection, 
recall that for the uniform metallic GSs of the conventional 
single-band Hubbard Hamiltonian, $\Sigma_{\sigma}^{\sc hf}({\Bf k}) 
= \frac{1}{\hbar} U n_{\bar\sigma}$, which is independent of ${\Bf k}$ 
and thus regular for {\sl all} ${\Bf k} \in {\rm 1BZ}$.

It is important to point out that ${\sf n}_{\sigma}({\Bf k})$ is 
singular ({\sl not} necessarily discontinuous) at {\sl all} ${\Bf k}$ 
satisfying Eq.~(\ref{e46}) above. 
\footnote{\label{f14}
The analysis underlying this statement is analogous to that 
presented in \protect\cite{BF02a} for the case of the single-band 
Hubbard Hamiltonian (see Eq.~(68) and the subsequent text in 
\protect\cite{BF02a}). }
However, since according to Eq.~(\ref{e53}) the satisfaction of 
Eq.~(\ref{e46}) at a given ${\Bf k}$ is {\sl not} sufficient for 
${\Bf k} \in{\cal S}_{{\sc f};\sigma}$, it follows that the ${\Bf k}$ 
that for a given direction $\hat{\bf e}_r$ determines the leading 
term in the large-$r$ asymptotic series of $\varrho_{\sigma}
(r \hat{\bf e}_r)$ (and of $\Sigma_{\sigma}^{\sc f}(r \hat{\bf e}_r)$)
is {\sl not} necessarily a point of ${\cal S}_{{\sc f};\sigma}$.
Since such a ${\Bf k}$ point satisfies Eq.~(\ref{e46}) but fails
to satisfy Eq.~(\ref{e52}), it follows that at this point 
${\sf n}_{\sigma}({\Bf k})$ is continuous (see the paragraph preceding 
that containing Eq.~(\ref{e54}) above) whereby the leading term in the 
large-$r$ asymptotic series of $\varrho_{\sigma}(r \hat{\bf e}_r)$ 
(and of $\Sigma_{\sigma}^{\sc f}(r \hat{\bf e}_r)$) acquires an 
anomalous exponent, leading to a stronger power-law decay in the 
mentioned asymptotic term in comparison with the case corresponding 
to $Z_{k \hat{\bf e}_r} \not= 0$ for $k \hat{\bf e}_r \in 
{\cal S}_{{\sc f};\sigma}$; with reference to the expressions in 
Eqs.~(\ref{e63}) and (\ref{e65}) above (where $k=k_{{\sc f};\sigma}$), 
the above-indicated anomalous exponent would read $3/2+\gamma$ and 
$5/2+\gamma$ respectively, with $\gamma > 0$.

\section*{\S~5.~Ground-state exchange and correlation potentials: 
a test case}
\label{s4}

A significant aspect of the result in Eq.~(\ref{e46}) is made explicit
as follows. For concreteness, here we consider the uniform GSs of 
systems defined on the continuum; these uniform GSs are necessarily 
isotropic. Since lattice models are excluded from our considerations 
in this Section (in a future work we explicitly deal with the more 
delicate case of lattice models), here we relax the definition for 
$n_{\sigma}$ introduced in Eq.~(\ref{e14}) above and introduce
\begin{equation}
\label{e66}
n_{\sigma} {:=} \frac{N_{\sigma}}{\Omega}.
\end{equation}
With $n {:=} n_{\sigma} + n_{\bar\sigma}$, the total concentration 
of the fermions in the $N$-particle uniform GS of the system, we 
define the spin magnetization fraction $m$ as follows
\begin{equation}
\label{e67}
m {:=} \frac{ n_{\sigma} - n_{\bar\sigma} }{n}.
\end{equation}
Denoting by $E(N,M)$ the GS total energy of the system described 
by the $\wh{H}$ in Eq.~(\ref{e1}), corresponding to the total number 
of particles $N \equiv \Omega n$ and the total spin magnetization 
$M \equiv \Omega (n_{\sigma}-n_{\bar\sigma})$, we define
\begin{equation}
\label{e68}
{\cal E}(n,m) {:=} \frac{1}{N}\, E(N,M), 
\end{equation}
which we assume to be a sufficiently smooth function of $n$ and $m$ 
in the thermodynamic limit \cite{AM81} (appendix B herein). In 
this limit, for a metallic GS we have 
\begin{equation}
\label{e69}
\varepsilon_{\sc f} = 
\frac{\partial}{\partial n} \big( n {\cal E}(n,m) \big).
\end{equation} 
From this we obtain
\begin{equation}
\label{e70}
\frac{\rm d}{{\rm d} n_{\alpha} }
\big(n {\cal E}(n,m)\big)
= \varepsilon_{\sc f} 
-(m \mp 1) \frac{\partial {\cal E}(n,m)}{\partial m},\;\; 
\alpha =
\Big\{
\begin{array}{l}
\sigma,\\
\bar\sigma. 
\end{array} 
\end{equation}
We now decompose ${\cal E}(n,m)$ as follows:
\begin{equation}
\label{e71}
{\cal E}(n,m) \equiv {\cal E}_{\rm k}(n,m) + 
{\cal E}_{\rm xc}(n,m),
\end{equation}
where ${\cal E}_{\rm k}(n,m)$, the `kinetic energy' contribution 
to ${\cal E}(n,m)$, corresponds to the GS of $\wh{\cal H}_0$ in 
Eq.~(\ref{e80}) below whose associated $n$ and $m$ are constrained 
to be the same as those pertaining to the exact GS of $\wh{H}$. Thus, 
Eq.~(\ref{e71}) {\sl defines} the exchange-correlation energy 
contribution ${\cal E}_{\rm xc}(n,m)$. By the assumption of the 
isotropy of the GSs under consideration and in consequence of a 
Luttinger theorem \cite{LW60,L60}, according to which the number 
of ${\Bf k}$ points enclosed by ${\cal S}_{{\sc f};\sigma}$ and 
${\cal S}_{{\sc f};\sigma}^{(0)}$ are equal, we have 
${\cal S}_{{\sc f};\sigma} = {\cal S}_{{\sc f};\sigma}^{(0)}$, 
$\forall\sigma$. From this and in analogy with the result in 
Eq.~(\ref{e70}) we have
\begin{equation}
\label{e72}
\frac{\rm d}{{\rm d} n_{\alpha}} 
\big(n {\cal E}_{\rm k}(n,m)\big)
= \varepsilon_{{\bF k}_{{\sc f};\alpha}}
-(m \mp 1) \frac{\partial {\cal E}_{\rm k}(n,m)}{\partial m},\;\; 
\alpha = \Big\{
\begin{array}{l}
\sigma,\\
\bar\sigma. 
\end{array} 
\end{equation}
Combining the expressions in Eqs.~(\ref{e70}), (\ref{e71}) and
(\ref{e72}), for the exchange-correlation potential $\mu_{{\rm xc};
\sigma}$, defined through
\begin{equation}
\label{e73}
\mu_{{\rm xc};\alpha} {:=}
\frac{\rm d}{{\rm d} n_{\alpha}}
\big( n {\cal E}_{\rm xc}(n,m) \big),\;\;\;
\alpha \in \{ \sigma,\bar\sigma\},
\end{equation}
we deduce that
\begin{equation}
\label{e74}
\mu_{{\rm xc};\alpha} = \varepsilon_{\sc f}
-\varepsilon_{{\bF k}_{{\sc f};\alpha}} - (m \mp 1)
\frac{\partial {\cal E}_{\rm xc}(n,m)}{\partial m},\;\;
\alpha =
\Big\{
\begin{array}{l}
\sigma,\\
\bar\sigma. 
\end{array} 
\end{equation}
By subtracting $\mu_{{\rm xc};\bar\sigma}$ from 
$\mu_{{\rm xc};\sigma}$, employing the expression in Eq.~(\ref{e74}),
we obtain
\begin{equation}
\label{e75}
\frac{\partial {\cal E}_{\rm xc}(n,m)}{\partial m}
= \frac{1}{2} \big(
\epsilon_{{\bF k}_{{\sc f};\sigma};\sigma} -
\epsilon_{{\bF k}_{{\sc f};\bar\sigma};\bar\sigma} \big),
\end{equation}
where 
\begin{equation}
\label{e76}
\epsilon_{{\bF k};\alpha} {:=} 
\varepsilon_{\bF k} + \mu_{{\rm xc};\alpha},\;\;\;
\alpha \in \{ \sigma,\bar\sigma \}.
\end{equation}
From Eq.~(\ref{e75}) we infer that in a paramagnetic state, where 
$\epsilon_{{\bF k};\sigma} \equiv \epsilon_{{\bF k};\bar\sigma}$, we 
must have
\begin{equation}
\label{e77}
\frac{\partial {\cal E}_{\rm xc}(n,m)}{\partial m}
=0 \;\;\;\mbox{\rm (for paramagnetic states).}
\end{equation}
From Eqs.~(\ref{e74}) and (\ref{e77}), for paramagnetic states we 
thus obtain
\begin{equation}
\label{e78}
\varepsilon_{\sc f} = \varepsilon_{{\bF k}_{{\sc f};\sigma}} +
\mu_{{\rm xc};\sigma}
\equiv \varepsilon_{\sc f}^{(0)} + \mu_{{\rm xc};\sigma},\;\;
\forall\sigma\;\;\;\;
({\Bf k}_{{\sc f};\sigma} \in {\cal S}_{{\sc f};\sigma}),\;\;
\end{equation}
which is a well-known theorem due to Seitz \cite{FS40}. 

A result which from the perspective of our present work
is of special interest is 
\begin{equation}
\label{e79}
\mu_{{\rm xc};\alpha} = 
\hbar\Sigma_{\alpha}^{\sc hf}({\Bf k}_{{\sc f};\alpha})
-(m\mp 1) \frac{\partial {\cal E}_{\rm xc}(n,m)}{\partial m},\;\;
\alpha =
\Big\{
\begin{array}{l}
\sigma,\\
\bar\sigma, 
\end{array} 
\end{equation}  
which is deduced through comparing Eqs.~(\ref{e46}) and (\ref{e74}).
For completeness, the non-interacting Hamiltonian
\begin{equation}
\label{e80}
\wh{\cal H}_0 {:=} \sum_{{\bF k},\sigma}
\epsilon_{{\bF k};\sigma} 
{\hat c}_{{\bF k};\sigma}^{\dag}
{\hat c}_{{\bF k};\sigma},
\end{equation}
with $\epsilon_{{\bF k};\sigma}$ as defined in Eq.~(\ref{e76}) above, 
is the so-called Kohn-Sham Hamiltonian \cite{KS65} encountered within 
the framework of the GS density-functional theory due to Hohenberg and 
Kohn \cite{HK64} (for a review see \cite{DG90}). 

By repeating the above considerations in terms of the GS energy per 
particle $\tilde{\cal E}(n)$ (an explicit function of solely the 
total concentration $n$ and not of the spin magnetization fraction 
$m$), and thus by introducing, analogously as in Eq.~(\ref{e71}) 
above, the energy functions $\tilde{\cal E}_{\rm k}(n)$ and 
$\tilde{\cal E}_{\rm xc}(n)$ satisfying 
$\tilde{\cal E}_{\rm k}(n)+\tilde{\cal E}_{\rm xc}(n) \equiv
\tilde{\cal E}(n)$, we arrive at ({\it cf}. Eq.~(\ref{e78}))
\begin{equation}
\label{e81}
\varepsilon_{\sc f} = \varepsilon_{\sc f}^{(0)} + \mu_{\rm xc},
\end{equation}
in which $\varepsilon_{\sc f}^{(0)}$ stands for the Fermi energy of 
the underlying paramagnetic non-interacting $N$-particle GS, and 
({\it cf}. Eq.~(\ref{e73}))
\begin{equation}
\label{e82}
\mu_{\rm xc} {:=} 
\frac{\rm d}{{\rm d} n}
\big(n \tilde{\cal E}_{\rm xc}(n) \big). 
\end{equation}
It is observed that, for paramagnetic GSs, Eq.~(\ref{e81}) coincides 
with Eq.~(\ref{e78}). In the case at hand, the non-interacting 
Kohn-Sham Hamiltonian $\wh{\cal H}_0$ in Eq.~(\ref{e80}) is defined 
in terms of the spin-degenerate energy dispersion ({\it cf}. 
Eq.~(\ref{e76}) above)
\begin{equation}
\label{e83}
\epsilon_{\bF k} {:=} \varepsilon_{\bF k} + \mu_{\rm xc},
\end{equation}
where $\mu_{\rm xc}$ is defined in Eq.~(\ref{e82}). In the light of
Eq.~(\ref{e81}), the result in Eq.~(\ref{e83}) expresses the fact 
that the energy eigenvalue corresponding to the highest occupied 
single-particle Kohn-Sham state coincides with the exact chemical 
potential. Following Eq.~(\ref{e77}), in the present case the 
counterpart of the expression in Eq.~(\ref{e79}) is as follows:
\begin{equation}
\label{e84}
\mu_{\rm xc} = 
\hbar\Sigma_{\sigma}^{\sc hf}({\Bf k}_{{\sc f};\sigma}),\;\;
\forall\sigma\;\;\;\;
({\Bf k}_{{\sc f};\sigma}\in {\cal S}_{{\sc f};\sigma}).
\end{equation}
Note that, since $\mu_{\rm xc}$ ($\mu_{{\rm xc};\sigma}$) is a 
potential, it is determined up to an additive constant, explicitly 
independent of $n$ ($n$ and $m$); the difference $\mu_{{\rm xc};
\sigma} -\mu_{{\rm xc};\bar\sigma}$ is, however, {\sl not} 
arbitrary.

We have employed the expression in Eq.~(\ref{e84}) specialized to 
the paramagnetic GS of the Coulomb-interacting uniform-electron-gas 
system for $d=3$, with $\varepsilon_{\bF k} = \hbar^2 
\|{\Bf k}\|^2/[2 m_{\rm e}]$, where $m_{\rm e}$ stands for the 
bare electron mass. Decomposing $\mu_{{\rm xc};\sigma}$ into 
exchange and correlation contributions, that is
\footnote{\label{f15}
Here and in the subsequent expressions we use the spin index 
$\sigma$ in order to avoid the confusion that often ensues on
suppression of $\sigma$; some workers use ${\sf n}({\Bf k})$ to 
denote ${\sf n}_{\sigma}({\Bf k})$ for paramagnetic GSs, while 
others employ ${\sf n}({\Bf k})$ to denote $\sum_{\sigma} 
{\sf n}_{\sigma}({\Bf k})$. }
\begin{equation}
\label{e85}
\mu_{{\rm xc};\sigma} = \mu_{{\rm x};\sigma} + 
\mu_{{\rm c};\sigma},
\end{equation}
in Hartree atomic units, where ${\sf g}$ is equal to unity, 
we have
\begin{eqnarray}
\label{e86}
\bar\mu_{{\rm x};\sigma} &=& 
\frac{\gamma_0}{\pi r_{\rm s}}
\int_0^{\infty} {\rm d}x\; 
x \ln\left|\frac{1-x}{1+x}\right|\,
{\sf n}_{\sigma}^{(0)}({\bar k}_{\sc f} x),\;\;\forall\sigma, \\
\label{e87}
\bar\mu_{{\rm c};\sigma} &=&
\frac{\gamma_0}{\pi r_{\rm s}}
\int_0^{\infty} {\rm d}x\; 
x \ln\left|\frac{1-x}{1+x}\right|\, \nonumber\\
& &\;\;\;\;\;\;\;\;\;\;\;\;\;\;\;\;\;\;
\times \big[ {\sf n}_{\sigma}({\bar k}_{\sc f} x) -
{\sf n}_{\sigma}^{(0)}({\bar k}_{\sc f} x) \big],\;\;\forall\sigma;
\end{eqnarray} 
here the bars (such as in ${\bar k}_{\sc f}$) indicate that the 
respective quantities are in Hartree atomic units. In 
Eqs.~(\ref{e86}) and (\ref{e87}), ${\sf n}_{\sigma}^{(0)}
({\bar k}_{\sc f} x)$ stands for the momentum distribution 
function pertaining to the GS of the non-interacting electron-gas 
system (equal to $1$ for $0\le x < 1$ and $0$ for $x > 1$) and 
$r_{\rm s}$ for the average inter-particle distance $r_{0}$ in 
units of the Bohr radius;
\begin{equation}
\label{e88}
\gamma_0 {:=} \left(\frac{9 \pi}{4}\right)^{1/3} = 1.919\dots,
\end{equation}
and ${\bar k}_{\sc f} \equiv \| {\bar {\Bf k}}_{{\sc f};\sigma}\|$
denotes the Fermi wave number in units of the inverse of the Bohr
radius, i.e.
\begin{equation}
\label{e89}
{\bar k}_{\sc f} = \frac{\gamma_0}{r_{\rm s}}.
\end{equation}
Making use of the standard result
\begin{equation}
\label{e90}
\int_0^1 {\rm d}x\; x \ln\left|\frac{1-x}{1+x}\right| = -1,
\end{equation}
from the expression in Eq.~(\ref{e86}) we trivially obtain
the well-known exact result
\begin{equation}
\label{e91}
\bar\mu_{{\rm x};\sigma} = \frac{-\gamma_0/\pi}{r_{\rm s}}
= \frac{-0.610887\dots}{r_{\rm s}}\;\;\;
\mbox{\rm (Hartree)}.
\end{equation}
We note that for $d=3$ (see Eq.~(\ref{e82}) above)
\begin{equation}
\label{e92}
\bar\mu_{{\rm xc};\sigma} = \bar{\!\tilde{\cal E}}_{\rm xc}
- \frac{r_{\rm s}}{3} \frac{\rm d}{{\rm d} r_{\rm s}}\;
\bar{\!\tilde{\cal E}}_{\rm xc}.
\end{equation}
With $\,\bar{\!\tilde{\cal E}}_{\rm xc} \equiv
\,\bar{\!\tilde{\cal E}}_{\rm x} + 
\,\bar{\!\tilde{\cal E}}_{\rm c}$ ({\it cf}. Eq.~(\ref{e85}) above), 
in which \cite{GMB57}
\begin{equation}
\label{e93}
\bar{\!\tilde{\cal E}}_{\rm x} = 
\frac{-3\gamma_0}{4\pi\, r_{\rm s}},
\end{equation}
from Eq.~(\ref{e92}) we identically reproduce the result in
Eq.~(\ref{e91}).

\begin{figure}[t!]
\protect
\centerline{
\psfig{figure=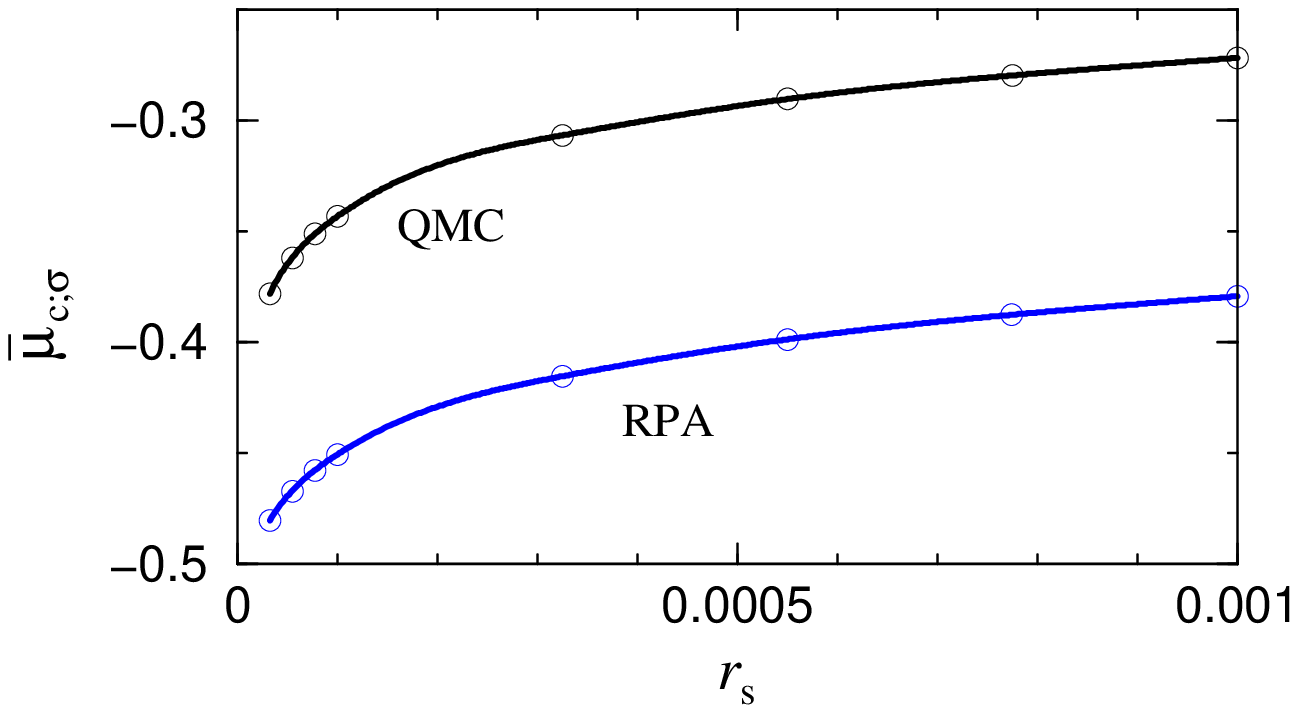,width=3.25in} }
\vskip 5pt
\caption{\label{fi1} \sf
The correlation potential $\bar\mu_{{\rm c};\sigma}$ (in Hartree 
atomic units) pertaining to the paramagnetic GS of the 
Coulomb-interacting uniform-electron-gas system for $d=3$ as 
calculated through the expression in Eq.~(\protect\ref{e87}) with 
${\sf n}_{\sigma} ({\bar k}_{\sc f} x)$ herein replaced by 
${\sf n}^{\sc rpa}_{\sigma}({\bar k}_{\sc f} x)$, due to Daniel 
and Vosko \protect\cite{DV60}, and the $\bar\mu_{{\rm c};\sigma}$ 
determined through $\bar\mu_{{\rm c};\sigma} = 
\bar{\!\tilde{\cal E}}_{\rm c} - (r_{\rm s}/3) 
{\rm d}\,\bar{\!\tilde{\cal E}}_{\rm c}/{\rm d} r_{\rm s}$
({\it cf}. Eq.~(\protect\ref{e92})) with 
$\bar{\!\tilde{\cal E}}_{\rm c}$ herein replaced by a parametrized
expression due to Vosko {\sl et al.} \protect\cite{VWN80} based 
on the quantum Monte Carlo results (curve QMC) due to Ceperley and Adler 
\protect\cite{CA80}. The open circles indicate the explicitly calculated 
data and the lines through these are cubic-splines interpolations. 
Explicit calculation shows that $\bar\mu_{{\rm c};\sigma}^{\sc qmc} 
\sim 0.0310907 \ln(r_{\rm s}) - 0.0570205$ for 
$r_{\rm s}\downarrow 0$. A least-squares fit of the expression 
$a \ln(r_{\rm s}) + b$ to curve QMC over the range presented yields 
$a=0.0308213$, $b=-0.0589913$; a similar least-squares fit to curve 
RPA over the range presented yields $a=0.0304144$, $b=-0.170157$. 
Explicit analysis of the eight presented data points for 
$\bar\mu_{{\rm c};\sigma}^{\sc rpa}$ and $\bar\mu_{{\rm c};
\sigma}^{\sc qmc}$ yields $\bar\mu_{{\rm c};\sigma}^{\sc qmc} - 
\bar\mu_{{\rm c};\sigma}^{\sc rpa} = 0.107\pm 0.002$, to be compared 
with $b\vert_{\sc qmc} - b\vert_{\sc rpa} = 0.111$. The constancy of 
$\bar\mu_{{\rm c};\sigma}^{\sc qmc} -\bar\mu_{{\rm c};\sigma}^{\sc rpa}$ 
over the entire range presented is consistent with the fact that the 
potential $\bar\mu_{{\rm c};\sigma}$ is determined up to an additive 
constant. For completeness, a somewhat more accurate, but vastly 
time-consuming, calculation of $\bar\mu_{{\rm c};\sigma}^{\sc rpa}$ 
at four different $r_{\rm s}$ values inside $[10^{-4},10^{-3}]$ 
reveals that while $\bar\mu_{{\rm c};\sigma}^{\sc qmc} 
-\bar\mu_{{\rm c};\sigma}^{\sc rpa}$ remains independent of 
$r_{\rm s}$, the absolute value of $b\vert_{\sc rpa}$ decreases 
slightly (by about $0.003$) with respect to the value presented 
above. For this reason, we do not consider $b\vert_{\sc rpa} 
= -0.170157$ as being the converged value. }
\end{figure}

\begin{figure}[t!]
\protect
\centerline{
\psfig{figure=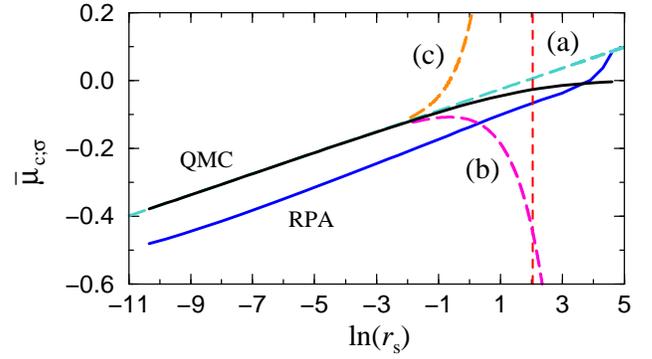,width=3.25in} }
\vskip 5pt
\caption{\label{fi2} \sf
The correlation potentials $\bar\mu_{{\rm c};\sigma}^{\sc rpa}$
and $\bar\mu_{{\rm c};\sigma}^{\sc qmc}$ over a larger range than 
in Fig.~1 (for details of the underlying calculations see the 
caption of Fig.~1). The vertical broken line at $\ln(r_{\rm s}
=7.769269\dots) = 2.050176\dots$ indicates the boundary beyond 
which $Z_{k_{{\sc f};\sigma}}^{\sc rpa} \equiv 
{\sf n}_{\sigma}^{\sc rpa}(k_{{\sc f};\sigma}^-)
-{\sf n}_{\sigma}^{\sc rpa}(k_{{\sc f};\sigma}^+)$ is negative,
indicative of the instability of the uniform paramagnetic GS of 
the system under consideration according to the RPA (see footnote
\protect\ref{f16}). Note the remarkable near-constant behaviour of 
$\bar\mu_{{\rm c};\sigma}^{\sc qmc}-\bar\mu_{{\rm c};\sigma}^{\sc rpa}$ 
over the range of validity of the RPA. That this aspect cannot
be due to smallness of higher-order correlation effects is
evident from the behaviour of finite-order asymptotic series of 
$\bar\mu_{{\rm c};\sigma}$, for $r_{\rm s}\downarrow 0$, at large 
values of $r_{\rm s}$, for which we employ (broken curves) 
$\bar\mu_{{\rm c};\sigma} \sim 0.0310907 \ln(r_{\rm s})
-0.0570205$ for curve (a), 
$\bar\mu_{{\rm c};\sigma} \sim 0.0310907 \ln(r_{\rm s}) -0.0570205 
-0.0585\, r_{\rm s}$ for curve (b) and 
$\bar\mu_{{\rm c};\sigma} \sim 0.0310907 \ln(r_{\rm s}) -0.0570205 
-0.0585\, r_{\rm s} +0.2794\, r_{\rm s}^{3/2}$ for curve (c). 
These asymptotic series correspond to $\bar\mu_{{\rm c};\sigma}^{\sc qmc}$. 
Our calculation of $\bar\mu_{{\rm c};\sigma}$ based on the expression 
in Eq.~(\protect\ref{e87}) with ${\sf n}_{\sigma}({\bar k}_{\sc f} x)$ 
therein replaced by a parametrized expression due to Gori-Giorgi and 
Ziesche \protect\cite{GGZ02} yields values between approximately 
$-5 \times 10^{-7}$ and $7\times 10^{-3}$ over the range $r_{\rm s}\, 
\protect\Ieq{\sim}{<}\, 12$ where this expression is asserted 
reliably to represent the momentum distribution function. The 
reason underlying this shortcoming is discussed in the text. }
\end{figure}

In order to display the significance of the expression for
$\bar\mu_{{\rm c};\sigma}$ in Eq.~(\ref{e87}), we employ the 
${\sf n}_{\sigma}(k)$ due to Daniel and Vosko \cite{DV60} 
evaluated within the framework of the random-phase approximation (RPA). 
\footnote{\label{f16}
Our numerical calculations reveal that for $r_{\rm s} \ge 
6.09887\dots$, ${\sf n}_{\sigma}^{\sc rpa}(k_{{\sc f};\sigma}^-) 
\le \frac{1}{2}$ (for $r_{\rm s} = 6.09887\dots$, 
${\sf n}_{\sigma}^{\sc rpa}(k_{{\sc f};\sigma}^+) = 
0.31760\dots$ so that for this $r_{\rm s}$ we have
$Z^{\sc rpa}_{{k}_{{\sc f};\sigma}} = 0.182397\dots$ and
$[{\sf n}_{\sigma}^{\sc rpa}(k_{{\sc f};\sigma}^-)
+ {\sf n}_{\sigma}^{\sc rpa}(k_{{\sc f};\sigma}^+)]/2
= 0.408801\dots$) and that for $r_{\rm s} \ge 7.769269\dots$,
${\sf n}_{\sigma}^{\sc rpa}(k_{{\sc f};\sigma}^-) \le 
{\sf n}_{\sigma}^{\sc rpa}(k_{{\sc f};\sigma}^+)$. The latter of 
course implies that, according to the RPA, the paramagnetic GS of 
the uniform electron-gas system should be unstable for $r_{\rm s} 
\ge 7.769269\dots$. For completeness, according to Bloch \cite{FB29} 
(see \cite{AM81}, pp. 682-684), within the framework of the 
Hartree-Fock approximation the latter GS is unstable towards a 
uniform ferromagnetic state for $r_{\rm s} \ge 5.4502186\dots$. 
This is suggestive of the possibility that the instability as
predicted by the RPA (which is an artefact of this approximation)
may also be one towards a ferromagnetic GS. }
Our numerical calculation of $\bar\mu_{{\rm c};\sigma}$ based 
on the expression in Eq.~(\ref{e87}) and the latter GS momentum
distribution function reproduces, to within the numerical accuracy 
of our calculation, the exact result \cite{GMB57}
\begin{equation}
\label{e94}
\bar\mu_{{\rm c};\sigma} \sim \frac{1-\ln 2}{\pi^2}\,
\ln(r_{\rm s})\;\;\; (\mbox{\rm Hartree})\;\;\;
\mbox{\rm for}\;\; r_{\rm s} \downarrow 0
\end{equation}
(Figs.~\ref{fi1} and \ref{fi2}). This result implies that for 
$r_{\rm s}\downarrow 0$ the {\sl integral} on the RHS of 
Eq.~(\ref{e87}) approaches zero like $r_{\rm s} \ln(r_{\rm s})$. 
The fact that the latter integral should be vanishing for 
$r_{\rm s} \downarrow 0$ can be understood by the observation that 
$r_{\rm s}$ is proportional to the coupling constant of interaction 
so that, for $r_{\rm s}\downarrow 0$, ${\sf n}_{\sigma}(k)
-{\sf n}_{\sigma}^{(0)}(k)$ should indeed approach zero for {\sl all}
$k$. However, the above-mentioned behaviour of the integral on the 
RHS of Eq.~(\ref{e87}) very crucially depends on the precise way in 
which ${\sf n}_{\sigma}(k)$ approaches ${\sf n}_{\sigma}^{(0)}(k)$ 
for $r_{\rm s}\downarrow 0$; even the slightest deviation of 
${\sf n}_{\sigma}(k)$ from the exact ${\sf n}_{\sigma}(k)$ for 
$r_{\rm s}\downarrow 0$ can significantly alter the behaviour of 
the latter integral for $r_{\rm s} \downarrow 0$ and consequently 
of $\bar\mu_{{\rm c};\sigma}$. We suggest therefore that any 
parametrized form of ${\sf n}_{\sigma}(k)$ should be required to 
reproduce a pre-determined $\bar\mu_{{\rm c};\sigma}$ through the 
expression in Eq.~(\ref{e87}). To illustrate the significance of
this aspect, by employing the expression in Eq.~(\ref{e92}) above 
we have determined $\bar\mu_{{\rm c};\sigma}$ through use of an 
interpolation expression for $\bar{\!\tilde{\cal E}}_{{\rm c};
\sigma}$ \cite{VWN80,VW80} based on quantum Monte Carlo calculations 
\cite{CA80} for the Coulomb-interacting uniform-electron-gas 
system at a number of densities. Comparison of this $\bar\mu_{{\rm c};
\sigma}$ with the $\bar\mu_{{\rm c};\sigma}$ obtained through 
Eq.~(\ref{e87}) in conjunction with ${\sf n}_{\sigma}^{\sc rpa}(k)$, 
reveals that to the numerical accuracy of our calculations the two 
results up to an unimportant additive constant (independent of 
$r_{\rm s}$) coincide for small values of $r_{\rm s}$ where 
${\sf n}_{\sigma}^{\sc rpa}(k)-{\sf n}_{\sigma}^{(0)}(k)$
constitutes the leading asymptotic contribution to 
${\sf n}_{\sigma}(k)-{\sf n}_{\sigma}^{(0)}(k)$ (see Fig.~\ref{fi1}). 
We have also determined $\bar\mu_{{\rm c};\sigma}$ by employing a very 
recent parametrized expression for ${\sf n}_{\sigma}(k)$, asserted by 
Gori-Giori and Ziesche \cite{GGZ02} to be valid in the range 
$r_{\rm s}\, \Ieq{\sim}{<}\, 12$, and observed {\sl no} resemblance
whatever between this $\bar\mu_{{\rm c};\sigma}$ and the aforementioned 
quantum-Monte-Carlo-based $\bar\mu_{{\rm c};\sigma}$ over the entire 
range $r_{\rm s}\, \Ieq{\sim}{<}\, 12$ (see caption of Fig.~\ref{fi2}). 
The reason for this lies in the fact that, even at $r_{\rm s}=1$, 
$\bar\mu_{{\rm c};\sigma}$ is smaller by one order of magnitude than 
$\bar\mu_{{\rm x};\sigma}$, with the disparity between these two 
quantities further increasing for smaller values of $r_{\rm s}$ (this 
as evidenced by the fact that $\bar\mu_{{\rm c};
\sigma}/\bar\mu_{{\rm x};\sigma} \to 0$ for $r_{\rm s} \downarrow 0$); 
it follows that no parameterization of ${\sf n}_{\sigma}(k)$ is likely 
to reproduce $\bar\mu_{{\rm c};\sigma}$ through the expression in 
Eq.~(\ref{e87}) unless this reproduction has explicitly been enforced 
in the determination of the parameters of ${\sf n}_{\sigma}(k)$.

\section*{\S~6.~Behaviour of 
$\lowercase{\sf n}_{\sigma}(\lowercase{\Bf k})$ for
$\lowercase{\Bf k}$ infinitesimally close to 
${\cal S}_{{\sc f};\sigma}$ }
\label{s5}

We now proceed with the investigation of the behaviour of
${\sf n}_{\sigma}({\Bf k})$ for ${\Bf k}$ approaching 
${\cal S}_{{\sc f};\sigma}$. Along the lines presented
in \cite{BF02a}, for ${\Bf k} \in {\cal S}_{{\sc f};\sigma}$
we have
\begin{equation}
\label{e95}
\Lambda_{\sigma}({\Bf k}) 
= \Gamma_{\sigma}({\Bf k}),
\end{equation}
where $\Lambda_{\sigma}({\Bf k})$ has been defined in 
Eq.~(\ref{e32}) above and
\begin{equation}
\label{e96}
\Gamma_{\sigma}({\Bf k}) {:=}
\frac{ \mu - \varepsilon_{{\bF k};\sigma}^{<} }
{ \varepsilon_{{\bF k};\sigma}^{>} - \mu}.
\end{equation}
Making use of the expressions in Eqs.~(\ref{e29}) and (\ref{e30}) 
above, we deduce that 
\begin{eqnarray}
\label{e97}
\Gamma_{\sigma}({\Bf k}) &\sim&
\frac{ - {\Bf a}({\Bf k}_{{\sc f};\sigma})
\cdot ({\Bf k} - {\Bf k}_{{\sc f};\sigma}) - {\sf g}\,
\zeta_{{\bF k};\sigma} }
{{\Bf a}({\Bf k}_{{\sc f};\sigma})
\cdot ({\Bf k} - {\Bf k}_{{\sc f};\sigma}) - {\sf g}\,
\Lambda_{\sigma}({\Bf k}) 
(\zeta_{{\bF k};\sigma} - \eta_{{\bF k};\sigma}) }, \nonumber\\
& &\;\;\;\;\;\;\;\;\;\;\;\;\;\;\;\;\;\;\;\;\;
\;\;\;\;\;\;\;\;\;\;\;\;\;\;\;\;\;\;\;\;\;\;\;\;
{\Bf k} \to {\cal S}_{{\sc f};\sigma},
\end{eqnarray}
where on account of the assumed stability of the GS of the system 
both the numerator and the denominator on the RHS must be non-negative. 
The fundamental difference between the expression on the RHS of 
Eq.~(\ref{e97}) and its counterpart specific to the uniform metallic 
GSs of the conventional single-band Hubbard Hamiltonian ({\it cf}. 
Eq.~(76) in \cite{BF02a}) arises from $\eta_{{\bF k};\sigma}$ which 
for the latter GSs is identically vanishing. Since we have earlier 
considered the Hubbard Hamiltonian in some detail \cite{BF02a}, in 
what follows, unless we explicitly indicate otherwise, $\eta_{{\bF k};
\sigma} \not\equiv 0$. As we shall see below, an identically 
non-vanishing $\eta_{{\bF k};\sigma}$ has far-reaching consequences 
for the behaviour of ${\sf n}_{\sigma}({\Bf k})$ for ${\Bf k}$ 
approaching ${\cal S}_{{\sc f};\sigma}$.

With reference to the expressions in Eqs.~(\ref{e28}) and (\ref{e26}) 
above, we consider the following behaviours for $\zeta_{{\bF k};\sigma}$ 
and $\eta_{{\bF k};\sigma}$ as ${\Bf k} \to {\Bf k}_{{\sc f};\sigma} 
\in {\cal S}_{{\sc f};\sigma}$: 
\begin{eqnarray}
\label{e98}
&&\vert\zeta_{{\bF k};\sigma}\vert\sim
\vert B({\Bf k}_{{\sc f};\sigma})\vert\,
\|{\Bf k}-{\Bf k}_{{\sc f};\sigma}\|^{\gamma},\\
\label{e99}
&&\vert\zeta_{{\bF k};\sigma}\vert\sim
\vert B({\Bf k}_{{\sc f};\sigma})\,
\ln\|{\Bf k}-{\Bf k}_{{\sc f};\sigma}\| \vert\,
\|{\Bf k}-{\Bf k}_{{\sc f};\sigma}\|^{\gamma},
\end{eqnarray}
\begin{eqnarray}
\label{e100}
&&\vert\eta_{{\bF k};\sigma}\vert\sim
\vert C({\Bf k}_{{\sc f};\sigma})\vert\,
\|{\Bf k}-{\Bf k}_{{\sc f};\sigma}\|^{\tau}, \\
\label{e101}
&&\vert\eta_{{\bF k};\sigma}\vert\sim
\vert C({\Bf k}_{{\sc f};\sigma})\,
\ln\|{\Bf k}-{\Bf k}_{{\sc f};\sigma}\| \vert\,
\|{\Bf k}-{\Bf k}_{{\sc f};\sigma}\|^{\tau},
\end{eqnarray}
where $B({\Bf k}_{{\sc f};\sigma})$, $\gamma$, $C({\Bf k}_{{\sc f};
\sigma})$ and $\tau$ are parameters. Below we consider a variety of
possibilities corresponding to various ranges of values for the 
parameters $\gamma$ and $\tau$. Since in the present case, in addition 
to the intervals in which $\gamma$ and $\tau$ lie ($0 < \gamma < 1$ or 
$\gamma \ge 1$, for instance), the precise relationship between $\gamma$ 
and $\tau$ (i.e. $\gamma < \tau$, $\gamma=\tau$ and $\gamma > \tau$) 
is significant, an exhaustive analysis of all possible cases will be 
too extensive to consider here. We shall therefore explicitly deal with 
a limited number of cases which are representative and further pave 
the way for future complementary considerations. Below we consider 
the following cases: $0 < \gamma < \tau < 1$; $0 <\tau <\gamma < 1$; 
$0 <\tau =\gamma < 1$; $\gamma = 1, \tau = 1$. For obvious reasons, 
when for instance Eq.~(\ref{e98}) applies, the cases corresponding 
to $0 < \gamma < 1$ and $\gamma=1$ should be considered separately 
whereas, when Eq.~(\ref{e99}) applies, we need to deal with one case 
only, corresponding to $0 < \gamma \le 1$. Similarly for 
Eqs.~(\ref{e100}) and (\ref{e101}) and $0 < \tau < 1$, $\tau=1$ and 
$0< \tau \le 1$ respectively. Note that in principle the values 
of $B$, $C$, $\gamma$ and $\tau$ can depend on whether ${\Bf k}$ 
approaches ${\cal S}_{{\sc f};\sigma}$ from the inside (outside) of 
the Fermi sea, in which case $B({\Bf k}_{{\sc f};\sigma})$, 
$C({\Bf k}_{{\sc f};\sigma})$, $\gamma$ and $\tau$ in Eqs.~(\ref{e98}) 
-- (\ref{e101}) should be denoted by $B({\Bf k}_{{\sc f};\sigma}^-)$, 
$C({\Bf k}_{{\sc f};\sigma}^-)$, $\gamma^-$ and $\tau^-$ 
($B({\Bf k}_{{\sc f};\sigma}^+)$, $C({\Bf k}_{{\sc f};\sigma}^+)$, 
$\gamma^+$ and $\tau^+$) respectively.

\subsection*{6.1.~The case when $\;\;0 < \gamma < 1$,
$\; 0 < \tau < 1$}
\label{s5a}

In this case from Eq.~(\ref{e97}) we obtain
\begin{equation}
\label{e102}
\Gamma_{\sigma}({\Bf k}) \sim 
\frac{1}{\Lambda_{\sigma}({\Bf k})}\,
\frac{\zeta_{{\bF k};\sigma}}
{\zeta_{{\bF k};\sigma}-\eta_{{\bF k};\sigma}},\;\;
{\Bf k} \to {\cal S}_{{\sc f};\sigma}.
\end{equation}
In the present case, the assumption with regard to the 
stability of the GS of the system under consideration
implies that 
\begin{equation}
\label{e103}
\zeta_{{\bF k};\sigma} < 0\;\;\;\mbox{\rm and}\;\;\;
\zeta_{{\bF k};\sigma} < \eta_{{\bF k};\sigma},\;\;\;
\mbox{\rm for}\;\; {\Bf k}\to {\cal S}_{{\sc f};\sigma}, 
\end{equation}
must hold; violation of these conditions imply that for ${\Bf k}$ 
sufficiently close to ${\cal S}_{{\sc f};\sigma}$, 
$\varepsilon_{{\bF k};\sigma}^{<}$ would increase above $\mu$ and
$\varepsilon_{{\bF k};\sigma}^{>}$ would decrease below $\mu$ 
respectively.

To proceed, we need to consider three different cases, corresponding 
to $\gamma < \tau$, to $\gamma > \tau$ and to $\gamma = \tau$. 
In the last case, it is relevant whether $\zeta_{{\bF k};\sigma}$
satisfies Eq.~(\ref{e98}) or Eq.~(\ref{e99}) and similarly whether 
$\eta_{{\bF k};\sigma}$ satisfies Eq.~(\ref{e100}) or Eq.~(\ref{e101}). 
The case corresponding to $\gamma=\tau=1$, with $\zeta_{{\bF k};\sigma}$ 
and $\eta_{{\bF k};\sigma}$ satisfying Eqs.~(\ref{e99}) and (\ref{e101}) 
respectively, concerns in particular (i.e. {\it a priori} not 
exclusively) systems of Coulomb-interacting fermions (see the 
paragraph subsequent to that containing Eq.~(\ref{e55}) above).

\subsubsection*{\bf 6.1.1.~$\;\;\; 0 < \gamma < \tau < 1$}
\label{s5aa}

For $\gamma < \tau$, from Eq.~(\ref{e102}) we obtain
\begin{equation}
\label{e104}
\Gamma_{\sigma}({\Bf k}) \sim 
\frac{1}{\Lambda_{\sigma}({\Bf k})},\;\;
{\Bf k} \to {\cal S}_{{\sc f};\sigma},
\end{equation}
which in conjunction with Eqs.~(\ref{e32}) and (\ref{e95}) yields
\begin{equation}
\label{e105}
{\sf n}_{\sigma}({\Bf k}_{{\sc f};\sigma}^{\mp})
= \frac{1}{2}.
\end{equation}
Note that in this case, where $\vert\zeta_{{\bF k};\sigma}\vert
> \vert\eta_{{\bF k};\sigma}\vert$ for ${\Bf k}$ sufficiently
close to ${\cal S}_{{\sc f};\sigma}$, satisfaction of the first 
requirement in Eq.~(\ref{e103}) above implies satisfaction of 
the second requirement. With
\begin{equation}
\label{e106}
{\sf n}_{\sigma}({\Bf k}_{{\sc f};\sigma})
{:=} \frac{1}{2} \big[
{\sf n}_{\sigma}({\Bf k}_{{\sc f};\sigma}^-) +
{\sf n}_{\sigma}({\Bf k}_{{\sc f};\sigma}^+) \big],
\end{equation}
from Eq.~(\ref{e105}) we obtain (see Eq.~(\ref{e41}) above)
\begin{equation}
\label{e107}
{\sf n}_{\sigma}({\Bf k}_{{\sc f};\sigma}) = \frac{1}{2},\;\;\;\;
Z_{{\bF k}_{{\sc f};\sigma}} = 0\;\;\;\; (\gamma < \tau).
\end{equation}
It follows that in the case under consideration the 
single-particle spectral function $A_{\sigma}({\Bf k}_{{\sc f};
\sigma};\varepsilon)$ is free from a quasi-particle peak at 
$\varepsilon=\varepsilon_{\sc f}$ (breakdown of the FL picture).

\subsubsection*{\bf 6.1.2.~$\;\;\; 0 < \tau < \gamma < 1$}
\label{s5ab}

For $\gamma > \tau$, from Eq.~(\ref{e102}) we have
\begin{equation}
\label{e108}
\Gamma_{\sigma}({\Bf k}) \sim 
\frac{-1}{\Lambda_{\sigma}({\Bf k})}\,
\frac{ \zeta_{{\bF k};\sigma} }
{\eta_{{\bF k};\sigma}} \sim +\infty,\;\;
{\Bf k} \to {\cal S}_{{\sc f};\sigma}.
\end{equation}
Since here $\vert\eta_{{\bF k};\sigma}\vert > \vert\zeta_{{\bF k};
\sigma}\vert$ for ${\Bf k}$ sufficiently close to ${\cal S}_{{\sc f};
\sigma}$, for such ${\Bf k}$ the stability conditions in 
Eq.~(\ref{e103}) imply that $\eta_{{\bF k};\sigma} > 0$; hence we 
have $+\infty$ on the RHS of Eq.~(\ref{e108}). The result in 
Eq.~(\ref{e108}) leads to
\begin{equation}
\label{e109}
{\sf n}_{\sigma}({\Bf k}_{{\sc f};\sigma}^{\mp}) = 1,\;\;\;\;
Z_{{\bF k}_{{\sc f};\sigma}} = 0\;\;\;\; (\gamma > \tau).
\end{equation}
In the case when such behaviour for ${\sf n}_{\sigma}({\Bf k})$ is 
unlikely, one is left to conclude that $0< \tau < \gamma < 1$ 
is equally unlikely to be satisfied for a realistic model of 
interacting fermion systems. It is interesting, however, to note
that (see our pertinent discussions in \S~1) ${\sf n}_{\sigma}
({\Bf k}_{{\sc f};\sigma}^{\mp}) = 1$ is not dissimilar to that 
observed in the experimentally determined ${\sf n}_{\sigma}({\Bf k})$ 
corresponding to Bi2212 and Bi2201, at the Fermi-surface points 
adjacent to the $\overline{M}$ points of the Brillouin zone (see 
Figs.~2(g) and (i) in \cite{CGD99}).

\subsubsection*{\bf 6.1.3.~$\;\;\;0 < \tau = \gamma \le 1\;\;$ 
\rm (includes Fermi liquids)}
\label{s5ac}

Here we explicitly deal with the instances where the expressions in 
Eqs.~(\ref{e99}) and (\ref{e101}) apply. As we have indicated earlier 
(see the paragraph following that including Eq.~(\ref{e55}) above), 
for the metallic states of fermions interacting through the long-range 
Coulomb potential, ${\Bf\nabla}_{\bF k}\Sigma_{\sigma}^{\sc hf}({\Bf k}) 
\equiv ({\sf g}/\hbar) {\Bf\nabla}_{\bF k}\vartheta_{{\bF k};\sigma}$ 
(see Eq.~(\ref{e17})) is logarithmically divergent for ${\Bf k}
={\Bf k}_{{\sc f};\sigma} \in {\cal S}_{{\sc f};\sigma}$. It follows 
therefore that the FL metallic states of Coulomb-interacting fermions 
fall into the category of systems dealt with in this Section.
\footnote{\label{f16a}
From the defining expression in Eq.~(\protect\ref{e25}) and the 
fact that $\eta_{{\bF k};\sigma} \sim 0$ for ${\Bf k} \to 
{\Bf k}_{{\sc f};\sigma} \in {\cal S}_{{\sc f};\sigma}$ (the latter 
rendering an assumption with regard to the boundedness of 
${\Bf\nabla}_{\bF k} {\sf n}_{\sigma}({\Bf k})$ for ${\Bf k} \to 
{\Bf k}_{{\sc f};\sigma}$ redundant) it follows that a possible
(logarithmic) divergence of ${\Bf\nabla}_{\bF k}\vartheta_{{\bF k};
\sigma}$ for ${\Bf k}\to {\Bf k}_{{\sc f};\sigma}$ directly implies an 
analogous divergence of ${\Bf\nabla}_{\bF k} \eta_{{\bF k};\sigma}$ 
for ${\Bf k}\to {\Bf k}_{{\sc f};\sigma}$. }

From the expressions in Eqs.~(\ref{e99}) and (\ref{e101}) it follows 
that to the leading order in $({\Bf k}-{\Bf k}_{{\sc f};\sigma})$ we 
can write 
\begin{equation}
\label{e110}
\eta_{{\bF k};\sigma} \sim \lambda_{\sigma}^{\mp}\,
\zeta_{{\bF k};\sigma},\;\;\;
{\Bf k} \in 
\Big\{
\begin{array}{l}
\mbox{\rm FS}_{\sigma}, \\
\overline{\rm FS}_{\sigma},
\end{array} 
\end{equation} 
where $\lambda_{\sigma}^{-}$ and $\lambda_{\sigma}^{+}$ stand for 
finite constants specific to ${\Bf k}={\Bf k}_{{\sc f};\sigma}$. 
Thus, in view of the expressions in Eqs.~(\ref{e99}) and (\ref{e101}), 
for $\Gamma_{\sigma}({\Bf k})$, as presented in Eq.~(\ref{e97}) above, 
we have
\begin{equation}
\label{e111}
\Gamma_{\sigma}({\Bf k}) \sim
\frac{ -{\sf g} \zeta_{{\bF k};\sigma} }
{-(1-\lambda_{\sigma}^{\mp}) {\sf g}
\Lambda_{\sigma}({\Bf k}) \zeta_{{\bF k};\sigma} } \equiv 
\frac{1}{(1-\lambda_{\sigma}^{\mp}) \Lambda_{\sigma}({\Bf k}) }.
\end{equation}
The requirement of the stability of the GS under consideration
implies that
\begin{equation}
\label{e112}
\lambda_{\sigma}^{\mp} \le 1.
\end{equation}
Solving Eq.~(\ref{e95}) for ${\Bf k}={\Bf k}_{{\sc f};\sigma}^{\mp}$,
${\Bf k}_{{\sc f};\sigma}\in {\cal S}_{{\sc f};\sigma}$, with 
$\Gamma_{\sigma}({\Bf k})$ therein replaced by the asymptotic 
expression on the RHS of Eq.~(\ref{e111}), we obtain
\begin{equation}
\label{e113}
\Lambda_{\sigma}({\Bf k}_{{\sc f};\sigma}^{\mp}) = 
\frac{1}{\sqrt{1-\lambda_{\sigma}^{\mp}} },
\end{equation}
which in conjunction with Eq.~(\ref{e32}) yields
\begin{equation}
\label{e114}
{\sf n}_{\sigma}({\Bf k}_{{\sc f};\sigma}^{\mp})
= \frac{1}{1 +\sqrt{1-\lambda_{\sigma}^{\mp}}}.
\end{equation}
From this and Eq.~(\ref{e106}) we have
\begin{equation}
\label{e115}
{\sf n}_{\sigma}({\Bf k}_{{\sc f};\sigma}) 
= \frac{1}{2} \Big(
\frac{1}{1+\sqrt{1-\lambda_{\sigma}^-}} +
\frac{1}{1+\sqrt{1-\lambda_{\sigma}^+}} \Big),
\end{equation}
which is {\sl not} necessarily equal to $\frac{1}{2}$. 

\begin{figure}[t!]
\protect
\centerline{
\psfig{figure=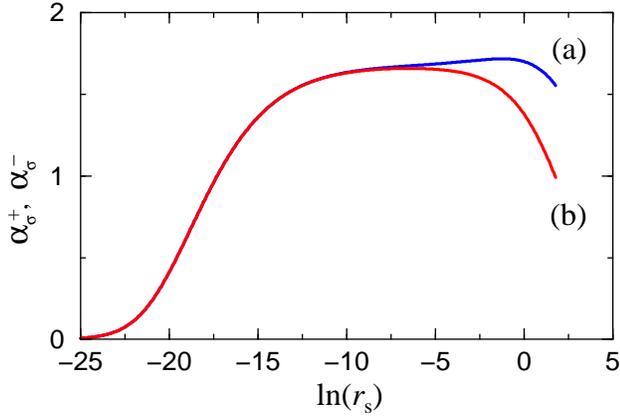,width=3.25in} }
\vskip 5pt
\caption{\label{fi3} \sf
The behaviours of $\alpha_{\sigma}^{-} {:=} 
[1 - {\sf n}_{\sigma}(k_{{\sc f};\sigma}^-)]/{\sf a}$ 
(curve (a)) and $\alpha_{\sigma}^{+} {:=} 
{\sf n}_{\sigma}(k_{{\sc f};\sigma}^+)/{\sf a}$ (curve 
(b)), where ${\sf a} {:=} r_{\rm s}/(\pi^2 \gamma_0)$, in which 
$\gamma_0$ is defined in Eq.~(\protect\ref{e88}); the results
presented here correspond to ${\sf n}_{\sigma}(k) \equiv 
{\sf n}_{\sigma}^{\sc rpa}(k)$ \protect\cite{DV60}. One observes that, 
although for sufficiently small $r_{\rm s}$, $\alpha_{\sigma}^{-}$ 
and $\alpha_{\sigma}^{+}$ approach towards the same value
$\alpha_{\sigma}$, in contrast with the finding by Daniel and Vosko 
\protect\cite{DV60} this value is {\sl not} equal to approximately 
$1.7$, a value to which $\alpha_{\sigma}^{-}$ and $\alpha_{\sigma}^{+}$ 
are indeed relatively close for $10^{-4}\protect\Ieq{\sim}{<}\, 
r_{\rm s}\,\protect\Ieq{\sim}{<}\, 1$. From the results depicted in 
this figure, one further observes that, for $r_{\rm s}\downarrow 0$ 
(explicitly, for $r_{\rm s}\, \protect\Ieq{\sim}{<}\, 10^{-4}$) 
${\sf n}_{\sigma}^{\sc rpa}(k_{{\sc f};\sigma})$, as defined according 
to Eq.~(\protect\ref{e106}), approaches $\frac{1}{2}$ more rapidly 
than would be expected from the above-mentioned finding in 
\protect\cite{DV60}. } 
\end{figure}

To gain insight into the quantitative values to be expected from 
$\lambda_{\sigma}^+$ and $\lambda_{\sigma}^-$, we recall that 
in the weak-coupling limit for the Coulomb-interacting 
uniform-electron-gas system (and thus in the paramagnetic phase 
of this system) one has 
\footnote{\label{f17}
Here we are relying on the fact that for $r_{\rm s}\downarrow 0$ 
the leading-order asymptotic behaviour of the exact  
${\sf n}_{\sigma}({\Bf k}) - {\sf n}_{\sigma}^{(0)}({\Bf k})$ 
is exactly reproduced by ${\sf n}_{\sigma}^{\sc rpa}({\Bf k}) - 
{\sf n}_{\sigma}^{(0)}({\Bf k})$, $\forall {\Bf k}$. } 
\begin{eqnarray}
\label{e116}
\left.
\begin{array}{l}
{\sf n}_{\sigma}({\Bf k}_{{\sc f};\sigma}^-) \sim 1 
-\alpha_{\sigma}^{-} {\sf a} \\ \\
{\sf n}_{\sigma}({\Bf k}_{{\sc f};\sigma}^+) \sim  
\alpha_{\sigma}^{+} {\sf a}
\end{array} \right\}\;\; \mbox{\rm for}\;\; 
r_{\rm s} \downarrow 0,
\end{eqnarray}
where 
\begin{equation}
\label{e117}
{\sf a} {:=} \frac{r_{\rm s}}{\pi^2 \gamma_0},
\end{equation}
in which $\gamma_0$ has been defined in Eq.~(\ref{e88}) above. 
The constants $\alpha_{\sigma}^{\pm}$ in Eq.~(\ref{e116}) are in 
principle bounded functions of $r_{\rm s}$; according to 
Daniel and Vosko \protect\cite{DV60}, within the framework of 
the RPA, $\alpha_{\sigma}^{-} \equiv \alpha_{\sigma}^{+} \approx 
1.7$. As can be observed from the results in Fig.~\ref{fi3}, 
$\alpha_{\sigma}^{\pm}$ are {\sl not} constants; although for 
$r_{\rm s}$ approaching zero, $\alpha_{\sigma}^{-}$ and 
$\alpha_{\sigma}^{+}$ tend towards the same value (which we denote 
by $\alpha_{\sigma}$), for sufficiently small $r_{\rm s}$, 
$\alpha_{\sigma}$ monotonically decreases towards zero, resulting 
in the fact that for $r_{\rm s} \downarrow 0$, ${\sf n}_{\sigma}
({\Bf k}_{{\sc f};\sigma})$, as defined in Eq.~(\ref{e106}) above, 
approaches $\frac{1}{2}$ more rapidly than asserted in \cite{DV60}.

Making use of the results in Eqs.~(\ref{e116}) and (\ref{e117}),
from Eq.~(\ref{e114}) we obtain
\begin{eqnarray}
\label{e118}
\left.
\begin{array}{l}
\displaystyle
\lambda_{\sigma}^- \sim 1 - 
\left(\frac{\alpha_{\sigma}^- {\sf a}}{1 -\alpha_{\sigma}^- 
{\sf a}}\right)^2
\sim 1 - (\alpha_{\sigma} {\sf a})^2\\ \\
\displaystyle
\lambda_{\sigma}^+ \sim 1 - 
\left(\frac{1 -\alpha_{\sigma}^+ {\sf a}}{\alpha_{\sigma}^+ 
{\sf a}}\right)^2 \sim 1 - \frac{1}{(\alpha_{\sigma} {\sf a})^2}
\end{array} \right\}\;\;\; r_{\rm s} \downarrow 0
\end{eqnarray}
where
\begin{equation}
\label{e119}
\alpha_{\sigma}^{\pm} \sim \alpha_{\sigma}\;\;\;
\mbox{\rm for}\;\;\; r_{\rm s}\downarrow 0. 
\end{equation}
We observe that in the regime of weak interaction, corresponding 
to $r_{\rm s} \ll 1$, $\lambda_{\sigma}^- \uparrow 1$ and 
$\lambda_{\sigma}^+ \downarrow -\infty$. We point out that similar 
expressions to Eqs.~(\ref{e114}), (\ref{e116}) and (\ref{e118}) 
apply to anisotropic metallic GSs, with the parameters therein 
dependent on the direction in the ${\Bf k}$ space along which 
${\cal S}_{{\sc f};\sigma}$ is approached.

From the above considerations it follows that in the weak-coupling
limit, corresponding to $0 \le \alpha_{\sigma} {\sf a} \ll 1$, for 
${\Bf k} \to {\cal S}_{{\sc f};\sigma}$ one must have (see 
Eqs.~(\ref{e22}) and (\ref{e23}) above)
\begin{equation}
\label{e120}
\frac{ \xi_{{\bF k};\sigma} 
-(\varepsilon_{\sc f}
-\varepsilon_{{\bF k}_{{\sc f};\sigma}})/{\sf g} }
{ \vartheta_{{\bF k};\sigma} 
-(\varepsilon_{\sc f}
-\varepsilon_{{\bF k}_{{\sc f};\sigma}})/{\sf g} } \sim
\left\{
\begin{array}{ll}
1 + \alpha_{\sigma} {\sf a}, & \mbox{\rm for}\;\; 
{\Bf k}\in \mbox{\rm FS}_{\sigma}, \\ \\
-\alpha_{\sigma} {\sf a}, & \mbox{\rm for}\;\; 
{\Bf k}\in \overline{\mbox{\rm FS}}_{\sigma}.
\end{array} \right.
\end{equation}
In arriving at this result we have made use of Eqs.~(\ref{e27}),
(\ref{e25}), (\ref{e110}), (\ref{e116}) and (\ref{e118}).

\subsection*{6.2.~The case when $\;\;\gamma=1$,
$\;\tau=1$: general}
\label{s5b}

Here we explicitly deal with the instances where the expressions in
Eq.~(\ref{e98}) and (\ref{e100}) apply. In the present case, for 
${\Bf k}\to {\Bf k}_{{\sc f};\sigma} \in {\cal S}_{{\sc f};\sigma}$ 
we have 
\footnote{\label{f18}
Here as in \cite{BF02a}, for ${\Bf k} \to {\Bf k}_{{\sc f};\sigma} 
\in {\cal S}_{{\sc f};\sigma}$ we assume (without loss of generality) 
that ${\Bf k} = {\Bf k}_{{\sc f};\sigma} \mp
\|{\Bf k}-{\Bf k}_{{\sc f};\sigma}\|\, 
\hat{\Bf n}({\Bf k}_{{\sc f};\sigma})$,
${\Bf k} \in {\rm FS}_{\sigma}, \overline{\rm FS}_{\sigma}$, 
where $\hat{\Bf n}({\Bf k}_{{\sc f};\sigma})$ stands for the
outward unit vector normal to ${\cal S}_{{\sc f};\sigma}$
at ${\Bf k}_{{\sc f};\sigma}$. Consequently, in what follows
${\Bf k}_{{\sc f};\sigma}^{\mp} {:=}
{\Bf k}_{{\sc f};\sigma} \mp \kappa\, 
\hat{\Bf n}({\Bf k}_{{\sc f};\sigma})$, with $\kappa \downarrow 0$.
See Eqs.~(88) and (89) in \cite{BF02a}. }
\begin{equation}
\label{e121}
\zeta_{{\bF k};\sigma} \sim 
{\Bf b}_{\sigma}({\Bf k}_{{\sc f};\sigma}^{\mp}) \cdot
({\Bf k}-{\Bf k}_{{\sc f};\sigma}),\;\;\;
{\Bf k} \in 
\Big\{
\begin{array}{l}
\mbox{\rm FS}_{\sigma}, \\
\overline{\rm FS}_{\sigma},
\end{array} 
\end{equation}
\begin{equation}
\label{e122}
\eta_{{\bF k};\sigma} \sim 
{\Bf c}_{\sigma}({\Bf k}_{{\sc f};\sigma}^{\mp}) \cdot
({\Bf k}-{\Bf k}_{{\sc f};\sigma}),\;\;\;
{\Bf k} \in 
\Big\{
\begin{array}{l}
\mbox{\rm FS}_{\sigma}, \\
\overline{\rm FS}_{\sigma}.
\end{array} 
\end{equation}
With
\begin{equation}
\label{e123}
b_{\sigma}^{\mp} {:=} 
{\Bf b}_{\sigma}({\Bf k}_{{\sc f};\sigma}^{\mp}) \cdot
\hat{\Bf n}({\Bf k}_{{\sc f};\sigma}),
\end{equation}
\begin{equation}
\label{e124}
d_{\sigma}^{\mp} {:=} 
{\Bf d}_{\sigma}({\Bf k}_{{\sc f};\sigma}^{\mp}) \cdot 
\hat{\Bf n}({\Bf k}_{{\sc f};\sigma}),
\end{equation}
in which
\begin{equation}
\label{e125}
{\Bf d}_{\sigma}({\Bf k}_{{\sc f};\sigma}^{\mp}) {:=}
{\Bf b}_{\sigma}({\Bf k}_{{\sc f};\sigma}^{\mp}) -
{\Bf c}_{\sigma}({\Bf k}_{{\sc f};\sigma}^{\mp}),
\end{equation}
from Eq.~(\ref{e97}) we obtain
\begin{equation}
\label{e126}
\Gamma_{\sigma}({\Bf k}) \sim 
\frac{- (a_{\sigma} + {\sf g}\, b_{\sigma}^{\mp})}
{ a_{\sigma} - {\sf g}\, \Lambda_{\sigma}({\Bf k})
d_{\sigma}^{\mp} },\;\;\;
{\Bf k} \to {\Bf k}_{{\sc f};\sigma} 
\in {\cal S}_{{\sc f};\sigma}.
\end{equation}
The assumed stability of the GS of the system under consideration 
implies that both the numerator and the denominator of the function 
on the RHS of Eq.~(\ref{e97}) must be non-negative. From this, 
considering 
\footnote{\label{f19}
Although in principle $a_{\sigma}=0$, ${\sf g}=0$ and 
$\Lambda_{\sigma}({\Bf k})=0$ are feasible (separately or in 
combination), the cases corresponding to these possibilities 
are most conveniently dealt with by considering the limits of 
the pertinent results, to be presented in the following, for 
$a_{\sigma}\downarrow 0$, ${\sf g}\downarrow 0$ and 
$\Lambda_{\sigma}({\Bf k})\downarrow 0$. }
$a_{\sigma} > 0$, ${\sf g} > 0$ and $\Lambda_{\sigma}({\Bf k}) > 0$, 
we arrive at the following conditions 
\begin{eqnarray}
\label{e127}
{} \left.
\begin{array}{ll}
\displaystyle
b_{\sigma}^- > -\frac{a_{\sigma}}{\sf g} < 0, 
&\;\;\;\;\;\mbox{\rm (I)}\\ \\
\displaystyle
d_{\sigma}^- >  \frac{a_{\sigma}}{{\sf g} 
\Lambda_{\sigma}^-} > 0, 
&\;\;\;\;\;\mbox{\rm (II)}\\ \\
\displaystyle
b_{\sigma}^+ < -\frac{a_{\sigma}}{\sf g} < 0, 
&\;\;\;\;\;\mbox{\rm (III)} \\ \\
\displaystyle
d_{\sigma}^+ < \frac{a_{\sigma}}{{\sf g} 
\Lambda_{\sigma}^+ } > 0,
&\;\;\;\;\;\mbox{\rm (IV)}
\end{array} \right.
\end{eqnarray}
where 
\begin{equation}
\label{e128}
\Lambda_{\sigma}^{\mp} {:=} 
\Lambda_{\sigma}({\Bf k}_{{\sc f};\sigma}^{\mp}).
\end{equation}

Making use of the expression in Eq.~(\ref{e126}) above, the equation 
for ${\cal S}_{{\sc f};\sigma}$ in Eq.~(\ref{e95}) reduces to the 
following quadratic equation for $\Lambda_{\sigma}^{\mp}$:
\begin{equation}
\label{e129}
{\sf g}\, d_{\sigma}^{\mp} (\Lambda_{\sigma}^{\mp})^2
- a_{\sigma} \Lambda_{\sigma}^{\mp} 
- (a_{\sigma} + {\sf g}\, b_{\sigma}^{\mp}) = 0.
\end{equation}
For interaction potentials with non-vanishing range, in general
$d_{\sigma}^{\mp} \not= b_{\sigma}^{\mp}$ (see Eq.~(\ref{e125})
above), whereby Eq.~(\ref{e129}) acquires a rich spectrum
of solutions. Considering the second-order equation
$a x^2 + b x + c = 0$, for which we have the solutions
$x_{\mp} \equiv -b/[2 a] \mp (b^2 - 4 a c)^{1/2}/[2 a]$,
below in analogy we denote the possible two solutions of 
Eq.~(\ref{e129}) by $\Lambda_{\sigma\pm}^{\mp}$. 
From Eq.~(\ref{e129}) we immediately obtain the following
results
\begin{eqnarray}
\label{e130}
&&\Lambda_{\sigma+}^{\mp} + \Lambda_{\sigma-}^{\mp}  
= \frac{a_{\sigma}}{{\sf g}\, d_{\sigma}^{\mp} },\\
\label{e131}
&&\Lambda_{\sigma+}^{\mp}\, \Lambda_{\sigma-}^{\mp}
= -\frac{ a_{\sigma} + {\sf g}\, b_{\sigma}^{\mp} }
{{\sf g}\, d_{\sigma}^{\mp} }.
\end{eqnarray}
We note that for $x_{\pm} = u \pm i v$, with $u$ and $v$ real, 
we have $x_+ x_- = u^2 + v^2 \ge 0$ from which it follows that
in cases where the RHS of Eq.~(\ref{e131}) is negative, the 
solutions (whether $\Lambda_{\sigma+}^+$ and $\Lambda_{\sigma-}^+$ 
{\sl or} $\Lambda_{\sigma+}^-$ and $\Lambda_{\sigma-}^-$) 
cannot be complex. In connection with $x_{\pm} = u \pm i v$,
we point out that, since ${\sf g}$, $a_{\sigma}$, $b_{\sigma}^{\mp}$ 
and $d_{\sigma}^{\mp}$ are real, the possible complex solutions
of Eq.~(\ref{e129}) must indeed occur in complex-conjugate pairs.

Below we separately consider $\Lambda_{\sigma}^{-}$ and
$\Lambda_{\sigma}^{+}$. In our following considerations
we shall encounter $c_{\sigma}^{\mp}$ which in analogy with 
$b_{\sigma}^{\mp}$ and $d_{\sigma}^{\mp}$ in Eqs.~(\ref{e123}) 
and (\ref{e124}) respectively are defined according to
\begin{equation}
\label{e132}
c_{\sigma}^{\mp} {:=}
{\Bf c}_{\sigma}({\Bf k}_{{\sc f};\sigma}^{\mp}) \cdot 
\hat{\Bf n}({\Bf k}_{{\sc f};\sigma}).
\end{equation}

\subsubsection*{\bf 6.2.1.~Considering 
${\sf n}_{\sigma}({\Bf k}_{{\sc f};\sigma}^-) \equiv
\Lambda_{\sigma}^{-}/(1+\Lambda_{\sigma}^-)$ }
\label{s5ba}

According to (II) in Eq.~(\ref{e127}) above, $d_{\sigma}^- > 0$. 
Since according to (I) in Eq.~(\ref{e127}), $a_{\sigma} + {\sf g}\, 
b_{\sigma}^{-} > 0$, in view of $d_{\sigma}^- > 0$ from
Eq.~(\ref{e131}) it follows that $\Lambda_{\sigma-}^{-}\, 
\Lambda_{\sigma+}^{-} < 0$. Thus, of the possible two solutions of 
Eq.~(\ref{e129}) for $\Lambda_{\sigma}^{-}$ (which by our above 
argument, presented subsequent to Eq.~(\ref{e131}), {\sl cannot} 
be complex), one is positive and one is negative; with reference
to Eq.~(\ref{e32}) above, it is the positive solution, the `physical' 
solution, that is of interest to our considerations that follow. 
Introducing
\begin{equation}
\label{e133}
{\cal L}_{\pm}(x,y) {:=} 
\frac{1}{2 x}\, \Big[ 1 \pm 
\big( 1 + 4 x (1 + y)\big)^{1/2} \Big],
\end{equation}
for the physical solution of Eq.~(\ref{e129}) for 
$\Lambda_{\sigma}^{-}$ we have
\begin{equation}
\label{e134}
\Lambda_{\sigma}^{-} = \Lambda_{\sigma+}^{-} \equiv 
{\cal L}_{+}(x_{\sigma}^-,y_{\sigma}^-),
\end{equation}
where 
\begin{equation}
\label{e135}
x_{\sigma}^- {:=} 
\frac{{\sf g}\, d_{\sigma}^-}{a_{\sigma}},\;\;\;\;\;
y_{\sigma}^- {:=} 
\frac{{\sf g}\, b_{\sigma}^-}{a_{\sigma}}
\equiv x_{\sigma}^- + \delta x_{\sigma}^-,
\end{equation}
in which (see Eq.~(\ref{e125}) above)
\begin{equation}
\label{e136}
\delta x_{\sigma}^- {:=} 
\frac{{\sf g}\, c_{\sigma}^-}{a_{\sigma}}.
\end{equation}
From (I) in Eq.~(\ref{e127}) we deduce the following two identical 
relationships
\begin{equation}
\label{e137}
y_{\sigma}^- > -1\; \Longleftrightarrow\;
1 + x_{\sigma}^- + \delta x_{\sigma}^- > 0. 
\end{equation}
From Eqs.~(\ref{e130}) and (\ref{e135}) and the fact that 
$\Lambda_{\sigma}^{-} = \Lambda_{\sigma+}^{-}$ (see Eq.~(\ref{e134}))
and $\Lambda_{\sigma-}^{-} < 0$, we further have
\begin{equation}
\label{e138}
x_{\sigma}^- = \frac{1}{\Lambda_{\sigma+}^- +
\Lambda_{\sigma-}^-},\;\;\;\;\;\;
x_{\sigma}^- > \frac{1}{\Lambda_{\sigma}^-}.
\end{equation}
Making use of the exact result 
\begin{equation}
\label{e139}
{\cal L}_{+}(x,x+\delta x)
= \frac{1}{2 x} \Big[ 1 + 2 (x +\frac{1}{2})
\Big(1 + \frac{x \delta x}{(x + 1/2)^2} \Big)^{1/2}\Big],
\end{equation}
from Eq.~(\ref{e134}) we deduce the following inequalities 
(based on $x_{\sigma}^- > 0$)
\begin{eqnarray}
\label{140}
&&\Lambda_{\sigma}^{-} \,\Ieq<> \, 1\, \nonumber\\
&&\;\;
\Longleftrightarrow \,
1 + 2 (x_{\sigma}^- +\frac{1}{2}) 
\Big(1 + \frac{ x_{\sigma}^- \delta x_{\sigma}^-}
{ (x_{\sigma}^- + 1/2)^2} \Big)^{1/2} 
\, \Ieq<>\, 2 x_{\sigma}^-,
\end{eqnarray}
which after some manipulations give rise to
\begin{equation}
\label{e141}
\Lambda_{\sigma}^{-} \,\Ieq<> \, 1\;
\Longleftrightarrow \;
\delta x_{\sigma}^- \,\Ieq<> \, -2.
\end{equation}
From Eqs.~(\ref{e136}), (\ref{e141}) and (\ref{e32}) we thus obtain
\begin{equation}
\label{e142}
c_{\sigma}^- \,\Ieq<>\, \frac{-2 a_{\sigma}}{\sf g}\;
\Longleftrightarrow \; \Lambda_{\sigma}^- \,\Ieq<>\, 1\;
\Longleftrightarrow \; 
{\sf n}_{\sigma}({\Bf k}_{{\sc f};\sigma}^-) \,\Ieq<>\,
\frac{1}{2}.
\end{equation}
Since $-2 a_{\sigma}/{\sf g} \le 0$, it follows that in the event 
$c_{\sigma}^{-} =0$ (as is the case with the single-band Hubbard 
Hamiltonian 
\footnote{\label{f20}
For instance, at half-filling, for $d=1$ and $\varepsilon_k = -2 t
\cos(k)$, $t > 0$ (here $k$ is in units of the inverse lattice 
constant so that at half-filling $\pm k_{{\sc f};\sigma} =\pm\pi/2$), 
in the limit of large $U/t$ one has the {\sl exact} result 
\protect\cite{MT77} (see also \protect\cite{CB88}) ${\sf n}_{\sigma}(k) 
\sim \frac{1}{2} + 4\ln(2)\cos(k) t/U$, $k \in [-\pi,\pi)$; it is 
seen that in this limit indeed ${\sf n}_{\sigma}(k) \ge \frac{1}{2}$ 
for $\vert k\vert \le k_{{\sc f};\sigma}$ and ${\sf n}_{\sigma}(k) 
\le \frac{1}{2}$ for $\vert k\vert \ge k_{{\sc f};\sigma}$. Similar 
behaviours obtain for $d=2$ (square lattice) and $d=3$ (simple cubic 
lattice) \protect\cite{MT77}. We point out that for $d=1$ and
$\varepsilon_k = -2 t \cos(k)$, $t > 0$, following the exact 
Bethe-Ansatz solution due to Lieb and Wu \protect\cite{LW68}, at 
half-filling the GS is insulating for {\sl all} $U > 0$, from which 
it follows that for the case at hand the absence of discontinuity in 
${\sf n}_{\sigma}(k)$ at $k=\pm\pi/2$ is {\sl not} specific to $U/t\to
\infty$, but to {\sl all} $U/t > 0$ (in fact, in this case and so 
long as $U/t > 0$, ${\rm d}^m{\sf n}_{\sigma}(k)/{\rm d}k^m$ can 
be shown to be bounded at $k=\pm\pi/2$ for {\sl any} finite value 
of $m$). Conversely, contrary to the statement in \protect\cite{MT77}, 
the absence of the usual discontinuity in ${\sf n}_{\sigma}({\Bf k})$ 
(with reference to the above case, at $k=\pm\pi/2$) does {\sl not} 
necessarily imply absence of Fermi surface and thus an insulating 
underlying GS. }
\cite{BF02a}) we have $\Lambda_{\sigma}^- \ge 1$ or 
${\sf n}_{\sigma}({\Bf k}_{{\sc f};\sigma}^-) \ge \frac{1}{2}$. The 
more general result in Eq.~(\ref{e142}), allowing for ${\sf n}_{\sigma}
({\Bf k}_{{\sc f};\sigma}^-) < \frac{1}{2}$, signifies the considerable 
influence that the (long) range of the interaction potential can have 
on the behaviour of ${\sf n}_{\sigma}({\Bf k}_{{\sc f};\sigma}^-)$.

\subsubsection*{\bf 6.2.2.~Considering 
${\sf n}_{\sigma}({\Bf k}_{{\sc f};\sigma}^+) \equiv
\Lambda_{\sigma}^{+}/(1+\Lambda_{\sigma}^+)$ }
\label{s5bb}

In contrast with $d_{\sigma}^-$ which is positive, $d_{\sigma}^+$ 
has no {\it a priori} sign; we know only that (see (IV) in 
Eq.~(\ref{e127}) above) $d_{\sigma}^+ < a_{\sigma}/({\sf g}\, 
\Lambda_{\sigma}^+)$. This necessitates us to consider separately 
the cases $d_{\sigma}^+ > 0$ and $d_{\sigma}^+ < 0$. Here we have 
however (see (III) in Eq.~(\ref{e127}) above) $-(a_{\sigma} + 
{\sf g}\, b_{\sigma}^+) > 0$. Thus following Eqs.~(\ref{e130}) and 
(\ref{e131}), for ${\sf g} > 0$ we have
\begin{equation}
\label{e143}
\Lambda_{\sigma+}^{+} + \Lambda_{\sigma-}^{+} 
\,\Ieq<>\, 0,\;\;
\Lambda_{\sigma+}^{+} \Lambda_{\sigma-}^{+} 
\,\Ieq<>\, 0 \;\Longleftrightarrow\;
d_{\sigma}^{+} \,\Ieq<>\, 0,
\end{equation}
that is, the two solutions of Eq.~(\ref{e129}) for $\Lambda_{\sigma}^{+}$ 
are positive when $d_{\sigma}^{+} > 0$ while they have different signs 
when $d_{\sigma}^{+} < 0$ (note that, by the argument presented 
following Eq.~(\ref{e131}) above, in the case where $d_{\sigma}^+ < 0$
the solutions {\sl cannot} be complex); in the latter case, the negative 
solution has a greater magnitude than the positive solution; in the 
former case ($d_{\sigma}^+ > 0$), where there are two positive 
solutions, we need to invoke the restriction $\Lambda_{\sigma}^{-} \ge 
\Lambda_{\sigma}^{+}$ (see Eq.~(\ref{e153}) below) in order to guarantee 
${\sf n}_{\sigma}({\Bf k}_{{\sc f};\sigma}^-) - {\sf n}_{\sigma}
({\Bf k}_{{\sc f};\sigma}^+) {=:} Z_{{\bF k}_{{\sc f};\sigma}} \ge 0$. 
In this connection, we note that for the case when $d_{\sigma}^{+} < 0$, 
where we have the single positive solution $\Lambda_{\sigma-}^{+}$ 
for $\Lambda_{\sigma}^{+}$, the condition $\Lambda_{\sigma}^{-} \ge 
\Lambda_{\sigma}^{+}$ is similarly {\sl not} trivially fulfilled so 
that it necessarily restricts the ranges of variation of (some of) 
the parameters of the problem at hand.

\subsubsection*{\bf 6.2.2.1.~The case when $\;\; d_{\sigma}^+ < 0$}
\label{s5bba}

Here, in view of the requirement $\Lambda_{\sigma}({\Bf k}) 
\ge 0$, $\Lambda_{\sigma}^{+}$ is to be identified with 
$\Lambda_{\sigma-}^{+}$ (following Eq.~(\ref{e143}), in the 
present case $\Lambda_{\sigma+}^{+} < 0$), that is
\begin{equation}
\label{e144}
\Lambda_{\sigma}^{+} = \Lambda_{\sigma-}^{+} \equiv
{\cal L}_{-}(x_{\sigma}^{+},y_{\sigma}^{+}),
\end{equation}
where ${\cal L}_{-}(x,y)$ has been defined in Eq.~(\ref{e133})
above, and
\begin{equation}
\label{e145}
x_{\sigma}^{+} {:=} 
\frac{ {\sf g}\, d_{\sigma}^{+} }{a_{\sigma} },\;\;\;\;
y_{\sigma}^{+} {:=} \frac{{\sf g}\, b_{\sigma}^{+} }
{ a_{\sigma} } \equiv x_{\sigma}^{+} 
+ \delta x_{\sigma}^{+},
\end{equation}
in which
\begin{equation}
\label{e146}
\delta x_{\sigma}^{+} {:=} 
\frac{{\sf g}\, c_{\sigma}^{+} }{ a_{\sigma} }.
\end{equation}
From (III) in Eq.~(\ref{e127}) we deduce the following identical
relationships ({\it cf}. Eq.~(\ref{e137}) above)
\begin{equation}
\label{e147}
y_{\sigma}^+ < -1\;
\Longleftrightarrow\; 1 + x_{\sigma}^+ + \delta x_{\sigma}^+ < 0.
\end{equation}
In addition, from Eqs.~(\ref{e130}) and (\ref{e145}) and the 
fact that $\Lambda_{\sigma}^{+} = \Lambda_{\sigma-}^{+}$ (see 
Eq.~(\ref{e144})) and $\Lambda_{\sigma+}^{+} < 0$, we have
\begin{equation}
\label{e148}
x_{\sigma}^+ = \frac{1}{\Lambda_{\sigma+}^+ +
\Lambda_{\sigma-}^+},\;\;\;\;\;\;
x_{\sigma}^+ > \frac{1}{\Lambda_{\sigma}^+}.
\end{equation}
Making use of the exact result
\begin{equation}
\label{e149}
{\cal L}_{-}(x,x+\delta x)
= \frac{1}{2 x} \Big[1 - 2 \vert x +\frac{1}{2}\vert
\Big( 1 + \frac{x\delta x}{(x + 1/2)^2} \Big)^{1/2} \Big],
\end{equation}
from Eq.~(\ref{e144}) we arrive at the following inequalities 
(based on $x_{\sigma}^{+} < 0$):
\begin{eqnarray}
\label{e150}
&&\Lambda_{\sigma}^{+} \,\Ieq<>\, 1 \, \nonumber\\
&&\;\;
\Longleftrightarrow\,
-1 + 2 \vert x_{\sigma}^{+} + \frac{1}{2}\vert
\Big( 1 + \frac{ x_{\sigma}^{+} \delta x_{\sigma}^{+}}
{(x_{\sigma}^{+} + 1/2)^2} \Big)^{1/2}
\,\Ieq<>\, -2 x_{\sigma}^{+},
\nonumber \\
\end{eqnarray}
which after some manipulations give rise to
\begin{equation}
\label{e151}
\Lambda_{\sigma}^{+} \,\Ieq<>\, 1\,
\Longleftrightarrow\,
\delta x_{\sigma}^{+} \,\Ieq><\, -2.
\end{equation}
From Eqs.~(\ref{e146}), (\ref{e151}) and (\ref{e32}) we thus
obtain
\begin{equation}
\label{e152}
c_{\sigma}^{+} \,\Ieq><\, \frac{-2 a_{\sigma} }{\sf g}\;
\Longleftrightarrow\;
\Lambda_{\sigma}^{+} \,\Ieq<>\, 1 \;
\Longleftrightarrow\;
{\sf n}_{\sigma}({\Bf k}_{{\sc f};\sigma}^+)
\,\Ieq<>\, \frac{1}{2}.
\end{equation}
The fact that here the condition ${\sf n}_{\sigma}({\Bf k}_{{\sc f};
\sigma}^+) > \frac{1}{2}$ is in principle feasible clearly signifies 
the considerable influence that a non-vanishing range of the
two-body interaction potential can have on the behaviour of 
${\sf n}_{\sigma}({\Bf k}_{{\sc f};\sigma}^+)$. The requirement 
${\sf n}_{\sigma}({\Bf k}_{{\sc f};\sigma}^-)
-{\sf n}_{\sigma}({\Bf k}_{{\sc f};\sigma}^+) {=:}
Z_{{\bF k}_{{\sc f};\sigma}} \ge 0$ implies the necessity for
the following condition to be satisfied:
\begin{equation}
\label{e153}
\frac{\Lambda_{\sigma}^-}{\Lambda_{\sigma}^+} \equiv 
\frac{{\cal L}_+(x_{\sigma}^-,y_{\sigma}^-)}
{{\cal L}_-(x_{\sigma}^+,y_{\sigma}^+)}
\ge 1\;\;\;
\mbox{\rm for}\;\;\;
\delta x_{\sigma}^-\, \Ieq><\, -2\; \wedge\;
\delta x_{\sigma}^+\, \Ieq<>\, -2.
\end{equation}
In connection with the above, we mention that the combination
$\delta x_{\sigma}^{-} > -2$, $\delta x_{\sigma}^{+} > -2$
corresponds to ${\sf n}_{\sigma}({\Bf k}_{{\sc f};\sigma}^-) > 
\frac{1}{2}$, ${\sf n}_{\sigma}({\Bf k}_{{\sc f};\sigma}^+) < 
\frac{1}{2}$ so that it does not imply any further restriction; on 
the other hand, the combination $\delta x_{\sigma}^{-} < -2$, 
$\delta x_{\sigma}^{+} < -2$ which corresponds to 
${\sf n}_{\sigma}({\Bf k}_{{\sc f};\sigma}^-) < \frac{1}{2}$, 
${\sf n}_{\sigma}({\Bf k}_{{\sc f};\sigma}^+) > \frac{1}{2}$, cannot 
be realized for stable GSs. It is for these reasons that only for the 
two combinations of $\delta x_{\sigma}^-$ and $\delta x_{\sigma}^+$ 
explicitly presented in Eq.~(\ref{e153}) can 
$\Lambda_{\sigma}^-/\Lambda_{\sigma}^+ \ge 1$ {\sl in principle} 
amount to a non-trivial condition; below we shall demonstrate
that indeed the latter {\sl cannot} be trivially satisfied. For 
completeness, $\delta x_{\sigma}^{-} = 0$ and $\delta x_{\sigma}^{+} = 0$ 
correspond to the general category of $\delta x_{\sigma}^- > -2$ and 
$\delta x_{\sigma}^+ > -2$ respectively for which, as indicated above, 
${\sf n}_{\sigma}({\Bf k}_{{\sc f};\sigma}^-) > \frac{1}{2}$ and 
${\sf n}_{\sigma}({\Bf k}_{{\sc f};\sigma}^+) < \frac{1}{2}$. In this 
connection, it is interesting to note that following Eqs.~(\ref{e146}), 
(\ref{e132}), (\ref{e122}), (\ref{e25}), (\ref{e26}) and (\ref{e17}), 
$\delta x_{\sigma}^{-} =\delta x_{\sigma}^{+} = 0$ corresponds to 
cases where (not exclusively) $\Sigma_{\sigma}^{\sc hf}({\Bf k})$ 
is independent of ${\Bf k}$ (such as is the case for the uniform 
metallic GSs of the single-band Hubbard Hamiltonian), so that with 
reference to the findings in \cite{BF02a}, the results ${\sf n}_{\sigma}
({\Bf k}_{{\sc f};\sigma}^-) > \frac{1}{2}$ and ${\sf n}_{\sigma}
({\Bf k}_{{\sc f};\sigma}^+) < \frac{1}{2}$, obtained here, were in 
fact to be expected. The fact that, for $\delta x_{\sigma}^{\mp} \equiv 
{\sf g} c_{\sigma}^{\mp}/a_{\sigma} \to 0$ one has ${\sf n}_{\sigma}
({\Bf k}_{{\sc f};\sigma}^-) > \frac{1}{2}$ and ${\sf n}_{\sigma}
({\Bf k}_{{\sc f};\sigma}^+) < \frac{1}{2}$ implies that, for interaction 
potentials of non-vanishing range, ${\sf g}/e_0$ must be greater 
than some {\sl finite} value ${\sf g}_{\rm c}/e_0$ in order for 
${\sf n}_{\sigma}({\Bf k}_{{\sc f};\sigma}^-) < \frac{1}{2}$ or 
${\sf n}_{\sigma}({\Bf k}_{{\sc f};\sigma}^+) > \frac{1}{2}$ to be 
realized ({\it cf}. footnote \ref{f16} above and footnote \ref{f23} 
below); here $e_0$ stands for an energy scale specific to the system 
under consideration.

We now demonstrate that the condition 
$\Lambda_{\sigma}^-/\Lambda_{\sigma}^+ \ge 1$ in Eq.~(\ref{e153}) 
{\sl cannot} be trivially satisfied; this we do through deducing
restricted intervals in which various parameters of the problem at 
hand (such as $x_{\sigma}^{\pm}$ and $y_{\sigma}^{\pm}$, and thereby 
$b_{\sigma}^{\pm}$, $c_{\sigma}^{\pm}$ and $d_{\sigma}^{\pm}$) have 
to be confined in order for the mentioned condition to be fulfilled.
Readers who wish to take the validity of this statement for granted, 
may omit the remaining part of this Section.

Making use of the property $x_{\sigma}^- > 0$ (see Eq.~(\ref{e135}) 
and (II) in Eq.~(\ref{e127}) above) we re-write the expression in 
Eq.~(\ref{e153}) as follows
\begin{equation}
\label{e154}
f(x_{\sigma}^-) \ge \alpha\, x_{\sigma}^-,
\end{equation}
where
\begin{equation}
\label{e155}
f(x) {:=}
1 + \big( 1 + 4 x ( 1 + x +\delta x) \big)^{1/2}, 
\end{equation}
\begin{equation}
\label{e156}
\alpha {:=} \frac{1}{x_{\sigma}^+}
\big[ 1 - \big( 1 + 4 x_{\sigma}^+ (1 + x_{\sigma}^+
+\delta x_{\sigma}^+)\big)^{1/2} \big].
\end{equation}
Above for clarity we have employed the short-hand notations 
$f(x_{\sigma}^-)$ and $\alpha$ for $f(x_{\sigma}^-,
\delta x_{\sigma}^-)$ and $\alpha(x_{\sigma}^+,\delta x_{\sigma}^+)$ 
respectively. When necessary, we shall employ the latter more
complete notations. We have
\begin{equation}
\label{e157}
f(x_{\sigma}^-) \sim 2 x_{\sigma}^- + (2+\delta x_{\sigma}^-)
{=:} g(x_{\sigma}^-)\;\;\;
\mbox{\rm for}\;\;\; x_{\sigma}^- \to \infty.
\end{equation} 
For the considerations that follow it is crucial to realize that 
in consequence of Eq.~(\ref{e137}) and (II) in Eq.~(\ref{e127}), 
$x_{\sigma}^-$ is bound to satisfy
\begin{equation}
\label{e158}
x_{\sigma}^- > {\tilde x}_{\sigma}^-\;\;\;
\mbox{\rm where}\;\;\;
{\tilde x}_{\sigma}^- {:=} \max\{0,-(1+\delta x_{\sigma}^-)\}.
\end{equation}
Consequently, unless indicated or implied otherwise, in what follows 
we implicitly assume that $x_{\sigma}^-$ satisfies the condition in 
Eq.~(\ref{e158}). We have
\begin{equation}
\label{e159}
f({\tilde x}_{\sigma}^-) = 2.
\end{equation}

The graphical representation of the functions on both sides of 
Eq.~(\ref{e154}) clearly reveals the conditions under which the 
inequality in Eq.~(\ref{e154}) (or Eq.~(\ref{e153})) is satisfied. 
One readily verifies that, for $-2 \le \delta x_{\sigma}^- \le 0$, 
$f(x_{\sigma}^-)$ possesses a global minimum, equal to $1+\big(1 - 
(1+\delta x_{\sigma}^-)^2\big)^{1/2}$, at $x_{\sigma}^- = - 
(1+\delta x_{\sigma}^-)/2$; for $\delta x_{\sigma}^- < -2$,
$f(x_{\sigma}^-)$ is complex-valued at $x_{\sigma}^- = - 
(1+\delta x_{\sigma}^-)/2$. It is straightforwardly shown that 
for $\delta x_{\sigma}^- < -2$ and $\delta x_{\sigma}^- > 0$,
$f(x_{\sigma}^-)$ approaches the linear asymptote $g(x_{\sigma}^-)$
(defined in Eq.~(\ref{e157}) above) from below, whereas for 
$-2 < \delta x_{\sigma}^- < 0$ it approaches $g(x_{\sigma}^-)$ 
from above. From these observations we deduce that the inequality 
in Eq.~(\ref{e154}) is {\sl always} violated when the following 
conditions are  satisfied:
\begin{eqnarray}
\label{e160}
&&\alpha \ge 2\; \wedge \; \delta x_{\sigma}^- < -2.
\end{eqnarray}
On the other hand, the relationship in Eq.~(\ref{e154}) is
{\sl always} satisfied when either
\begin{equation}
\label{e161}
\alpha \le 2 \;\wedge  \delta x_{\sigma}^- > -2
\end{equation}
or
\begin{equation}
\label{e162}
\alpha \le \frac {2}{-(1+\delta x_{\sigma}^-)}\; \wedge\; 
\delta x_{\sigma}^- < -2.
\end{equation}
Note that, for $\delta x_{\sigma}^- \le -2$, we have
$-2/(1+\delta x_{\sigma}^-) \in (0,2]$.

One trivially obtains the following for $\alpha \equiv
\alpha(x_{\sigma}^+,\delta x_{\sigma}^+)$
\begin{equation}
\label{e163}
\alpha \,\Ieq<>\, 2 \;\;\;
\mbox{\rm for}\;\;\;
\delta x_{\sigma}^+ \,\Ieq><\, -2.
\end{equation}
In this connection, note that in general (see Eq.~(\ref{e147}) above)
\begin{equation}
\label{e164}
x_{\sigma}^+ < -(1+\delta x_{\sigma}^+).
\end{equation}
In the present case where $x_{\sigma}^+ < 0$ (owing to $d_{\sigma}^+ 
< 0$ considered here; see Eq.~(\ref{e145}) above), $x_{\sigma}^+$ is 
bound to satisfy
\begin{equation}
\label{e165}
x_{\sigma}^+ < \min\{0, -(1+\delta x_{\sigma}^+)\}.
\end{equation}
Combining the above results, we conclude that the inequality
in Eq.~(\ref{e154}) (or Eq.~(\ref{e153})) is violated when the 
following conditions are satisfied:
\begin{eqnarray}
\label{e166}
\delta x_{\sigma}^- < -2 \wedge\,
\delta x_{\sigma}^+ < -2. 
\end{eqnarray}
These conditions are precisely those we excluded for stable GSs 
in our discussions following Eq.~(\ref{e153}) above. 

Considering the case corresponding to ({\it cf.} Eq.~(\ref{e153}) above) 
$\delta x_{\sigma}^- > -2$ and $\delta x_{\sigma}^+ < -2$ (the latter 
according to Eq.~(\ref{e163}) implying $\alpha > 2$), from our above 
considerations we deduce that the condition in Eq.~(\ref{e154}) (or, 
equivalently, that in Eq.~(\ref{e153})) is satisfied provided that 
(see Eq.~(\ref{e158}) above)
\begin{equation}
\label{e167}
{\tilde x}_{\sigma}^- < x_{\sigma}^- <
\frac{2 \alpha + 4(1+\delta x_{\sigma}^-)}
{\alpha^2 - 4},\;\;\;\;\;\;\;\;\alpha > 2.
\end{equation}
As for the case corresponding to ({\it cf.} Eq.~(\ref{e153}) above) 
$\delta x_{\sigma}^- < -2$ and $\delta x_{\sigma}^+ > -2$ (the latter 
implying $\alpha < 2$; see above, and in particular Eq.~(\ref{e163})), 
from our above considerations we similarly deduce that the condition 
in Eq.~(\ref{e154}) (or, equivalently, that in Eq.~(\ref{e153})) is 
satisfied provided that
\begin{equation}
\label{e168}
x_{\sigma}^- \ge \frac{-4 (1+\delta x_{\sigma}^-) - 2\alpha}
{4 -\alpha^2},\;\;\;\;\;
\frac{2}{-(1+\delta x_{\sigma}^-)} \le \alpha < 2.
\end{equation}
Note that for $\alpha=-2/(1+\delta x_{\sigma}^-)$ the RHS of 
the above inequality for $x_{\sigma}^-$ is, as expected, equal 
to $-(1+\delta x_{\sigma}^-)\equiv 2/\alpha$; the condition
considered here, namely $\delta x_{\sigma}^- < -2$, implies that 
$-(1+\delta x_{\sigma}^-) > 1$.

\subsubsection*{\bf 6.2.2.2.~The case when $\;\; d_{\sigma}^+ > 0$}
\label{s5bbb}

As we have indicated above, in the present case both solutions 
$\Lambda_{\sigma-}^{+}$ and $\Lambda_{\sigma+}^{+}$ of 
Eq.~(\ref{e129}) are positive, with $\Lambda_{\sigma-}^{+} \le 
\Lambda_{\sigma+}^{+}$. For these solutions we have
\begin{equation}
\label{e169}
\Lambda_{\sigma\pm}^{+} 
= {\cal L}_{\pm}(x_{\sigma}^{+},y_{\sigma}^{+}),
\end{equation}
where $x_{\sigma}^{+}$ and $y_{\sigma}^{+}$ have been defined in 
Eq.~(\ref{e145}) above. It is readily verified that
\begin{eqnarray}
\label{e170}
\left. \begin{array}{l}
\Lambda_{\sigma+}^{+}\,\Ieq<>\, 1 \\ \\
\Lambda_{\sigma-}^{+}\,\Ieq><\, 1
\end{array} \right\} \;\;\; \delta x_{\sigma}^{+}\,\Ieq<>\, -2.
\end{eqnarray}
In view of the fact that $\Lambda_{\sigma-}^{+} \le 
\Lambda_{\sigma+}^{+}$, it follows that we solely need to consider 
the case
\begin{equation}
\label{e171}
0 < \Lambda_{\sigma-}^{+} \le 1 \le \Lambda_{\sigma+}^{+}
\;\;\;\mbox{\rm for}\;\;\;
\delta x_{\sigma}^{+} \ge -2,
\end{equation}
which in fact shows the necessity for
\begin{equation}
\label{e172}
c_{\sigma}^{+} \ge \frac{-2 a_{\sigma} }{\sf g}.
\end{equation} 
According to Eq.~(\ref{e142}), for $c_{\sigma}^{-} < 
-2 a_{\sigma}/{\sf g}$, ${\sf n}_{\sigma}({\Bf k}_{{\sc f};\sigma}^-)
< \frac{1}{2}$ so that $\Lambda_{\sigma+}^{+}$ (which according to 
Eq.~(\ref{e152}) is greater than unity for $c_{\sigma}^{+} > -2 
a_{\sigma}/{\sf g}$ and thus following Eq.~(\ref{e32}) corresponds to 
${\sf n}_{\sigma}({\Bf k}_{{\sc f};\sigma}^+) > \frac{1}{2}$) is {\sl not} 
permitted. It follows that $\Lambda_{\sigma-}^{+}$ is the appropriate 
solution to Eq.~(\ref{e129}) for $\Lambda_{\sigma}^{+}$; for this 
solution to correspond to a non-negative $Z_{{\bF k}_{{\sc f};
\sigma}}$, it is required that the following condition be satisfied:
\begin{equation}
\label{e173}
\frac{\Lambda_{\sigma-}^{+}}{\Lambda_{\sigma}^{-}} \equiv 
\frac{{\cal L}_-(x_{\sigma}^+,y_{\sigma}^+)}
{{\cal L}_+(x_{\sigma}^-,y_{\sigma}^-)} \le 1\;\;\;
\mbox{\rm for}\;\;\;
\delta x_{\sigma}^- \le -2\; \wedge\;
\delta x_{\sigma}^+ \ge -2.
\end{equation}
Here $x_{\sigma}^-$ naturally satisfies the inequality in
Eq.~(\ref{e158}) above. On the other hand, since in this Section 
we deal with $d_{\sigma}^+ > 0$, from Eqs.~(\ref{e145}) and 
(\ref{e147}) we have ({\it cf}. Eq.~(\ref{e165}) above)
\begin{equation}
\label{e174}
0 < x_{\sigma}^+ < -(1+\delta x_{\sigma}^+).
\end{equation}
The inequalities in Eqs.~(\ref{e171}) and (\ref{e174}) restrict the 
range of variation in $\delta x_{\sigma}^+$ as follows:
\begin{equation}
\label{e175}
-2 \le \delta x_{\sigma}^+ < -(1+x_{\sigma}^+) < -1.
\end{equation}
The condition $\Lambda_{\sigma-}^+/\Lambda_{\sigma}^- \le 1$ in 
Eq.~(\ref{e173}) can be cast into the form $f(x_{\sigma}^-) \ge 
\alpha\, x_{\sigma}^-$ presented in Eq.~(\ref{e154}) above, with 
$f(x)$ and $\alpha$ as defined in Eqs.~(\ref{e155}) and (\ref{e156}) 
respectively. In spite of this, one should realize that here 
$x_{\sigma}^+ > 0$ (following $d_{\sigma}^+ > 0$; see 
Eq.~(\ref{e145}) above), whereas in our previous considerations 
$x_{\sigma}^+ < 0$. 

As in the case of Eq.~(\ref{e153}), below we show that the
condition $\Lambda_{\sigma-}^{+}/\Lambda_{\sigma}^{-} \le 1$
in Eq.~(\ref{e173}) {\sl cannot} be trivially satisfied. 
Readers who wish to take the validity of this statement for 
granted, may omit the following, up to and including
Eq.~(\ref{e179}) below.

Along the same lines as in the case of $d_{\sigma}^+ < 0$, we 
deduce that the inequality in Eq.~(\ref{e173}) is {\sl always} 
violated when
\begin{equation}
\label{e176}
\alpha \ge 2\,\wedge\, \delta x_{\sigma}^- < -2.
\end{equation}
On the other hand, the inequality in Eq.~(\ref{e173}) is {\sl always} 
satisfied when
\begin{equation}
\label{e177}
\alpha \le \frac{2}{-(1+\delta x_{\sigma}^-)}\,\wedge\,
\delta x_{\sigma}^- < -2.
\end{equation}
From the defining expression for $\alpha$ in Eq.~(\ref{e156}) it 
can be readily deduced that for $0 < x_{\sigma}^+ < -(1+\delta 
x_{\sigma}^+)$ (see Eq.~(\ref{e174}) above), $\alpha$ can take
non-negative values only for $-2 < \delta x_{\sigma}^+ < -1$. 
In fact we have
\begin{equation}
\label{e178}
0 \le \alpha \le 2\;\;
\mbox{\rm for}\, -2 \le \delta x_{\sigma}^+ \le -1\,
\wedge\, 0 \le x_{\sigma}^+ \le -(1+\delta x_{\sigma}^+);
\end{equation}
$\alpha=2$ applies only for $\delta x_{\sigma}^+=-2$ and 
$x_{\sigma}^+ \in [0,1]$. Assuming that $x_{\sigma}^+$ and 
$\delta x_{\sigma}^+$ are such that $\alpha \ge 
-2/(1+\delta x_{\sigma}^-)$, it is readily deduced that the 
inequality in Eq.~(\ref{e173}) is satisfied provided that
\begin{equation}
\label{e179}
x_{\sigma}^- \ge \frac{-4 (1+\delta x_{\sigma}^-) - 2\alpha}
{4 -\alpha^2},\;\;\;\;\;
\frac{2}{-(1+\delta x_{\sigma}^-)} \le \alpha < 2.
\end{equation}

Following Eq.~(\ref{e142}), for $c_{\sigma}^{-} > -2 a_{\sigma}/{\sf g}$
(or, equivalently, $\delta x_{\sigma}^- > -2$), ${\sf n}_{\sigma}
({\Bf k}_{{\sc f};\sigma}^-) > \frac{1}{2}$ so that, unless
$\Lambda_{\sigma+}^{+} > \Lambda_{\sigma}^-$ (which entails
$Z_{{\bF k}_{{\sc f};\sigma}} < 0$), $\Lambda_{\sigma+}^{+}$ cannot be 
{\it a priori} discarded. Evidently, since here $\Lambda_{\sigma-}^{+} 
> 0$, in the event $\Lambda_{\sigma+}^{+} \le \Lambda_{\sigma}^-$, one 
will be confronted with a situation where there are two possible values 
to be taken by ${\sf n}_{\sigma}({\Bf k}_{{\sc f};\sigma}^+)$. Which 
of the two is the correct value for ${\sf n}_{\sigma}({\Bf k}_{{\sc f};
\sigma}^+)$ is a matter that has to be decided on the basis of a 
further consistency test. Below we shall present an example of such 
a test but, before doing so, we derive the conditions to be satisfied 
for the occurrence of $\Lambda_{\sigma+}^{+} \le \Lambda_{\sigma}^-$. 
Readers who do not wish to consider this aspect, may omit the 
following, up to the last paragraph of this Section.

It can be easily shown that for $x_{\sigma}^- > 0$ (which holds 
on account of (II) in Eq.~(\ref{e127}) above) we have
\begin{equation}
\label{e180}
\Lambda_{\sigma+}^{+} \le \Lambda_{\sigma}^-\,
\Longleftrightarrow\,
f(x_{\sigma}^-) \ge \beta\, x_{\sigma}^-,
\end{equation}
where $f(x)$ has been defined in Eq.~(\ref{e155}) above and
\begin{equation}
\label{e181}
\beta {:=} \frac{1}{x_{\sigma}^+}\,
f(x_{\sigma}^+,\delta x_{\sigma}^+).
\end{equation}
It is readily verified that, for $x_{\sigma}^+ \ge 0$, $\beta \ge 2$.
For $x_{\sigma}^+ > 0$ (which holds on account of our present 
assumption, namely $d_{\sigma}^+ > 0$) one equivalently has
\begin{equation}
\label{e182}
\Lambda_{\sigma+}^{+} \le \Lambda_{\sigma}^-\,
\Longleftrightarrow\,
f(x_{\sigma}^+) \le \beta'\, x_{\sigma}^+,
\end{equation}
where
\begin{equation}
\label{e183}
\beta' {:=} \frac{1}{x_{\sigma}^-}\,
f(x_{\sigma}^-,\delta x_{\sigma}^-).
\end{equation}
It is easily deduced that for $x_{\sigma}^- \ge 0$, $\beta'\ge 2$.
On account of $\delta x_{\sigma}^- > -2$ (see the text following 
Eq.~(\ref{e179}) above), along the lines of our earlier 
considerations we deduce
\begin{equation}
\label{e184}
f(x_{\sigma}^-) \ge \beta\, x_{\sigma}^-
\Longleftrightarrow\,
x_{\sigma}^- \le 
\frac{2\beta + 4 (1+\delta x_{\sigma}^-)}
{\beta^2 -4}.
\end{equation}
Similarly
\begin{equation}
\label{e185}
f(x_{\sigma}^+) \le \beta'\, x_{\sigma}^+
\Longleftrightarrow\,
x_{\sigma}^+ \ge 
\frac{2\beta' + 4 (1+\delta x_{\sigma}^+)}
{{\beta'}^2 -4}.
\end{equation}
Note that from the defining expressions in Eqs.~(\ref{e181}) and 
(\ref{e183}) and the expressions in Eqs.~(\ref{e169}) and (\ref{e134}) 
respectively (see also Eq.~(\ref{e133}) above) we have
\begin{equation}
\label{e186}
\beta = 2 \Lambda_{\sigma+}^+,\;\;\;\;
\beta'= 2 \Lambda_{\sigma+}^-\, (\equiv 2 \Lambda_{\sigma}^-).
\end{equation}

In order to establish the conditions under which $\Lambda_{\sigma+}^{+} 
\le \Lambda_{\sigma}^-$ is realized (if at all), it is natural first to
establish the condition(s) required for the realization of
$\Lambda_{\sigma+}^{+} = \Lambda_{\sigma}^-$. For $\Lambda_{\sigma+}^{+}
= \Lambda_{\sigma}^-$ to materialize, in which case $\beta = \beta'$ 
(following Eq.~(\ref{e186})), we must have 
\begin{equation}
\label{e187}
x_{\sigma}^- = 
\frac{2\beta + 4 (1+\delta x_{\sigma}^-)}
{\beta^2 -4}\;\wedge\;
x_{\sigma}^+ =
\frac{2\beta + 4 (1+\delta x_{\sigma}^+)}
{\beta^2 -4}.
\end{equation}
On the other hand, these expressions cannot be meaningful unless 
$y_{\sigma}^- > -1$ (or $x_{\sigma}^- > {\tilde x}_{\sigma}^-$; 
see Eqs.~(\ref{e137}) and (\ref{e158}) above) and $y_{\sigma}^+ < -1$
(see Eq.~(\ref{e147}) above), from which it follows that (note that 
$\beta > 2$)
\begin{equation}
\label{e188}
\delta x_{\sigma}^- \ge -1 -\frac{2}{\beta}\;\wedge\;
-2 \le \delta x_{\sigma}^+ \le -1 -\frac{2}{\beta}.
\end{equation}
For $\beta > 2$, these inequalities are more restrictive than 
$\delta x_{\sigma}^- \ge -2$ and $-2 \le \delta x_{\sigma}^+ \le 
-1$ respectively. Note in particular that the range of variation 
in $\delta x_{\sigma}^+$ as required for the satisfaction of 
$\Lambda_{\sigma+}^{+} = \Lambda_{\sigma}^-$ becomes narrower, 
the closer to $2$ that $\beta$ approaches. The inequalities in 
Eq.~(\ref{e188}) imply the following intervals for $x_{\sigma}^-$ 
and $x_{\sigma}^+$ corresponding to $\beta = \beta'$ (or 
$\Lambda_{\sigma+}^+ = \Lambda_{\sigma}^-$)
\begin{equation}
\label{e189}
x_{\sigma}^- \ge \frac{2}{\beta}\;\wedge\;
\frac{2}{\beta + 2} \le x_{\sigma}^+ \le \frac{2}{\beta}.
\end{equation}
One observes that in the event $x_{\sigma}^- = x_{\sigma}^+$, there is 
only one possibility for $\Lambda_{\sigma+}^{+} = \Lambda_{\sigma}^-$ 
to be realized, namely when $x_{\sigma}^- = x_{\sigma}^+ = 2/\beta$. 
In view of the results in Eqs.~(\ref{e138}) and (\ref{e148}), 
$x_{\sigma}^- = x_{\sigma}^+$ in combination with $\Lambda_{\sigma+}^{+} 
= \Lambda_{\sigma}^-$ (note that $\Lambda_{\sigma}^- \equiv 
\Lambda_{\sigma+}^-$ since $\Lambda_{\sigma-}^- \le 0$) imply
$\Lambda_{\sigma-}^{+} = \Lambda_{\sigma-}^- = 0$. Since the choice 
of $\Lambda_{\sigma-}^+$ for $\Lambda_{\sigma}^+$ (which is the 
unrivalled choice in cases where $d_{\sigma}^+ < 0$) in the case when
$\Lambda_{\sigma-}^+=0$ implies that ${\sf n}_{\sigma}({\Bf k}_{{\sc f};
\sigma}^+) = 0$, and this in turn amounts to the highly unusual 
condition for the single-particle spectral function, namely
$A_{\sigma}({\Bf k}_{{\sc f};\sigma}^+;\varepsilon) \equiv 0$ for 
{\sl all} $\varepsilon <\mu$, we conclude that, in the event 
$x_{\sigma}^- = x_{\sigma}^+$ {\it and} $\Lambda_{\sigma+}^{+} =
\Lambda_{\sigma}^-$, $\Lambda_{\sigma}^+$ is to be identified with 
$\Lambda_{\sigma+}^+$. If follows that, in such event, 
$Z_{{\bF k}_{{\sc f};\sigma}} = 0$. Note that the inequalities in 
Eq.~(\ref{e189}) show that $x_{\sigma}^- \ge x_{\sigma}^+$ which in 
consequence of the results in Eq.~(\ref{e187}) implies that, for 
$\Lambda_{\sigma+}^{+} = \Lambda_{\sigma}^-$ to be realized, it is 
necessary that $\delta x_{\sigma}^- \ge \delta x_{\sigma}^+$;
the equality $x_{\sigma}^- =x_{\sigma}^+$ implies the equality 
$\delta x_{\sigma}^- = \delta x_{\sigma}^+$, which following the 
inequalities in Eq.~(\ref{e188}) is only possible if 
$\delta x_{\sigma}^- = \delta x_{\sigma}^+ = 2/\beta$.

It can be shown that $f(x,\delta x)/x$ (and therefore $\beta$ and 
$\beta'$ for $x=x_{\sigma}^+, \delta x = \delta x_{\sigma}^+$ and 
$x=x_{\sigma}^-, \delta x = \delta x_{\sigma}^-$, respectively;
see Eqs.~(\ref{e181}) and (\ref{e183})) is a monotonically 
decreasing function of $x$ (here for $x > 0$) and a 
monotonically increasing function of $\delta x$ (here for 
$\delta x$ satisfying the appropriate inequality in Eq.~(\ref{e188}) 
above). These properties are properly reflected in the asymptotic 
expression
\begin{equation}
\label{e190}
\frac{1}{x} f(x,\delta x) \sim 2 + \frac{2 +\delta x}{x}\;\;\;
\mbox{\rm for}\;\;\; x\to \infty,
\end{equation}
from which one further observes that so long as $\delta x > -2$,
$f(x,\delta x)/x$ (and thus $\beta$ and $\beta'$) never attains
the value $2$ for any finite $x$.

From the above properties it follows that, by starting from the set 
of values for $x_{\sigma}^-$, $\delta x_{\sigma}^-$, $x_{\sigma}^+$ 
and $\delta x_{\sigma}^+$  corresponding to $\Lambda_{\sigma+}^{+} 
= \Lambda_{\sigma}^-$, one can achieve the inequality 
$\Lambda_{\sigma+}^{+} < \Lambda_{\sigma}^-$ by (if possible; see 
the inequalities in Eq.~(\ref{e189}) above) increasing $x_{\sigma}^+$ 
and/or (if possible) decreasing $x_{\sigma}^-$ while maintaining the 
values of $\delta x_{\sigma}^+$ and $\delta x_{\sigma}^-$ corresponding 
to $\Lambda_{\sigma+}^{+} = \Lambda_{\sigma}^-$ (see Eq.~(\ref{e186}) 
above according to which $\beta = 2 \Lambda_{\sigma+}^+$ and $\beta' 
= 2 \Lambda_{\sigma}^-$). Alternatively, $\Lambda_{\sigma+}^{+} < 
\Lambda_{\sigma}^-$ can be achieved by maintaining the initial values 
of $x_{\sigma}^+$ and $x_{\sigma}^-$ while (if possible) increasing 
$\delta x_{\sigma}^-$ and/or decreasing $\delta x_{\sigma}^+$ from
their initial values. Evidently, these choices are {\sl not} unique, 
and one can obtain (if at all possible) $\Lambda_{\sigma+}^{+} < 
\Lambda_{\sigma}^-$ by other apparent combinations of changes in the 
parameters corresponding to the condition $\Lambda_{\sigma+}^{+} = 
\Lambda_{\sigma}^-$. 

We now deal with a specific case which illustrates in how far the 
condition $\Lambda_{\sigma+}^{+} < \Lambda_{\sigma}^-$ can be 
considered as feasible; it turns out that for the specific case that 
we consider, $\Lambda_{\sigma+}^{+} < \Lambda_{\sigma}^-$ is {\sl never} 
realized. In \S~6.3 below we investigate FL metallic states 
corresponding to two-body interaction potentials that are of shorter 
range than the Coulomb potential and arrive at two specific conditions 
(see Eqs.~(\ref{e195}) and (\ref{e196}) below) which result in 
$x_{\sigma}^+ = -{\sf g} b_{\sigma}^-/(a_{\sigma} \Lambda_{\sigma}^+)$ 
and $x_{\sigma}^- = -{\sf g} b_{\sigma}^+/(a_{\sigma} 
\Lambda_{\sigma}^-)$. From these we deduce that
\begin{equation}
\label{e191}
\delta x_{\sigma}^- = -x_{\sigma}^- - 
\Lambda_{\sigma}^+ x_{\sigma}^+,\;\;\;
\delta x_{\sigma}^+ = -x_{\sigma}^+ - 
\Lambda_{\sigma}^- x_{\sigma}^-,
\end{equation}
which lead to
\begin{equation}
\label{e192}
\delta x_{\sigma}^- - \delta x_{\sigma}^+ 
= (\Lambda_{\sigma}^- - 1) x_{\sigma}^- 
- (\Lambda_{\sigma}^+ - 1) x_{\sigma}^+.
\end{equation}
From this it follows that, for $\Lambda_{\sigma}^-
= \Lambda_{\sigma}^+$ and in the event $\Lambda_{\sigma}^+$
were to be identified with $\Lambda_{\sigma+}^+ \equiv \beta/2$,
through employing the expressions for $x_{\sigma}^-$ and
$x_{\sigma}^+$ in Eq.~(\ref{e187}) one would have 
\begin{equation}
\label{e193}
\delta x_{\sigma}^- - \delta x_{\sigma}^+
= \frac{2}{\beta + 2} (\delta x_{\sigma}^- - \delta x_{\sigma}^+);
\end{equation}
this and the fact that $\beta > 2$ imply that
\begin{equation}
\label{e194}
\delta x_{\sigma}^- = \delta x_{\sigma}^+ =  -1 -\frac{2}{\beta}.
\end{equation}
In arriving at the result on the RHS we have made use of the inequalities 
in Eq.~(\ref{e188}) above, which uniquely determine the only common value 
that $\delta x_{\sigma}^-$ and $\delta x_{\sigma}^+$ can take in the case 
under consideration. From the expressions in Eq.~(\ref{e187}) we observe 
that for $\delta x_{\sigma}^- = \delta x_{\sigma}^+ = -1 -2/\beta$, indeed 
$x_{\sigma}^- = x_{\sigma}^+ = 2/\beta$ and consequently, through 
Eq.~(\ref{e191}), $\Lambda_{\sigma}^- = \Lambda_{\sigma}^+ = \beta/2$. 
For completeness, here $\Lambda_{\sigma}^- =\Lambda_{\sigma+}^-$, 
$\Lambda_{\sigma}^+ = \Lambda_{\sigma+}^+$ and $\Lambda_{\sigma-}^- 
= \Lambda_{\sigma-}^- = 0$. We observe that, because in the present 
case $x_{\sigma}^-$ and $x_{\sigma}^+$ achieve respectively the 
highest and the lowest values that are compatible with the 
inequalities in Eq.~(\ref{e189}) (similarly and equivalently, for 
$\delta x_{\sigma}^-$ and $\delta x_{\sigma}^+$ according to the 
inequalities in Eq.~(\ref{e188}) above), it is {\sl not} possible to 
realize the strict inequality $\Lambda_{\sigma+}^+ < \Lambda_{\sigma}^-$ 
by appropriately altering the parameters (as described above) 
$x_{\sigma}^{\pm}$, $\delta x_{\sigma}^{\pm}$ corresponding to 
$\Lambda_{\sigma+}^+ = \Lambda_{\sigma}^-$.

\subsection*{6.3.~The case when $\;\;\gamma=1$,
$\;\tau=1$: Fermi liquids}
\label{s5c}

As we have indicated earlier (\S~3), for metallic states of 
fermions interacting through the long-range Coulomb potential,
$\vartheta_{{\bF k};\sigma}$ is {\sl not} differentiable in the 
neighbourhood of ${\cal S}_{{\sc f};\sigma}$. For this reason, 
considerations in this Section should be viewed as {\sl not} 
concerning the metallic GSs of the Coulomb-interacting fermion 
systems; the treatment appropriate to the latter systems is 
presented in \S~6.1.3 above.

Along the lines of reasoning as in \cite{BF02a} concerning FL metallic 
states, we deduce that for such states we must have (compare with 
Eqs.~(105) and (106) in \cite{BF02a})
\begin{eqnarray}
\label{e195}
&& b_{\sigma}^- = -\Lambda_{\sigma}^+ d_{\sigma}^+,\\
\label{e196}
&& b_{\sigma}^+ = -\Lambda_{\sigma}^- d_{\sigma}^-.
\end{eqnarray} 
In the case of the uniform GSs of the Hubbard Hamiltonian, where 
$b_{\sigma}^{\pm} = d_{\sigma}^{\pm}$, the results in 
Eqs.~(\ref{e195}) and (\ref{e196}) lead to $\Lambda_{\sigma}^+ 
\Lambda_{\sigma}^- = 1$ and consequently to ${\sf n}_{\sigma}
({\Bf k}_{{\sc f};\sigma}) = \frac{1}{2}$ for ${\Bf k}_{{\sc f};\sigma}\in 
{\cal S}_{{\sc f};\sigma}$ ({\it cf}. Eq.~(110) in \cite{BF02a}). 
For two-body interaction potentials which have longer range than 
the contact potential in the conventional Hubbard Hamiltonian, 
${\Bf c}_{\sigma}({\Bf k}_{{\sc f};\sigma}^{\pm})$ cannot be 
identically vanishing for {\sl all} ${\Bf k}_{{\sc f};\sigma} \in 
{\cal S}_{{\sc f};\sigma}$ so that, in general, $b_{\sigma}^{\pm} 
\not= d_{\sigma}^{\pm}$ (see Eqs.~(\ref{e123}) -- (\ref{e125}) 
above) whereby it is {\sl not} necessary that for FL metallic states 
${\sf n}_{\sigma}({\Bf k}_{{\sc f};\sigma}) = \frac{1}{2}$ should hold. 
Further, in contrast with the case of the Hubbard on-site 
interaction for which under all conditions considered in \cite{BF02a} 
one has ${\sf n}_{\sigma}({\Bf k}_{{\sc f};\sigma}^-) \ge \frac{1}{2}$ 
and ${\sf n}_{\sigma}({\Bf k}_{{\sc f};\sigma}^+) \le \frac{1}{2}$, 
even for FL metallic states considered in this Section both 
${\sf n}_{\sigma}({\Bf k}_{{\sc f};\sigma}^-) \le \frac{1}{2}$ and 
${\sf n}_{\sigma}({\Bf k}_{{\sc f};\sigma}^+) \ge \frac{1}{2}$, 
satisfying, however, the necessary stability condition ${\sf n}_{\sigma}
({\Bf k}_{{\sc f};\sigma}^-)-{\sf n}_{\sigma}({\Bf k}_{{\sc f};
\sigma}^+) {=:} Z_{{\bF k}_{{\sc f};\sigma}} \ge 0$, are feasible. 
Below we explicitly demonstrate the feasibility of ${\sf n}_{\sigma}
({\Bf k}_{{\sc f};\sigma}^-) \le \frac{1}{2}$.

From Eqs.~(\ref{e125}), (\ref{e132}) and (\ref{e196}) we obtain
\begin{equation}
\label{e197}
c_{\sigma}^- \equiv b_{\sigma}^- - d_{\sigma}^-
= b_{\sigma}^- + \frac{1}{\Lambda_{\sigma}^-}\, b_{\sigma}^+.
\end{equation}
From (I) and (III) in Eq.~(\ref{e127}) we have $b_{\sigma}^- > 
-a_{\sigma}/{\sf g}$ and $b_{\sigma}^+ < -a_{\sigma}/{\sf g}$
respectively so that we can write
\begin{equation}
\label{e198}
b_{\sigma}^- = -\frac{\lambda_{\sigma}^- a_{\sigma}}{\sf g},\;
\lambda_{\sigma}^- < 1;\;\;\;
b_{\sigma}^+ = -\frac{\lambda_{\sigma}^+ a_{\sigma}}{\sf g},\;
\lambda_{\sigma}^+ > 1.
\end{equation}
On account of the expressions in Eqs.~(\ref{e197}) and (\ref{e198})
it follows that, for
\begin{equation}
\label{e199}
\Lambda_{\sigma}^- \le \frac{\lambda_{\sigma}^+}
{2 - \lambda_{\sigma}^-} 
\end{equation}
one achieves the condition $c_{\sigma}^- \le -2 a_{\sigma}/{\sf g}$ 
required for $\Lambda_{\sigma}^- \le 1$ (see Eq.~(\ref{e142}) above) 
and thus ${\sf n}_{\sigma}({\Bf k}_{{\sc f};\sigma}^-) \le \frac{1}{2}$. 
Evidently, the consistency of the condition $\Lambda_{\sigma}^- \le 1$ 
with that in Eq.~(\ref{e199}) implies the further requirement that
\begin{equation}
\label{e200}
1 < \lambda_{\sigma}^+ \le 2 -\lambda_{\sigma}^-,
\end{equation}
which encompasses both $\lambda_{\sigma}^- < 1$ and $\lambda_{\sigma}^+ 
> 1$ indicated in Eq.~(\ref{e198}). It follows that, for two-body 
interaction potentials of non-vanishing range (corresponding to 
$c_{\sigma}^- \not=0$), ${\sf n}_{\sigma}({\Bf k}_{{\sc f};\sigma}^-) 
\le \frac{1}{2}$ is feasible even for FL metallic GSs considered in 
this Section.

\section*{\S~7.~Summary and concluding remarks}
\label{s6}

In this paper we have considered some exact properties of the 
uniform metallic GSs of systems of fermions for arbitrary spatial 
dimensions interacting through arbitrary isotropic pair potentials 
that are capable of being Fourier transformed; thus, for instance, 
systems of fermions interacting through the Lennard-Jones $6-12$ 
potential lie outside the domain of applicability of the 
considerations in this paper, however those interacting through 
the Aziz {\sl et al.} potential \cite{AZ79} lie inside.
\footnote{\label{f21}
Both of these potentials have been used for describing the 
inter-atomic interaction of ${}^3$He atoms, with the Aziz
{\sl et al.} potential as being the more superior; for a 
discussion of these and of a number of other available pair 
potentials see \protect\cite{AZ79} and \protect\cite{CCK77}. }
Our present work concerns natural extension of previous work 
\cite{BF02a} in which we have explicitly dealt with the conventional 
single-band Hubbard Hamiltonian where the interaction is operative 
only between particles of opposite spin at the same lattice site. 

In the present paper we have explicitly demonstrated that, for 
${\Bf k}$ approaching the Fermi surface ${\cal S}_{{\sc f};\sigma}$ 
of the interacting metallic GS and $\varepsilon$ approaching the 
Fermi energy $\varepsilon_{\sc f}$ of the interacting system, the 
self-energy $\Sigma_{\sigma}({\Bf k};\varepsilon)$ continuously 
approaches the exact Hartree-Fock self-energy 
$\Sigma_{\sigma}^{\sc hf}({\Bf k})$. On the basis of this and of 
an exact expression for the deviation of $\Sigma_{\sigma}({\Bf k};
\varepsilon_{\sc f})$ from $\Sigma_{\sigma}^{\sc hf}({\Bf k})$, 
obtained from the Kramers-Kr\"onig relation expressing the real part 
of the self-energy in terms of the imaginary part, we have deduced 
the subsidiary equation $S_{\sigma}({\Bf k}) = 0$ (Eqs.~(\ref{e52})
and (\ref{e51})), which in conjunction with the quasi-particle equation 
in terms of $\Sigma_{\sigma}^{\sc hf}({\Bf k})$ (Eq.~(\ref{e46})) fully 
determines ${\cal S}_{{\sc f};\sigma}$ (see Eq.~(\ref{e53})). Thus, 
whereas the latter-mentioned quasi-particle equation in terms of 
$\Sigma_{\sigma}^{\sc hf}({\Bf k})$ is satisfied for all ${\Bf k}$ on 
${\cal S}_{{\sc f};\sigma}$, the reverse does {\sl not} necessarily 
obtain; the failure to satisfy $S_{\sigma}({\Bf k})=0$, in spite of 
satisfying the quasi-particle equation in terms of 
$\Sigma_{\sigma}^{\sc hf}({\Bf k})$, implies that for the ${\Bf k}$ 
at hand the quasi-particle equation in terms of $\Sigma_{\sigma}
({\Bf k};\varepsilon)$ (Eq.~(\ref{e45})) is not fulfilled; following 
the arguments presented in \cite{BF02a}, the set of all such ${\Bf k}$ 
points constitutes the pseudogap regions of what otherwise would be 
constituent parts of the Fermi surface ${\cal S}_{{\sc f};\sigma}$. 
Amongst others, the GS momentum distribution function
${\sf n}_{\sigma}({\Bf k})$, though singular, is continuous
for ${\Bf k}$ transposed from inside to outside the underlying 
Fermi sea through pseudogap regions. For ${\Bf k} \in 
{\cal S}_{{\sc f};\sigma}$ the expression corresponding to 
$S_{\sigma}({\Bf k})=0$ amounts to a sum rule concerning 
${\rm Im}[\Sigma_{\sigma}({\Bf k};\varepsilon)]$ (Eq.~(\ref{e54})); 
in this capacity, it can be of considerable relevance both for 
the purpose of both constraining theoretical models for 
$\Sigma_{\sigma}({\Bf k};\varepsilon)$ and determining the 
consistency of experimental results that are directly related with 
$\Sigma_{\sigma}({\Bf k};\varepsilon)$, or ${\rm Im}[\Sigma_{\sigma}
({\Bf k};\varepsilon)]$ (such as those corresponding to photoemission 
and inverse-photoemission experiments). Owing to the generality of 
our considerations in this paper, the above-mentioned results equally 
apply to the uniform metallic GSs of the conventional single-band 
Hubbard Hamiltonian, a fact that can be explicitly verified. 

In connection with the pseudogap phenomenon as observed in, for
example, the angle-resolved photoemission data concerning the 
normal states of the cuprate compounds in the underdoped regime 
\cite{AGL96,MDL96,MRN98,AI99,DSH02}, it is relevant to mention that on 
general grounds it can be shown \cite{BF02b,BF02a} (see in particular 
\S~3.4 in \cite{BF02b}) that this phenomenon cannot be associated 
with the {\sl uniform} GSs of systems for which ${\Bf k}$ is defined 
over the unbounded reciprocal space;
\footnote{\label{f22}
In these cases, {\sl indirect} gap is {\sl not} excluded. Further, 
for ferromagnetically ordered {\sl uniform} GSs one encounters 
{\sl direct} spin gap even in models where ${\Bf k}$ is defined 
over the unbounded reciprocal space. It is interesting to note 
that the approximate solution of the Hubbard model, the so-called 
Hubbard-I solution \protect\cite{JHI63}, predicts an {\sl indirect} 
(Mott-Hubbard) gap (see Fig.~1 in \cite{JHI63}). }
thus the single-band Hubbard Hamiltonian (for which the reciprocal
space is confined to the 1BZ of the underlying Bravais lattice), 
even in cases where the interaction is operative solely amongst 
particles of opposite spin on the same lattice site, is special in 
that, through the periodicity over the 1BZ of the corresponding 
functions of ${\Bf k}$, it takes account of Umklapp processes, 
albeit in some restricted form; this aspect is vital for the 
possibility of a {\sl uniform} metallic GS of the single-band 
Hubbard Hamiltonian supporting the pseudogap phenomenon. With 
reference to the observations in \cite{BF02a}, for such states 
${\cal S}_{{\sc f};\sigma}^{(0)} \backslash {\cal S}_{{\sc f};
\sigma}$ is non-empty and, in view of our observations in the 
present work, $S_{\sigma}({\Bf k}) \not=0$ for {\sl all} ${\Bf k}\in 
{\cal S}_{{\sc f};\sigma}^{(0)} \backslash {\cal S}_{{\sc f};\sigma}$. 

For the uniform and isotropic GSs of the uniform-electron-gas system, 
where on account of isotropy and of the Luttinger theorem concerning 
the `volume' enclosed by ${\cal S}_{{\sc f};\sigma}$ \cite{LW60,L60} one 
has ${\cal S}_{{\sc f};\sigma}\equiv {\cal S}_{{\sc f};\sigma}^{(0)}$, 
through making use of $\Sigma_{\sigma}({\Bf k};\varepsilon) \sim 
\Sigma_{\sigma}^{\sc hf}({\Bf k})$ for ${\Bf k}$ approaching 
${\cal S}_{{\sc f};\sigma}^{(0)}$ and $\varepsilon\to
\varepsilon_{\sc f}$, we have deduced (\S~5) exact expressions for 
the exchange potential $\mu_{{\rm x};\sigma}$ and correlation 
potential $\mu_{{\rm c};\sigma}$, contributing to the 
exchange-correlation potential $\mu_{{\rm xc};\sigma} \equiv 
\mu_{{\rm x};\sigma} + \mu_{{\rm c};\sigma}$ which plays a central 
role within the framework of the GS density-functional theory 
\cite{HK64,KS65,DG90}. Our expression for $\mu_{{\rm x};\sigma}$, 
specialized to systems of electrons interacting through the Coulomb 
potential for $d=3$, identically reproduces the well-known exact result 
for the exchange potential. Making use of the RPA for ${\sf n}_{\sigma}
({\Bf k})$ corresponding to the paramagnetic GSs of the 
uniform-electron-gas system (note that, for systems of 
Coulomb-interacting fermions, ${\sf n}_{\sigma}^{\sc rpa}({\Bf k})-
{\sf n}_{\sigma}^{(0)}({\Bf k})$ reproduces the leading-order 
contribution to the exact ${\sf n}_{\sigma}({\Bf k})- 
{\sf n}_{\sigma}^{(0)}({\Bf k})$ in the high-density limit), we 
have numerically calculated $\mu_{{\rm c};\sigma}$ which in the 
high-density limit, and up to densities corresponding to the 
dimensionless Wigner-Seitz radius $r_{\rm s}\,\Ieq{\sim}{<}\, 1$, 
is in excellent agreement with the $\mu_{{\rm c};\sigma}$ deduced 
from the quantum Monte Carlo results for the GS correlation energy 
(see Figs.~\ref{fi1} and \ref{fi2}).

We have explicitly shown that interactions of non-zero range are of 
non-trivial consequence to a number of GS and excited-states 
properties of metallic systems. For instance, whereas for a strictly 
contact-type two-body interaction potential ${\sf n}_{\sigma}
({\Bf k}_{{\sc f};\sigma}^{-}) \ge \frac{1}{2}$ and ${\sf n}_{\sigma}
({\Bf k}_{{\sc f};\sigma}^{+}) \le \frac{1}{2}$ for ${\Bf k}_{{\sc f};
\sigma}\in {\cal S}_{{\sc f};\sigma}$ \cite{BF02a}, for two-body 
interaction potentials of non-vanishing range in principle both 
${\sf n}_{\sigma}({\Bf k}_{{\sc f};\sigma}^-) < \frac{1}{2}$ and 
${\sf n}_{\sigma}({\Bf k}_{{\sc f};\sigma}^+) > \frac{1}{2}$ are 
feasible (in both cases, subject to the condition ${\sf n}_{\sigma}
({\Bf k}_{{\sc f};\sigma}^-)-{\sf n}_{\sigma}({\Bf k}_{{\sc f};
\sigma}^+) \ge 0$, implied by the assumed stability of the 
underlying GSs). This aspect is borne out by the quantum Monte 
Carlo results for ${\sf n}_{\sigma}({\Bf k})$ pertaining to the 
uniform and isotropic GSs of the Coulomb-interacting electron-gas
system for $d=2$ \cite{SC97} (see Fig.~5.9 herein) in the 
low-density regime 
\footnote{\label{f23}
We recall that, for $d=3$ and Coulomb-interacting fermions,
${\sf n}_{\sigma}^{\sc rpa}({\Bf k}_{{\sc f};\sigma}^-)$ submerges 
below $\frac{1}{2}$ for densities corresponding to $r_{\rm s} > 
6.09887\dots$ (see footnote \protect\ref{f16} above); the Monte Carlo 
results by Ortiz and Ballone \protect\cite{OB94} on the other hand 
show that, for $r_{\rm s} = 10$, ${\sf n}_{\sigma}({\Bf k}_{{\sc f};
\sigma}^-)\approx 0.69$. Interestingly, for $d=2$, according to the 
quantum Monte Carlo results by Conti \protect\cite{SC97}, for $r_{\rm s} 
=10$ one has ${\sf n}_{\sigma}({\Bf k}_{{\sc f};\sigma}^-) \approx 
0.55$ and for $r_{\rm s} = 20$, ${\sf n}_{\sigma}({\Bf k}_{{\sc f};
\sigma}^-) \approx 0.42$. We observe that, for sufficiently 
large interaction strength, indeed the non-vanishing range of the 
interaction potential (here the Coulomb potential) gives rise to 
${\sf n}_{\sigma}({\Bf k}_{{\sc f};\sigma}^-) < \frac{1}{2}$. }
and of ${}^3$He in the normal liquid state at sufficiently large 
ambient pressures \cite{CCK77,MFF86,PSdWH97,MSF97,MFF98}. In the 
light of this and of the strict inequality ${\sf n}_{\sigma}
({\Bf k}_{{\sc f};\sigma}^-) \ge \frac{1}{2}$ applicable to the 
${\sf n}_{\sigma}({\Bf k})$ pertaining to the uniform GSs of the 
conventional single-band Hubbard Hamiltonian \cite{BF02a}, employed in 
studies of ${}^3$He \cite{AB78,DV84,VWA87}, as well as to the uniform 
GSs of the Stoner Hamiltonian (involving contact interaction but defined 
on the continuum; likewise employed in studies of ${}^3$He within the 
framework of the paramagnon theory \cite{BS66,DE66,BE68,JHB71,LV78};
for a review of spin-fluctuation theories see \cite{PS85}), one 
observes that such Hamiltonians cannot be capable of reliably 
describing the normal liquid state of ${}^3$He. In this connection, 
it is in place to refer to a series of papers by Hirsch (see
\cite{JEH97} and references herein) in which the viewpoint has 
been consistently advocated that many (collective) phenomena for whose 
rationalization traditionally the conventional single-band Hubbard 
Hamiltonian has been invoked should in fact be accounted for by means 
of `extended' Hubbard Hamiltonians (such as considered in
\cite{MRRT88,JEH90,HC90,JEH92b,XZY93,DR94,JEH97,CG97,YY98,SCCG98,%
MM00,JEHM00,HWL01,VV01,AHH02,CNP02,KK03}, or in 
\cite{RAB71,BP74,CC75,RFK81}), involving site off-diagonal (to be 
contrasted with {\sl on-site}) interaction terms (including, e.g., the 
nearest-neighbour exchange). In \cite{JEH97}, Hirsch referred to such 
phenomena as itinerant ferromagnetism \cite{JEH85,JEH89a} (see also 
\cite{BF02a}), 
\footnote{\label{f24}
Concerning the role of interactions of longer range than the on-site 
repulsion in promoting specifically partial ferromagnetism, we refer 
the reader to the phase diagram in Fig.~4 of \protect\cite{VV01} which 
illustrates, albeit on the basis of a mean-field approximation, the 
consequence of a nearest-neighbour {\sl repulsive} interaction energy 
$V$ (note that in the above-mentioned phase diagram the $U$ axis 
intersects the $V$ axis at some {\sl positive} value of $V$). In this 
connection we mention that according to the analysis in 
\protect\cite{BF02a}, in the case of the conventional single-band 
Hubbard Hamiltonian, partially polarized uniform ferromagnetic states 
are often barred from qualifying as eigenstates through a kinematic 
constraint (see Eq.~(63) in \protect\cite{BF02a}); it can be readily 
verified that for longer-range interaction potentials no such 
{\it a priori} kinematic constraint exists, this in consequence of 
the ${\Bf k}$ dependence of $\Sigma_{\sigma}^{\sc hf}({\Bf k})$ in 
the case of the latter potentials. }
heavy-fermion superconductivity \cite{PWA84,OR84}, high-temperature 
superconductivity in the cuprate compounds \cite{HL88} (see also 
\cite{DSW00}), and superfluidity in liquid ${}^3$He \cite{JEH97}.
\footnote{\label{f25}
We do not unreservedly subscribe to the assignments in
\protect\cite{JEH97} quoted above. For instance, in our opinion it 
is not the phenomenon of superconductivity in heavy-fermion 
compounds that has been rationalized by means of the conventional 
Hubbard model (explicitly, in \protect\cite{UR86}, to which in 
\protect\cite{JEH97} is referred, one does {\sl not} explicitly rely 
on the Hubbard Hamiltonian). Rather, in considering the heavy-fermion 
compounds and superconductivity \protect\cite{PWA84,OR84,VT84,BQ85,UR86} 
in these, one has in essence solely relied on the so-called 
`almost-localized Fermi-liquid' picture \protect\cite{AB78,DV84} 
of the charge carriers; the latter has its root in the Brinkman-Rice 
\protect\cite{BR70} scenario concerning the nature of the metallic 
states of the Hubbard Hamiltonian (for $d=3$) at half-filling for the 
on-site $U$ approaching from below a critical value $U_{\rm c}$, for 
which, according to the Gutzwiller {\sl Ansatz} for the GS wavefunction 
of the Hubbard Hamiltonian combined with the Gutzwiller approximation 
\protect\cite{MG63,MG65}, the jump in the GS momentum distribution 
function at ${\cal S}_{{\sc f};\sigma}$ vanishes and thereby the 
inverse of the quasi-particle effective mass (hence the qualification 
`almost localized'). For completeness, we mention that the Gutzwiller 
{\sl approximation} (to be distinguished from the Gutzwiller 
{\sl Ansatz}) is a well-known source of uncertainty. For instance, 
the exact results corresponding to the Gutzwiller {\sl Ansatz} for the 
GS wavefunction of the single-band Hubbard Hamiltonian for $d=1$ 
\protect\cite{MV88} reveal that at half-filling the above-mentioned 
jump in the momentum distribution function does {\sl not} vanish at 
any finite value of $U$; in contrast, according to the Gutzwiller 
{\sl approximation} for $d=1$, this jump vanishes for 
$U_{\rm c}/t \approx 10$ (see Fig.~13 in \protect\cite{MV88}); these 
results should in turn be contrasted with the exact result indicated 
in footnote \protect\ref{f20} above. Finally, insofar 
as the {\sl normal} states of the heavy-fermion compounds are 
concerned, although the considerations in \protect\cite{RUOR85} are 
based on the Gutzwiller Ansatz (together with the Gutzwiller 
approximation) for the GS wavefunction of the Hubbard Hamiltonian,
the work in \protect\cite{RU85}, based on the periodic Anderson
Hamiltonian, leaves no doubt that the analogies proposed in 
\protect\cite{RUOR85} have indeed been based on ``intuitive 
grounds''. }

Since for the (correlated) GS kinetic energy we have 
$\sum_{{\bF k},\sigma} \varepsilon_{\bF k} {\sf n}_{\sigma}({\Bf k})$, 
by continuity it follows that in general the GS kinetic energy of 
systems in which ${\sf n}_{\sigma}({\Bf k}_{{\sc f};\sigma}^-) < 
\frac{1}{2}$, ${\Bf k}_{{\sc f};\sigma}\in {\cal S}_{{\sc f};\sigma}$, 
exceeds that of comparable systems in which ${\sf n}_{\sigma}
({\Bf k}_{{\sc f};\sigma}^-) \ge \frac{1}{2}$ (note that 
${\sf n}_{\sigma}({\Bf k}) \ge 0$ and $\sum_{{\bF k},\sigma} 
{\sf n}_{\sigma}({\Bf k}) = N$, independent of interaction). 
We thus observe a direct relationship between the range of the 
two-body interaction potential and the magnitude of the corresponding 
GS kinetic energy, with the latter tending to increase (in the manner 
specified above) in consequence of increasing the range of the 
interaction potential. The existence of such a direct relationship 
between the range of the interaction potential and the value 
of the GS kinetic energy should be of relevance to a better 
understanding of the physics of the cuprate superconductors for which 
there is growing evidence indicating that in the normal state the 
kinetic-energy contribution to the GS total energy is relatively 
larger than in conventional metals and that superconductivity in 
the former is driven by a consequent decrease in the kinetic 
energy of these systems upon entering into the superconducting state. 
In the light of this and of our theoretical finding, we believe that 
any theoretical modelling of in particular the cuprate compounds is
required to take account of the non-zero range of the interparticle 
interaction potential. 

In addition to establishing the above-mentioned impact that a non-zero 
range of interaction has on the behaviour of ${\sf n}_{\sigma}({\Bf k})$ 
in the vicinity of ${\cal S}_{{\sc f};\sigma}$, we have explicitly 
exposed (\S~6) some distinctive features associated with the actual 
range of the interaction potential; that is whether this range is sharply 
limited to a finite inter-particle distance or it extents to arbitrary 
large distances. For instance, for interaction potentials $v({\Bf r}
-{\Bf r}')$ that possess power-law decay for $\|{\Bf r}-{\Bf r}'\| 
\to \infty$, the singular behaviour of ${\sf n}_{\sigma}({\Bf k})$ in 
the neighbourhood of any ${\Bf k}$ directly implies a singular 
behaviour in the Fock part $\Sigma_{\sigma}^{\sc f}({\Bf k})$
of the self-energy in the neighbourhood of the same ${\Bf k}$; the 
two singularities are, however, of different types. This observation is 
specifically significant for the ${\Bf k}$ points in the neighbourhood 
of ${\cal S}_{{\sc f};\sigma}$, where ${\sf n}_{\sigma}({\Bf k})$ is 
unconditionally singular (and not necessarily discontinuous). From our 
considerations it follows that, for systems of fermions interacting 
through a potential $v({\Bf r}-{\Bf r}')$ decaying, for instance, like 
$1/\|{\Bf r}-{\Bf r}'\|^m$ for $\|{\Bf r}-{\Bf r}'\|\to \infty$, a 
discontinuity in ${\sf n}_{\sigma}({\Bf k})$ at a {\sl general} 
${\Bf k}={\Bf k}_{{\sc f};\sigma} \in {\cal S}_{{\sc f};\sigma}$ 
implies a logarithmic divergence at ${\Bf k}_{{\sc f};\sigma}$ in 
the $m$th derivative (in the direction of the radius vector 
${\Bf k}_{{\sc f};\sigma}$) of $\Sigma_{\sigma}^{\sc f}({\Bf k})$.

We believe that a detailed examination of ${\sf n}_{\sigma}({\Bf k})$ 
determined experimentally (for ${\Bf k}$ in the close vicinity of 
${\cal S}_{{\sc f};\sigma}$), in conjunction with the appropriate 
expressions for this function as presented in this paper, can shed 
light on some essential aspects of the two-body interaction potential 
to be employed within the framework of an extended single-band 
Hubbard Hamiltonian. The details presented in this paper can also 
with advantage be employed to constrain (approximate) theoretical 
frameworks concerning correlation functions (such as the self-energy 
$\Sigma_{\sigma}({\Bf k};\varepsilon)$) pertaining to the metallic 
GSs of the many-body Hamiltonians considered in this paper.

%________________________
\vspace{1mm}
\section*{Acknowledgements}
\label{s7}
With pleasure I thank 
Dr R.~J. Needs and 
Professor G. Senatore 
for discussions concerning the quantum Monte Carlo data for the 
ground-state momentum distribution function of liquid ${}^3$He; 
I am further indebted to Professor Senatore for kindly making 
Ref.~\cite{SC97} available to me prior to publication. With 
appreciation I thank Spinoza Institute for hospitality and support.

%________________________
\vspace{0.6cm}
\centerline{\bf ---------------------}
\vspace{-4.42mm}
\centerline{\bf ------------------------------}
\vspace{-4.42mm}
\centerline{\bf ---------------------}

\begin{appendix}

\section{Estimation of error in $S_{\sigma}(\lowercase{\Bf k})$
within the framework of a finite-order many-body perturbation 
theory for $\Sigma_{\sigma}(\lowercase{\Bf k};\varepsilon)$}
\label{s8}

In the defining expression for $S_{\sigma}({\Bf k})$ in Eq.~(\ref{e51}) 
we encounter ${\rm Im}[\Sigma_{\sigma}({\Bf k};\varepsilon_{\sc f}\pm
\varepsilon)]$ in which for definiteness we assume $\Sigma_{\sigma}
({\Bf k};\varepsilon)$, $\varepsilon \in (-\infty,\infty)$, to have 
been calculated within the framework of a finite-order perturbation 
theory based on skeleton self-energy diagrams (for the definition
see \cite{LW60}) evaluated in terms of the {\sl exact} single-particle 
Green function. With the latter assumption we wish to estimate the 
error in $S_{\sigma}({\Bf k})$ as arising from the truncation of the 
set of skeleton self-energy diagrams and not from the further 
approximation corresponding to using an approximate single-particle 
Green function.

We proceed by writing $S_{\sigma}({\Bf k})$ as follows
\begin{equation}
\label{ea1}
S_{\sigma}({\Bf k}) = S_{\sigma}^{(1)}({\Bf k};\Delta)
+ S_{\sigma}^{(2)}({\Bf k};\Delta),
\end{equation}
where
\begin{equation}
\label{ea2}
S_{\sigma}^{(1)}({\Bf k};\Delta) {:=}
\frac{1}{\pi} \int_0^{\Delta} 
\frac{{\rm d}\varepsilon}{\varepsilon}\;
{\rm Im}[\Sigma_{\sigma}({\Bf k};\varepsilon_{\sc f}+\varepsilon) 
+\Sigma_{\sigma}({\Bf k};\varepsilon_{\sc f}-\varepsilon) ],
\end{equation}
\begin{equation}
\label{ea3}
S_{\sigma}^{(2)}({\Bf k};\Delta) {:=}
\frac{1}{\pi} \int_{\Delta}^{\infty} 
\frac{{\rm d}\varepsilon}{\varepsilon}\;
{\rm Im}[\Sigma_{\sigma}({\Bf k};\varepsilon_{\sc f}+\varepsilon) 
+\Sigma_{\sigma}({\Bf k};\varepsilon_{\sc f}-\varepsilon) ],
\end{equation}
in which $\Delta > 0$ stands for an energy sufficiently large with 
respect to a relevant energy scale in the system under consideration 
\cite{BF02b} so that a finite-order large-$\vert\varepsilon\vert$ 
asymptotic series for ${\rm Im}[\Sigma_{\sigma}({\Bf k};
\varepsilon_{\sc f} \pm\varepsilon)]$ is accurate for 
$\vert\varepsilon\vert \ge \Delta$. Later in our considerations we shall 
consider the case in which the latter series is formally summed to all 
orders whereby it is possible to effect the limit $\Delta\downarrow 0$. 
To avoid unnecessary technical complication, in what follows we assume 
that the two-body interaction potential $v({\Bf r}-{\Bf r}')$ is of 
shorter range than the Coulomb potential and moreover that the 
single-particle excitation spectrum $\varepsilon_{\bF k}$ is 
unbounded from above so that the single-particle spectral 
function $A_{\sigma}({\Bf k};\varepsilon)$ is of unbounded support. 
For such cases we have \cite{BF02b}
\begin{equation}
\label{ea4}
{\rm Im}[\Sigma_{\sigma}({\Bf k};\varepsilon)] \sim
\frac{\Xi_{\sigma}({\Bf k})}{\varepsilon} + 
\frac{\Pi_{\sigma}({\Bf k})}{\varepsilon^2} + \dots\;\;
\mbox{\rm for}\;\;\; \vert\varepsilon\vert\to\infty, 
\end{equation}
where 
\footnote{\label{f26}
For the stability of the GS of the system under consideration it is 
required that $\Xi_{\sigma}({\Bf k}) \ge 0$, $\forall {\Bf k},\sigma$
\cite{BF02b}. } 
\begin{eqnarray}
\label{ea5}
\Xi_{\sigma}({\Bf k}) &{:=}& 
\frac{-1}{\pi} \int_0^{\infty} {\rm d}\varepsilon'\;
\big\{ {\rm Re}[\Sigma_{\sigma}({\Bf k};\varepsilon') +
\Sigma_{\sigma}({\Bf k};-\varepsilon')] \nonumber\\
& &\;\;\;\;\;\;\;\;\;\;\;\;\;\;\;\;\;\;\;
-2 \Sigma_{\sigma;\infty_0}({\Bf k}) \big\},\\
\label{ea6}
\Pi_{\sigma}({\Bf k}) &{:=}& 
\frac{-1}{\pi} \int_0^{\infty} {\rm d}\varepsilon'\;
\big\{ \varepsilon'
{\rm Re}[\Sigma_{\sigma}({\Bf k};\varepsilon') -
\Sigma_{\sigma}({\Bf k};-\varepsilon')] \nonumber\\
& &\;\;\;\;\;\;\;\;\;\;\;\;\;\;\;\;\;\;\;
-2 \Sigma_{\sigma;\infty_1}({\Bf k}) \big\}.
\end{eqnarray}
In Eqs.~(\ref{ea5}) and (\ref{ea6}) the real-valued functions 
$\Sigma_{\sigma;\infty_m}({\Bf k})$, $m=0,1$, are the coefficients 
of the first two leading terms in the large-$\vert z\vert$ asymptotic 
series of $\wt{\Sigma}_{\sigma}({\Bf k};z)$, with ${\rm Im}(z)
\not=0$, that is
\begin{equation}
\label{ea7}
\wt{\Sigma}_{\sigma}({\Bf k};z) \sim
\Sigma_{\sigma;\infty_0}({\Bf k}) + 
\frac{\Sigma_{\sigma;\infty_1}({\Bf k})}{z} + 
\dots,\;\; \vert z\vert\to\infty.
\end{equation}
Here $\wt{\Sigma}_{\sigma}({\Bf k};z)$ stands for the analytic
continuation of $\Sigma_{\sigma}({\Bf k};\varepsilon)$ into the
physical Riemann sheet of the complex $z$ plane, for which we have 
$\Sigma_{\sigma}({\Bf k};\varepsilon) = \lim_{\eta\downarrow 0} 
\wt{\Sigma}_{\sigma}({\Bf k};\varepsilon \pm i\eta)$, $\varepsilon\,
\Ieq<>\, \mu$. We point out that, even for cases where $\Sigma_{\sigma;
\infty_m}({\Bf k})$ is bounded for all finite values of $m$, use of 
a complex $z$ in Eq.~(\ref{ea7}) is necessary when the series on 
the RHS of this equation is taken account of to some infinite order, 
because $\wt{\Sigma}_{\sigma}({\Bf k};z)$ possesses a branch-cut 
discontinuity along the real $\varepsilon$ axis. It can be shown 
that \cite{BF99,BF02b}
\begin{equation}
\label{ea8}
\Sigma_{\sigma;\infty_0}({\Bf k}) \equiv 
\Sigma_{\sigma}^{\sc hf}({\Bf k}). 
\end{equation}

We write
\begin{eqnarray}
\label{ea9}
\Xi_{\sigma}({\Bf k}) &\equiv& \Xi_{\sigma}^{(1)}({\Bf k};\Delta)
+ \Xi_{\sigma}^{(2)}({\Bf k};\Delta), \\
\label{ea10}
\Pi_{\sigma}({\Bf k}) &\equiv& \Pi_{\sigma}^{(1)}({\Bf k};\Delta)
+ \Pi_{\sigma}^{(2)}({\Bf k};\Delta), 
\end{eqnarray}
where $\Xi_{\sigma}^{(1)}({\Bf k};\Delta)$, 
$\Xi_{\sigma}^{(2)}({\Bf k};\Delta)$, 
$\Pi_{\sigma}^{(1)}({\Bf k};\Delta)$ and 
$\Pi_{\sigma}^{(2)}({\Bf k};\Delta)$ stand in relation to 
$\Xi_{\sigma}({\Bf k})$ and $\Pi_{\sigma}({\Bf k})$, 
defined in Eqs.~(\ref{ea5}) and (\ref{ea6}), as 
$S_{\sigma}^{(1)}({\Bf k})$ and
$S_{\sigma}^{(2)}({\Bf k})$ in Eqs.~(\ref{ea2}) and (\ref{ea3})
in relation to $S_{\sigma}({\Bf k})$,
defined in Eq.~(\ref{e51}).
From the defining expressions for the functions on the 
RHSs of Eqs.~(\ref{ea9}) and (\ref{ea10}), making use of the
asymptotic series for $\wt{\Sigma}_{\sigma}({\Bf k};z)$
in Eq.~(\ref{ea7}), for $\Delta \to \infty$ one straightforwardly 
obtains
\begin{eqnarray}
\label{ea11}
\Xi_{\sigma}^{(2)}({\Bf k};\Delta) &\sim&
-\frac{2\Sigma_{\sigma;\infty_2}({\Bf k})}{\pi\Delta} -
\frac{2\Sigma_{\sigma;\infty_4}({\Bf k})}{3\pi\Delta^3} - \dots,\\
\label{ea12}
\Pi_{\sigma}^{(2)}({\Bf k};\Delta) &\sim&
-\frac{2\Sigma_{\sigma;\infty_3}({\Bf k})}{\pi\Delta} -
\frac{2\Sigma_{\sigma;\infty_5}({\Bf k})}{3\pi\Delta^3} - \dots.
\end{eqnarray}
From the above results we deduce
\begin{eqnarray}
\label{ea13}
&&{\rm Im}[\Sigma_{\sigma}({\Bf k};\varepsilon_{\sc f}+\varepsilon)+
\Sigma_{\sigma}({\Bf k};\varepsilon_{\sc f}-\varepsilon)] 
\nonumber\\
&&\;\;\;\;\;\;
\sim 2 \Big[ \Pi_{\sigma}^{(1)}({\Bf k};\Delta) 
-\varepsilon_{\sc f}\,
\Xi_{\sigma}^{(1)}({\Bf k};\Delta) \nonumber\\
&&\;\;\;\;\;\;\;\;\;\;\;\;\;
+\frac{2}{\pi\Delta} \big( \Sigma_{\sigma;\infty_3}({\Bf k})
-\varepsilon_{\sc f}\,
\Sigma_{\sigma;\infty_2}({\Bf k}) \big) \Big]
\frac{1}{\varepsilon^2}, \nonumber\\
&&\;\;\;\;\;\;\;\;\;\;\;\;\;\;\;\;\;\;\;\;\;\;\;\;\;\;\;\;
\;\;\;\;\;\;\;\;\;\;\;\;\;\;\;\;\;\;\;\;\;
\vert\varepsilon\vert,\, \Delta \to \infty.
\end{eqnarray}
Making use of this expression,
from the defining expression for $S_{\sigma}^{(2)}({\Bf k})$ in 
Eq.~(\ref{ea3}) we obtain
\begin{eqnarray}
\label{ea14}
&&S_{\sigma}^{(2)}({\Bf k};\Delta) \sim
\frac{ \big[ \Pi_{\sigma}^{(1)}({\Bf k};\Delta) 
-\varepsilon_{\sc f}\,
\Xi_{\sigma}^{(1)}({\Bf k};\Delta) \big]/\pi}{\Delta^2} \nonumber\\
&&+\frac{2 \big[\Sigma_{\sigma;\infty_3}({\Bf k})
-\varepsilon_{\sc f}\,
\Sigma_{\sigma;\infty_2}({\Bf k}) \big]/\pi^2}{\Delta^3} +\dots,
\;\Delta\to \infty.
\end{eqnarray}
It is interesting to note the occurrence in Eq.~(\ref{ea14}) (in 
particular in the second term on the RHS) of $\varepsilon_{\sc f}$, 
the energy with respect to which the energies of the lowest-lying 
single-particle excitations are measured, in conjunction with 
coefficients that determine the behaviour of $\Sigma_{\sigma}
({\Bf k};\varepsilon)$ at large values of $\vert\varepsilon\vert$. 

The expression on the RHS of Eq.~(\ref{ea14}) is {\sl not} a 
well-ordered asymptotic series, as the first term involves 
contributions to terms of higher order in $1/\Delta$ than 
$1/\Delta^2$. Since $\big(\Pi_{\sigma}^{(1)}({\Bf k};\Delta)
-\Pi_{\sigma}({\Bf k}) \big) = o(1)$ and 
$\big(\Xi_{\sigma}^{(1)}({\Bf k};\Delta) -\Xi_{\sigma}({\Bf k}) \big) 
= o(1)$ for $\Delta\to\infty$, to the leading-order in $1/\Delta$ 
we have
\begin{equation}
\label{ea15}
S_{\sigma}^{(2)}({\Bf k};\Delta) \sim
\frac{ \big[ \Pi_{\sigma}({\Bf k}) 
-\varepsilon_{\sc f}\,
\Xi_{\sigma}({\Bf k}) \big]/\pi}{\Delta^2}\;\;
\mbox{\rm for}\;\; \Delta\to \infty.
\end{equation}
Note that by replacing $\Pi_{\sigma}({\Bf k})$ and $\Xi_{\sigma}
({\Bf k})$ on the RHS of this expression by the expressions on 
the RHSs of Eqs.~(\ref{ea9}) and (\ref{ea10}) respectively and using 
the leading-order terms in the asymptotic series on the RHSs of 
Eqs.~(\ref{ea11}) and (\ref{ea12}) respectively, we recover the 
expression on the RHS of Eq.~(\ref{ea14}).
 
From our perspective, the most important aspect of the asymptotic 
expression in Eq.~(\ref{ea14}) is the explicit dependence on
$\Sigma_{\sigma;\infty_2}({\Bf k})$ and $\Sigma_{\sigma;\infty_3}
({\Bf k})$ of the second term on the RHS. This in view of the fact
that according to the analysis in \cite{BF02b}, the reproduction of 
the exact $\Sigma_{\sigma;\infty_2}({\Bf k})$ requires calculation 
of $\Sigma_{\sigma}({\Bf k};\varepsilon)$ to third order and that 
of $\Sigma_{\sigma;\infty_3}({\Bf k})$ to fourth order in the 
perturbation series in terms of skeleton self-energy diagrams, with 
these in turn evaluated in terms of the {\sl bare} particle-particle 
interaction function and the exact interacting single-particle Green 
function. Thus the second term on the RHS of Eq.~(\ref{ea14}) associated 
with the $\Sigma_{\sigma}({\Bf k};\varepsilon)$ evaluated in terms of the 
full set of first- and second-order skeleton diagrams deviates from its 
exact counterpart. In connection with the result in Eq.~(\ref{ea15})
(or Eq.~(\ref{ea14})), it is important to realize that since 
$S_{\sigma}^{(1)}({\Bf k};\Delta)$ and $S_{\sigma}^{(2)}({\Bf k};
\Delta)$ as defined in Eqs.~(\ref{ea2}) and (\ref{ea3}) respectively 
exist for $\Delta \downarrow 0$, and since $S_{\sigma}^{(1)}({\Bf k};
\Delta) \to 0$ for $\Delta\downarrow 0$, one can in principle recover 
the entire $S_{\sigma}({\Bf k})$ by evaluating the large-$\Delta$ 
asymptotic series of $S_{\sigma}^{(2)}({\Bf k};\Delta)$ to all orders 
in $1/\Delta$ and, subsequent to its full summation, taking the limit 
$\Delta\downarrow 0$. Although this approach is of no practical use, 
the fact that the formal infinite-order large-$\Delta$ asymptotic 
series of $S_{\sigma}^{(2)}({\Bf k};\Delta)$ is capable of exactly 
reproducing $S_{\sigma}({\Bf k})$ is significant in that it sheds 
light on the way (as in Eq.~(\ref{ea14}) above) in which the 
coefficients $\{ \Sigma_{\sigma;\infty_m}({\Bf k}) \}$ of the 
large-$\vert\varepsilon\vert$ asymptotic series of $\Sigma_{\sigma}
({\Bf k};\varepsilon)$ contribute to those of the former asymptotic 
series; by the further knowledge that for $m\ge 1$ {\sl all} skeleton 
self-energy diagrams from order $2$ up to and including order $m+1$ 
contribute to $\Sigma_{\sigma;\infty_m}({\Bf k})$ \cite{BF02b}, one 
gains insight into the consequences for $S_{\sigma}({\Bf k})$ of 
employing a finite-order perturbation series for $\Sigma_{\sigma}
({\Bf k};\varepsilon)$. The same line of reasoning applies to 
$\Xi_{\sigma}^{(2)}({\Bf k};\Delta)$ and $\Pi_{\sigma}^{(2)}({\Bf k};
\Delta)$ from whose complete large-$\Delta$ asymptotic series the 
exact $\Xi_{\sigma}({\Bf k})$ and $\Pi_{\sigma}({\Bf k})$ respectively 
can be deduced; these asymptotic series are evidently fully determined 
by $\{ \Sigma_{\sigma;\infty_m}({\Bf k})\}$ (see Eqs.~(\ref{ea11}) 
and (\ref{ea12}) above). In this connection note that according to 
Eq.~(\ref{ea15}) the leading-order term in the large-$\Delta$ asymptotic 
series of $S_{\sigma}^{(2)}({\Bf k})$ is in addition to
$\varepsilon_{\sc f}$ fully determined by $\Xi_{\sigma}({\Bf k})$ and 
$\Pi_{\sigma}({\Bf k})$. Concerning $\varepsilon_{\sc f}$, in view of 
Eqs.~(\ref{e46}) and (\ref{ea8}) one observes that this constant is 
determined by the first-order (skeleton) self-energy diagram; its 
exact determination is, however, dependent on the knowledge of the 
exact interacting single-particle Green function $G_{\sigma}({\Bf k};
\varepsilon)$ or, what in this context is the same, 
${\sf n}_{\sigma}({\Bf k})$.

The inaccuracy in a calculated $S_{\sigma}({\Bf k})$, which in 
practice arises from use of both a finite set of (skeleton) 
self-energy diagrams and an approximate single-particle Green 
function (either a non-interacting single-particle Green function 
or one determined self-consistently), combined with the inaccuracy 
in $\Sigma_{\sigma}^{\sc hf}({\Bf k})$ arising from use of an 
approximate single-particle Green function (or ${\sf n}_{\sigma}
({\Bf k})$; see Eqs.~(\ref{e12}), (\ref{e16}) and (\ref{e17})), 
leads through Eqs.~(\ref{e45}) and (\ref{e50}) to deviation of a 
calculated Fermi surface from its exact counterpart 
${\cal S}_{{\sc f};\sigma}$. For the uniform paramagnetic metallic 
GSs of the single-band Hubbard Hamiltonian where 
$\Sigma_{\sigma}^{\sc hf}({\Bf k})$ is equal to the exactly-known 
constant value $U n/(2\hbar)$, in which $n=N/N_{\sc l}$ (the total number 
of fermions per lattice site), the inaccuracy in a calculated Fermi 
surface based on the use of Eq.~(\ref{e45}) is entirely attributable 
to that in the underlying $S_{\sigma}({\Bf k})$ (see Eq.~(\ref{e50})).
The apparent violation (albeit very slightly) of the exact result 
${\cal S}_{{\sc f};\sigma} \subseteq {\cal S}_{{\sc f};\sigma}^{(0)}$, 
deduced in \cite{BF02a}, by the ${\cal S}_{{\sc f};\sigma}$ 
(pertaining to the uniform paramagnetic GS of the single-particle 
Hubbard Hamiltonian, with $U/t=4$) determined from a self-consistently 
evaluated $\Sigma_{\sigma}({\Bf k};\varepsilon)$ (based on the 
second-order expansion of $\Sigma_{\sigma}({\Bf k};\varepsilon)$ 
\cite{HN99}) is thus attributable to the fact that {\sl none} of 
$\Sigma_{\sigma;\infty_m}$, $m\ge 2$, corresponding to a second-order 
perturbation expansion of $\Sigma_{\sigma}({\Bf k};\varepsilon)$ 
can be correct \cite{BF02b}. For our criticism of non-selfconsistently 
calculated Fermi surfaces pertaining to the metallic states of the 
single-band Hubbard Hamiltonian we refer the reader to \cite{BF02a}.
Note the principal advantage of the expression for ${\cal S}_{{\sc f};
\sigma}$ in Eq.~(\ref{e53}) in comparison with that in Eq.~(\ref{e45}); 
following the evident fact that $\Sigma_{\sigma}^{\sc hf}({\Bf k})$ 
is in principle exactly determined by the first-order self-energy 
operator (i.e. leaving aside the necessity for the knowledge of the 
exact ${\sf n}_{\sigma}({\Bf k})$), use of the expression in 
Eq.~(\ref{e53}) in conjunction with a first-order expression for the 
self-energy operator guarantees the exact reproduction of the skeleton 
of the underlying Fermi surface (if not the entire Fermi surface).
\hfill $\Box$

\end{appendix}

%________________________

%________________________

% 1.
\pagebreak
%\clearpage

% 2.
\pagebreak
%\clearpage

% 3.
\pagebreak
%\clearpage

\end{document}